\documentclass[review]{elsarticle}
\usepackage{graphicx}
\usepackage{amssymb}
\usepackage{amsmath}
\usepackage{stackengine}
\usepackage{mathrsfs}
\usepackage{mathtools}
\usepackage{float}
\usepackage{subcaption}
\usepackage{caption}
\usepackage{makecell}
\usepackage{xcolor}
\usepackage{lineno}
\usepackage[]{hyperref}
\usepackage[ruled,vlined]{algorithm2e}
\usepackage[scale=0.7]{geometry}
\modulolinenumbers[2]

\journal{Journal of ++++++}

\begin{document}

\begin{frontmatter}

\title{Aerodynamic Shape Optimization of unsteady flapping wings}

\author[a]{Hamid R. Karbasian\corref{mycorrespondingauthor}}
\cortext[mycorrespondingauthor]{Corresponding author}
\ead{hkarbasian@smu.edu}

\author[b]{Brian C. Vermeire}

\address[a]{Artificial Intelligence for Design (AI4D) Laboratory, Department of Mechanical Engineering, Southern Methodist University, Dallas, TX, 75275}
\address[b]{Department of Mechanical, Industrial and Aerospace Engineering, Concordia University, Montreal, Canada.}

\begin{abstract}
This study introduces a new approach for building surrogate models, in the form of Reduced Order Models (ROMs), for PDE-constrained optimization. A physics-constrained data-driven framework is developed to transform large-scale non-linear optimization problems into ROM-constrained optimization problems. Unlike conventional methods, which suffer from instability of the forward sensitivity function, the proposed approach projects optimization problems to system dynamics optimization problems in Hilbert space to increase stability, diminish memory requirements, and lower computational cost. The present approach is applied to shape optimization of a 2D flapping wing section at $Re=2,400$, where the non-linear interaction of the wing with fluid includes chaotic flow structures. It is shown that the proposed approach can significantly enhance the time-averaged thrust force and propulsive efficiency by 37\% and 40\%, respectively. We also tested this approach for a 3D flapping wing and the results reveal the potential of the proposed approach for more complex optimal designs in the aerospace industry.
\end{abstract}


\end{frontmatter}
\linenumbers

\section{Introduction}
\label{sec:intro}
Unsteady flows around moving bodies are among the most challenging design problems in fluid dynamics, particularly when the flow is chaotic. For example, birds oscillate their wings and produce complicated flow structures, which help generate more lift than a stationary wing. This type of problem involves dynamic stall, a phenomenon that also matters for wind turbines, helicopters, and aircraft take-off and landing. 
From a mathematical standpoint, optimizing such wing (or blade) is challenging because of the chaotic, turbulent flow and the complex vortical structures in their wake. Consequently, there is still no robust and well-established framework to optimize such systems.

Sensitivity analysis, which computes the derivatives of an output of a dynamical system with respect to its inputs, is the core of gradient-based design, optimization, control, and uncertainty quantification \cite{jameson1988aerodynamic, le1986variational, gunzburger2003perspectives, najm2009uncertainty}. Conventional sensitivity analysis, such as the adjoint method, is widely used for steady-state problems \cite{skinner2018state}, and is routinely coupled with low-fidelity Reynolds-Averaged Navier--Stokes (RANS), Unsteady RANS (URANS), and potential-flow solvers. Higher-fidelity simulations, such as Large Eddy Simulation (LES), Direct Numerical Simulation (DNS), and Detached Eddy Simulation (DES), are becoming more prevalent owing to algorithmic improvements and advances in computing power. Unlike RANS, LES and DNS resolve the fine structures that carry the chaotic behaviour of turbulent flows \cite{blonigan2014least}. However, conventional sensitivity analysis breaks down when applied to time-averaged quantities of such chaotic systems \cite{blonigan2017toward}. Because of the sensitive dependence of turbulent flows on initial conditions, commonly referred to as the \textit{Butterfly Effect} \cite{blonigan2014least}, small perturbations grow along at least one unstable mode with a positive Lyapunov exponent. The instantaneous flow then diverges, finite time-averaging introduces noise into the objective function, and the computed sensitivities become unusable \cite{wang2014least}. Therefore, gradient-based optimization with high-fidelity CFD is currently impractical for chaotic flows, and gradient-based aerodynamic design remains largely limited to RANS, which keeps the sensitivity functions stable \cite{slotnick2014cfd}.

\begin{figure}
	\centering
	\begin{subfigure}[h]{0.45\textwidth}
		\centering
		\includegraphics[width=1\textwidth]{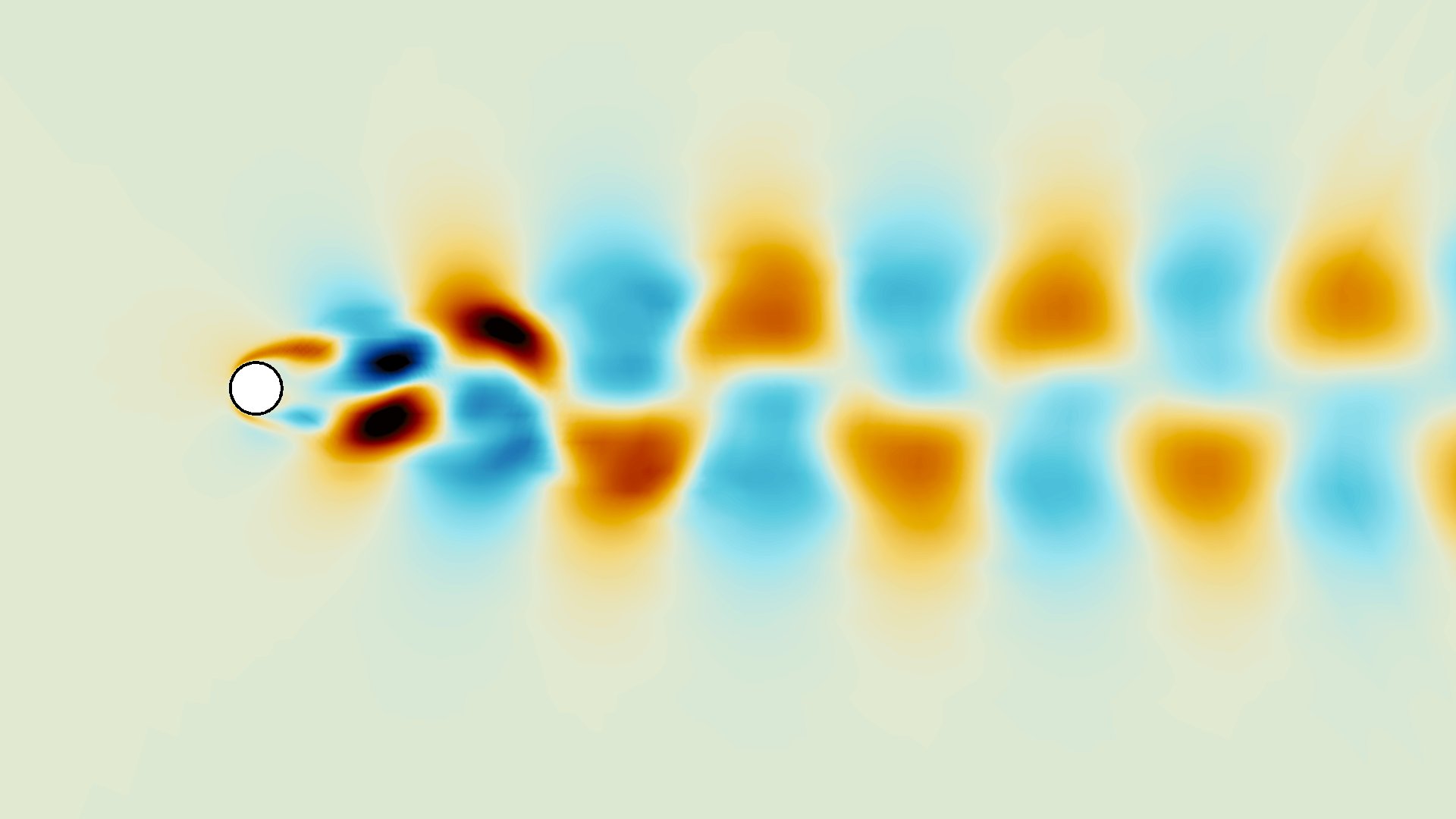}
		\caption{Beginning of analysis}
		\label{fig:cyl-example_a}
	\end{subfigure}
	\begin{subfigure}[h]{0.45\textwidth}
		\centering
		\includegraphics[width=1\textwidth]{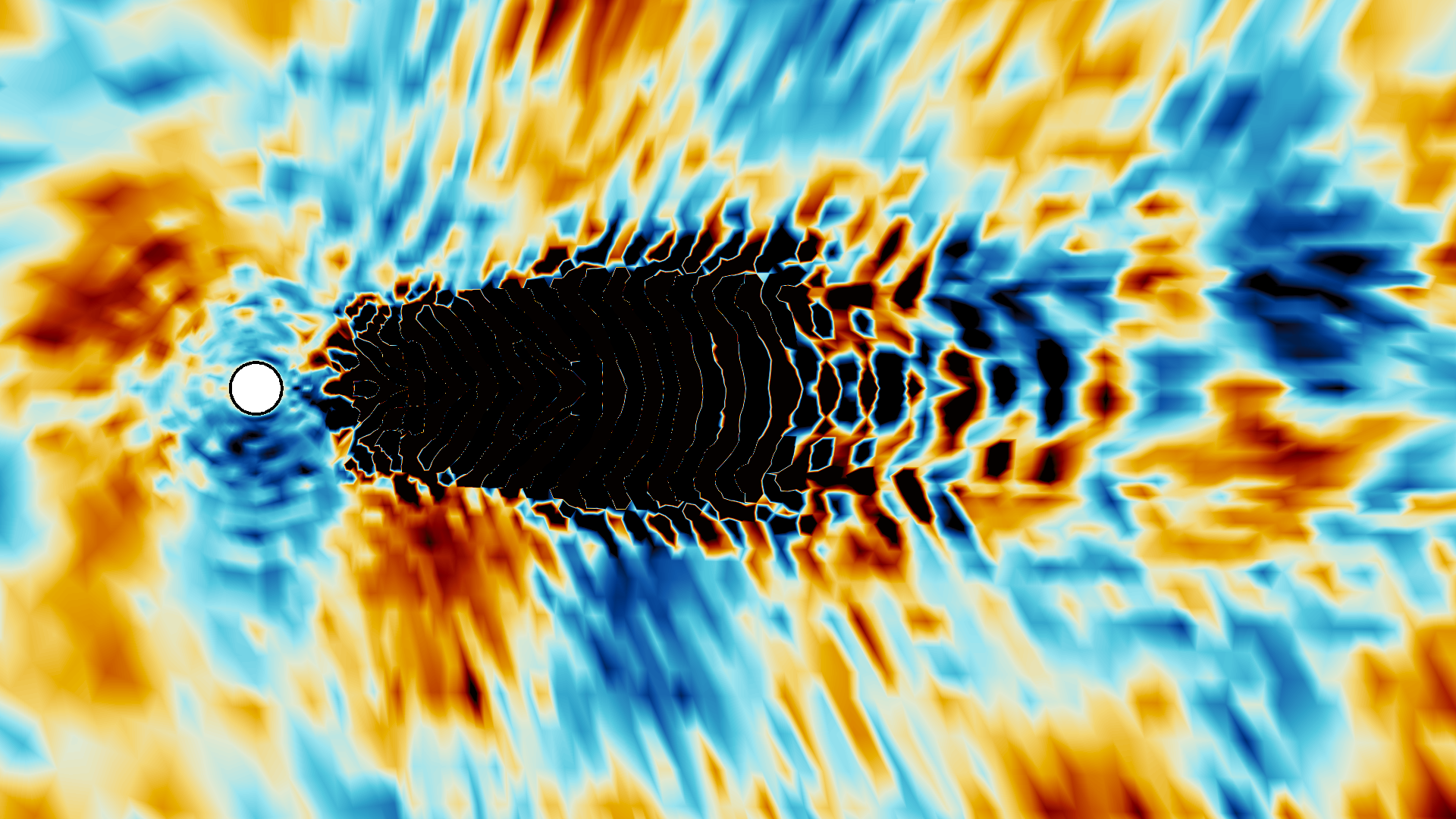}
		\caption{End of analysis}
		\label{fig:cyl-example_b}
	\end{subfigure}
	\caption{Contours of sensitivity of $x$-momentum with respect to $Re$ at the beginning and end of sensitivity analysis.}
	\label{fig:cyl-example}
\end{figure}

Even for periodic flows, forward sensitivities grow in time, because a perturbation of a design parameter shifts the period of the solution. Figure \ref{fig:cyl-example} illustrates this for the flow past a circular cylinder at $Re=150$, showing the sensitivity of the $x$-momentum with respect to the Reynolds number. At the beginning of the analysis, the sensitivity is confined to the near wake (Figure \ref{fig:cyl-example_a}), whereas by the end of the analysis the error has propagated through the whole domain (Figure \ref{fig:cyl-example_b}). For chaotic flows, this growth is exponential, and the problem is correspondingly more severe.

Two families of methods have been used to optimize such systems. The first is gradient-free methods, which do not require sensitivities and are therefore unaffected by their divergence. Genetic Algorithms (GA) \cite{yamazaki2008aerodynamic, castillo2008soft} and heuristic search methods, such as Generalized Pattern Search (GPS) \cite{bogani2009generalized} and Mesh Adaptive Direct Search (MADS) \cite{audet2006mesh, audet2010mesh, marsden2007trailing}, are popular examples. MADS is particularly promising, since each design cycle only requires the values of the objective function at several design points \cite{audet2010mesh}, and it has been combined with LES \cite{marsden2007trailing}. However, unlike the adjoint method, the number of simulations per design cycle grows with the number of design parameters, so gradient-free optimization with high-fidelity solvers is limited to relatively few parameters. The second family restores meaningful gradients for chaotic systems. Among these, Least Squares Shadowing (LSS) \cite{wang2014least} computes bounded sensitivities of time-averaged quantities by seeking a perturbed trajectory that remains close to the reference one, and it has been applied to turbulent and separated flows \cite{blonigan2014least, blonigan2018least}. However, LSS couples all time steps in a single Karush--Kuhn--Tucker (KKT) system, so the Jacobians at every time step must be stored simultaneously. For a large-scale simulation, this requires terabytes of memory, which has confined LSS to relatively small problems.

Reduced-order models (ROMs) offer a route to lower this cost. Projection-based ROMs approximate the state in a low-dimensional subspace, typically spanned by Proper Orthogonal Decomposition (POD) modes, and solve projected governing equations in that subspace. Least-squares Petrov--Galerkin (LSPG) projection minimizes the full-order residual at each time step and is more accurate and robust than Galerkin projection for nonlinear fluid problems \cite{carlberg2017galerkin}, while hyper-reduction further lowers the cost of evaluating the projected terms \cite{carlberg2013gnat}. For optimization, ROMs have been used as surrogates within PDE-constrained frameworks, where enriching the basis with sensitivity information makes it accurate for gradient computations \cite{carlberg2008compact, carlberg2011low, huynh2007reduced}, and where the basis is updated progressively as the design evolves \cite{zahr2015progressive}. To the authors' knowledge, however, ROMs have not yet been used to make shadowing-based sensitivity analysis affordable, and there is no reliable framework for unsteady aerodynamic optimization in the presence of chaotic flow.

The present study removes this barrier by solving the LSS problem in a low-dimensional subspace built from data, which enables gradient-based optimization of highly non-linear, chaotic flows. The contributions are: (1) an LSS formulation projected onto an LSPG reduced-order model, so that the size of the KKT system scales with the reduced dimension $r$ rather than the full dimension $n_{u}$; (2) a physics-constrained, data-driven procedure that enriches the reduced basis with steady and unsteady sensitivity snapshots; and (3) a demonstration on shape optimization of a 2D flapping wing and kinematics optimization of a 3D flapping wing, where the memory required for sensitivity analysis drops from about 1.5\,TB to about 4\,MB. The remainder of this paper is organized as follows. Section \ref{sec:opt-setup-2d-flappingwing} introduces the governing equations and the PDE-constrained optimization problem. Sections \ref{sec:least square shadowing} and \ref{sec:lagrange multiplier} present the LSS formulation and its KKT system. Section \ref{sec:lspg} projects this system onto an LSPG reduced-order model, and Section \ref{sec:learning-strategy} describes the data-driven construction of the reduced basis. Section \ref{sec:overview} summarizes the complete framework, Section \ref{sec:results} presents the optimization results, and Section \ref{sec:conclusions} concludes the paper.

\subsection{Kinematics and aerodynamic characteristics}
\label{subsec:kinematics}
In flapping wing problems, the reduced frequency, $k_{a}$, is a non-dimensional parameter, representing a relation between the oscillational velocity of the wing and free-stream velocity, $u_{\infty}$. We can define this reduced frequency as $k_{a}=\pi f_{a} c/u_{\infty}$, where $f_{a}$ is the flapping frequency, and $c$ is the chord length. We also define the Reynolds number according to the chord length of the wing as $Re=\rho_{\infty}c u_{\infty}/\mu$, where $\rho_{\infty}$ and $\mu$ are free-stream fluid density and viscosity, respectively.

\begin{figure}[H]
	\centering
	\includegraphics[width=0.22\textwidth]{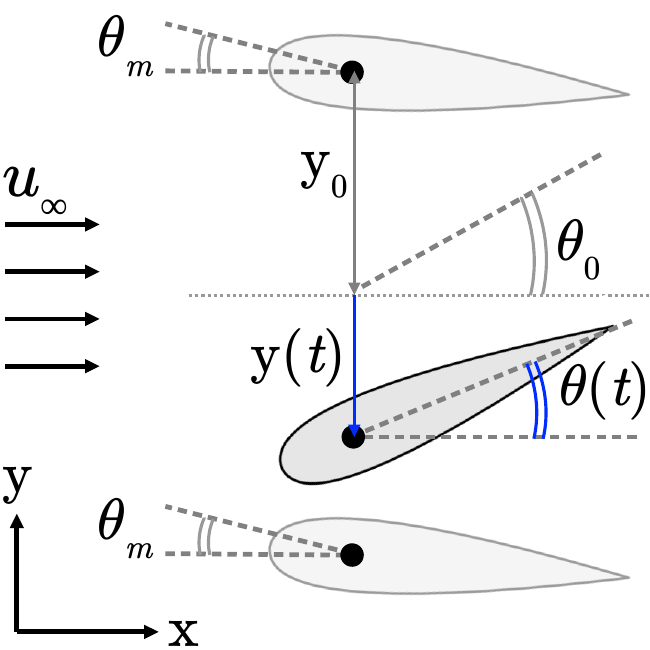}
	\caption{Kinematics of a flapping wing in cartesian coordinates. }
	\label{fig-flapp-sketch}
\end{figure}

In the present work, the kinematics of a wing section, $\mathcal{X}(t, \mathcal{S})=\{x \in \mathbb{R}^{4} \hspace{3pt} | \hspace{3pt} x \subset \mathscr{D} \}$ is considered in cartesian coordinates, where $\mathcal{S}$ is a set of design parameters, and $\mathscr{D}$ presents design space. The translation and pitching functions of a wing can be defined as
\begin{equation}
	\mathcal{X}(t, \mathcal{S}) = 
	\begin{bmatrix} 
		\text{x}(t)\\
		\text{y}(t)\\
		\text{z}(t)\\
		\theta(t)\\
	\end{bmatrix}
	=
	\begin{bmatrix} 
		0\\
		\text{y}_{0} \sin(2\pi f_{a}t)\\
		0\\
		\theta_{m} + \theta_{0} \sin(2\pi f_{a}t + \theta_{s})\\
	\end{bmatrix}
	\label{eq1}
\end{equation}
where, $\text{y}_{0}$ is the plunging amplitude, normally represented in non-dimensional form, $\text{y}_{0}/c$. Additionally, $\theta_{m}$ denotes the mean pitching angle, $\theta_{0}$ is the pitching amplitude, and $\theta_{s}$ is the shift angle between the plunging and pitching motions. The pivot point/line is located at the center of pressure $\text{x}_{cp}=0.25c$ to reduce the effect of aerodynamic loads on the pitching moment. Figure \ref{fig-flapp-sketch} shows a schematic of the flapping wing that uses both translation and pitching functions. A combination of the up-stroke and down-stroke phases for one flapping cycle is given by $T^{*}=1/f_{a}$. 

To investigate the aerodynamic performance of a flapping wing, aerodynamic loads are non-dimensionalized in the form of lift and thrust coefficients
\begin{equation}
	\begin{split}
		C_{L} &=\frac{F_{L}}{0.5\rho_{\infty}c u_{\infty}^{2}},\\
		C_{T} &=\frac{F_{T}}{0.5\rho_{\infty}c u_{\infty}^{2}},\\
	\end{split}
	\label{eq-aerodyn}
\end{equation}
where $F_{L}$ and $F_{T}$ are lift and thrust, respectively. Additionally, the power consumed by the flapping wing is approximated by the power coefficient
\begin{equation}
	C_{P} \approx \frac{F_{L}\dot{\text{y}}(t)}{0.5\rho_{\infty}c u_{\infty}^{3}},
	\label{eq-cp}
\end{equation}
where $\dot{\text{y}}(t)$ denotes translational velocity of the wing in $\text{y}$ direction. Here, we neglect the power consumed by pitching moment, since the pivot point is located at the center of pressure. Finally, using Eq. (\ref{eq-aerodyn}) and Eq. (\ref{eq-cp}), the propulsive efficiency is given by
\begin{equation}
	E_{p} = \frac{\overline{C_{T}}}{\overline{C_{P}}}.
	\label{eq-peff}
\end{equation}

\section{Governing Dynamical System}
\label{sec:opt-setup-2d-flappingwing}
An arbitrary high-dimensional Full-Order Model (FOM), represents the governing equations of a dynamical system in nature. The evolution of this FOM can be represented by a system of PDEs, with known initial and boundary conditions, such that the dynamical evolution of attractors happens in a smooth time domain $\mathbb{T}$. We consider a FOM with dimension of $n_{u}$, evolving with time $t \in \mathbb{T}$ with the initial value of $\textbf{\textit{u}}_{0} \in \mathbb{R}^{n_{u}}$ at $t=0$. This FOM can be described as a set of primal PDEs in space with some possible residual due to spatial discretization
\begin{equation}
	\textbf{\textit{r}} = \frac{\partial \textbf{\textit{u}}}{\partial t} + \nabla.\textbf{\textit{f}}(\textbf{\textit{u}}, \nabla \textbf{\textit{u}}, \mathcal{S}) =0,\quad \textbf{\textit{u}}(0)=\textbf{\textit{u}}_{0}, \quad t \in \mathbb{T}, 
	\label{eq:dynsys}
\end{equation}
where, $\textbf{\textit{u}}:\mathbb{T} \rightarrow \mathbb{R}^{n_{u}}$ is the state vector with dimension of $n_{u}$, and $\nabla.\textbf{\textit{f}}$ denotes divergence of the flux $\textbf{\textit{f}}: \mathbb{R}^{n_{u}} \rightarrow \mathbb{R}^{n_{u}}$, and $\textbf{\textit{r}}:\mathbb{R}^{n_{u}} \rightarrow \mathbb{R}^{n_{u}}$ is the residual. Additionally, $\mathcal{S}=[s_{1}, s_{2},\dots, s_{n_{s}}]^{\top} \subset \mathscr{D}$ is a set of design parameters in the design space $\mathscr{D} \in \mathbb{R}^{n_{s}\times n_{s}}$, where $n_{s}$ is the number of design parameters. Here $x^{\top}$ denotes transpose of variable $x$. Additionally, the dynamical behaviour of the primal PDE evolves over $\mathbb{T}$, such that $\nabla.\textbf{\textit{f}}: \textbf{\textit{u}}\times \mathbb{T} \rightarrow \textbf{\textit{u}}$.

\subsection{PDE-constrained optimization}
\label{subsec:pde-opt}
In PDE-constrained optimization, the objective function, $\mathcal{J}(\textbf{\textit{u}}, \mathcal{S}): \mathbb{R}^{n_{u}}\rightarrow \mathbb{R}$, is constrained by the primal PDEs. Since the time-averaged value of the objective function, $\overline{\mathcal{J}}$, is of interest in engineering problems, primal PDEs are considered in the range of $[0, T] \in \mathbb{T}$, where $T \in \mathbb{R}_{+}$ denotes the final solution time. Here $\overline{x}$ denotes the time-averaged of $x$. Additionally, in order to reduce the magnitude of $\overline{\mathcal{J}}$, the optimizer seeks better design parameters in the design space, $\mathcal{S}=[s_{1}, s_{2}, \dots s_{n_{s}}]^{\top} \subset \mathscr{D}$, at each design cycle. This optimization problem can be defined as
\begin{equation}
	\begin{split}
		\underset{\tiny \makecell{\textbf{\textit{u}} \in \mathbb{R}^{n_{u}}\\ \mathcal{S} \in \mathscr{D}}  }{\text{minimize}} \quad &\overline{\mathcal{J}} = \frac{1}{T} \int_{0}^{T}\mathcal{J}(\textbf{\textit{u}},t,\mathcal{S}) dt,\\
		\text{subject to} \quad &\frac{\partial \textbf{\textit{u}}}{\partial t} + \nabla.\textbf{\textit{f}}(\textbf{\textit{u}}, \nabla \textbf{\textit{u}}, \mathcal{S}) =0, \quad t \in \mathbb{T},\\
		&\textbf{C}(\textbf{\textit{u}}, \mathcal{S}) \leq 0,\\
	\end{split}
	\label{eq:pde-min-cont1}
\end{equation}
where $\textbf{C}$ represents constraint functions that can be dependent on either the state vector or design parameters. For optimization, we need to find the sensitivity of the state vector with respect to the set of design parameters, $\frac{\partial \textbf{\textit{u}}}{\partial \mathcal{S}}: \mathbb{R}^{n_{u}} \times \mathbb{T} \rightarrow \mathbb{R}^{n_{u}\times n_{s}}$, which can be obtained by dual representation of the primal PDEs
\begin{equation}
	\frac{d}{d \mathcal{S}} \bigg(\frac{\partial \textbf{\textit{u}}}{\partial t} + \nabla.\textbf{\textit{f}}\bigg)=0, \quad t \in \mathbb{T}.
	\label{eq:dual-pde-form}
\end{equation}
Let us suppose that the primal PDEs are the Navier-Stokes equations. In this case, a set of the dual PDEs is the ``forward sensitivity" function that describes the evolution of the sensitivity solution with respect to the design parameters over $\mathbb{T}$. Therefore, the forward sensitivity function, in the form of dual PDEs, is given by
\begin{equation}
	\frac{\partial \textbf{\textit{v}}}{\partial t} + \frac{\partial \nabla.\textbf{\textit{f}}}{\partial \textbf{\textit{u}}} \textbf{\textit{v}} + \frac{\partial \nabla.\textbf{\textit{f}}}{\partial \mathcal{S}}=0, \quad \textbf{\textit{v}}(0)=\textbf{\textit{v}}_{0}, \quad t \in \mathbb{T},
	\label{eq:get_v1}
\end{equation}
where, $\textbf{\textit{v}}: \mathbb{R}^{n_{u}} \times \mathbb{T} \rightarrow \mathbb{R}^{n_{u}\times n_{s}}$ is the sensitivity solution, and $\textbf{\textit{v}}_0 \in \mathbb{R}^{n_{u}\times n_{s}}$ is the initial condition. Moreover, Eq. (\ref{eq:get_v1}) illustrates that the sensitivity solution has its own specific dynamical features, which are connected to those of the state vector in Eq. (\ref{eq:dynsys}). Additionally, Eq. (\ref{eq:get_v1}) is advanced in time with $\textbf{\textit{v}}\times0 \rightarrow \textbf{\textit{v}}_{0}$ as the initial condition at time $0 \in \mathbb{T}$. Therefore, $\textbf{\textit{v}}$ is often called the forward sensitivity solution.

According to Eq. (\ref{eq:pde-min-cont1}), in order to find the time-averaged sensitivity of the objective function with respect to $\mathcal{S}$, the gradient of the objective function is computed by
\begin{equation}
	\frac{\delta \mathcal{J}}{\delta \mathcal{S}}
	=\Big[\frac{\partial \mathcal{J}}{\partial \textbf{\textit{u}}}\Big]^{\top} \textbf{\textit{v}}
	+ \frac{\partial \mathcal{J}}{\partial \mathcal{S}},
	\label{eq:grad_obj}
\end{equation}
where $\frac{\partial \mathcal{J}}{\partial \textbf{\textit{u}}}: \mathbb{R}^{n_{u}} \rightarrow \mathbb{R}^{n_{u}\times n_{s}}$. Substituting Eq. (\ref{eq:grad_obj}) into Eq. (\ref{eq:pde-min-cont1}), we can obtain the time-averaged sensitivity of the objective function by
\begin{equation}
	\overline{\frac{\Delta \mathcal{J}}{\Delta \mathcal{S}}}
	=\frac{1}{\textit{T}}\int_{0}^{\textit{T}} \bigg(
	\Big[\frac{\partial \mathcal{J}}{\partial \textbf{\textit{u}}}\Big]^{\top} \textbf{\textit{v}}
	+ \frac{\partial \mathcal{J}}{\partial \mathcal{S}} \bigg) dt \quad
	\text{s.t.} \quad \frac{d}{d \mathcal{S}} \bigg(\frac{\partial \textbf{\textit{u}}}{\partial t} + \nabla.\textbf{\textit{f}}\bigg) =0.
	\label{eq:final_grad_sensitivity} 
\end{equation}
Therefore, the final formulation of a PDE-constrained optimization using the forward sensitivity function is
\begin{equation}
	\begin{split}
		\underset{\tiny \makecell{\textbf{\textit{u}}, \textbf{\textit{v}} \in \mathbb{R}^{n_{u}}\\ \mathcal{S} \in \mathscr{D}} }{\text{minimize}} \quad &\overline{\mathcal{J}} = \frac{1}{T} \int_{0}^{T}\mathcal{J}(\textbf{\textit{u}},t,\mathcal{S}) dt,\\
		\text{subject to} \quad &\frac{\partial \textbf{\textit{u}}}{\partial t} + \nabla.\textbf{\textit{f}} =0, \quad t \in \mathbb{T},\\
		&\frac{\partial \textbf{\textit{v}}}{\partial t} + \frac{\partial \nabla.\textbf{\textit{f}}}{\partial \textbf{\textit{u}}} \textbf{\textit{v}} + \frac{\partial \nabla.\textbf{\textit{f}}}{\partial \mathcal{S}}
		=0, \quad \textbf{\textit{v}}(0)= \textbf{\textit{v}}_{0}, \quad t \in \mathbb{T},\\
		&\textbf{C}(\textbf{\textit{u}}, \mathcal{S}) \leq 0.\\
	\end{split}
	\label{eq:pde-min-cont2}
\end{equation}
The objective of this study is to show how to compute $\overline{\frac{\Delta \mathcal{J}}{\Delta \mathcal{S}}}$, which is the primary challenge in sensitivity analysis and uncertainty quantifications of chaotic problems. Note that the minimization set in Eq. (\ref{eq:pde-min-cont2}) requires a Hessian matrix, which can be approximated via a Quasi-Newton method series, such as Broyden-Fletcher-Goldfarb-Shanno (BFGS), and Conjugate Gradient (CG), developed for gradient-based optimizations.

\section{Least Squares Shadowing (LSS)}
\label{sec:least square shadowing}
\begin{figure}
	\centering
	\begin{subfigure}[b]{0.4\textwidth}
		\centering
		\includegraphics[width=0.9\textwidth]{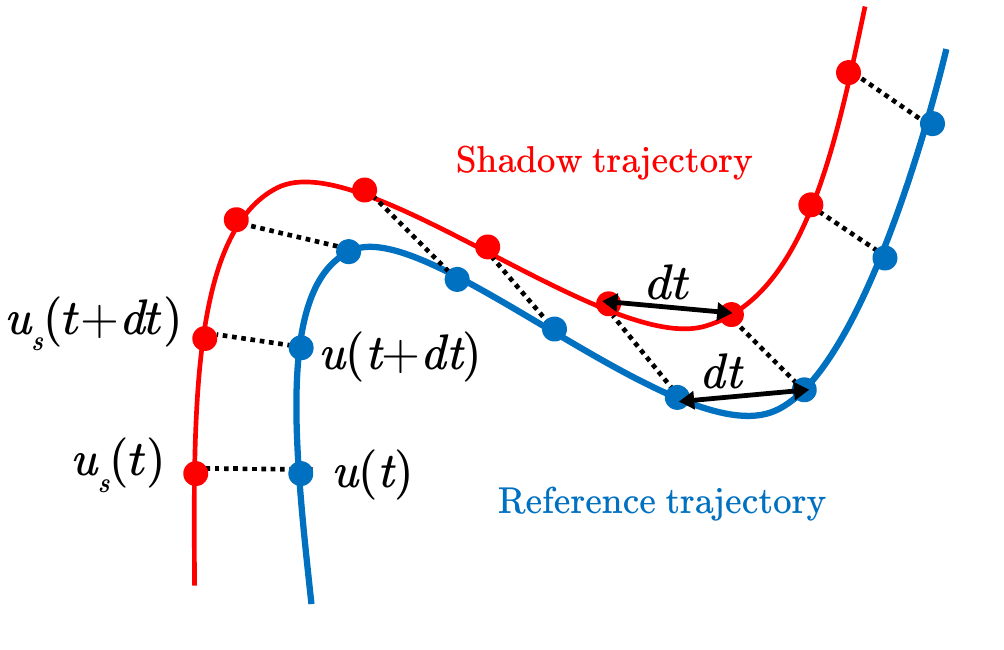}
		\caption{Without time dilation}
		\label{fig:lss-traja}
	\end{subfigure}
	\begin{subfigure}[b]{0.4\textwidth}
		\centering
		\includegraphics[width=0.9\textwidth]{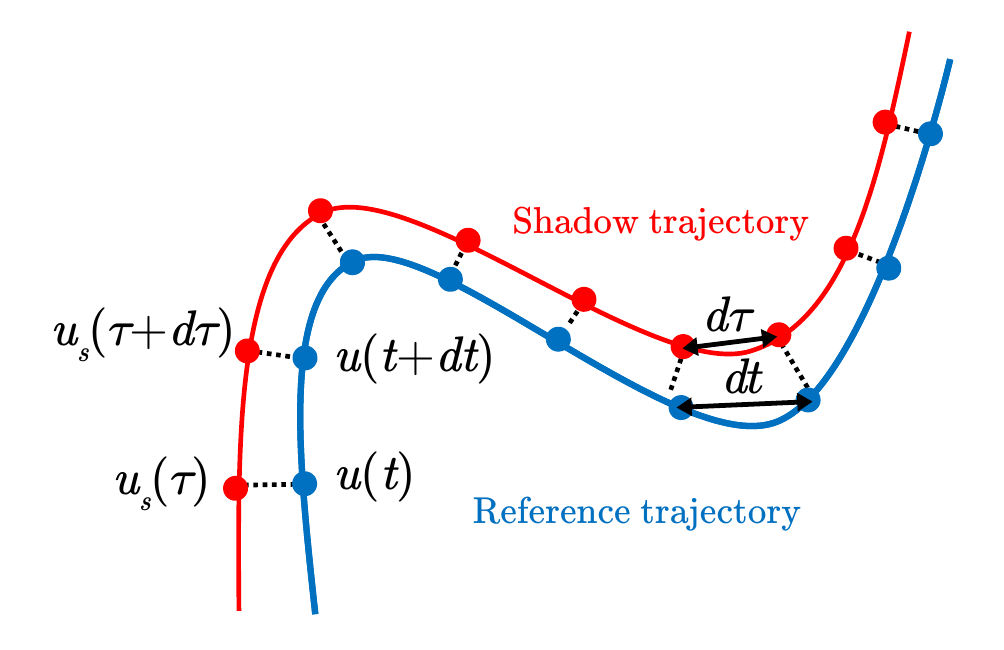}
		\caption{With time dilation}
		\label{fig:lss-trajb}
	\end{subfigure}
	\caption{Schematic of shadow and reference trajectories.}
	\label{fig:lss-traj}
\end{figure}

Least Squares Shadowing (LSS), proposed by Wang et al.~\cite{wang2014least}, is an efficient method to evaluate sensitivities of chaotic dynamical systems \cite{wang2014least}. LSS leverages \textit{shadowing lemma}, and ergodicity of dynamical system to solve these sensitivities \cite{blonigan2014least}. According to the \textit{shadowing lemma}, there is one perturbed solution for any ergodic dynamical system, such that this perturbed solution sticks as close as possible to the reference one over time domain $\mathbb{T}$. Figure \ref{fig:lss-traj} shows a schematic of a shadow solution that tracks a reference trajectory for two different conditions. In this figure, $\textbf{\textit{u}}: \mathbb{T} \rightarrow \mathbb{R}^{n_{u}}$ is a reference solution, evolving with physical time $t \in \mathbb{T}$, and $dt \in \mathbb{R}_{+}$ denotes the time-step. Additionally, $\tau \in \mathbb{T}$ is a time transformation for a perturbed dynamical system, and $d\tau \in \mathbb{R}_{+}$ is the time-step to advance shadow solution with time, $\textbf{\textit{u}}_{s}: \mathbb{T} \rightarrow \mathbb{R}^{n_{u}}$. In Figure \ref{fig:lss-traja}, shadow and reference trajectories have the same discrete time-steps, $d\tau=dt$, yielding low accuracy results for LSS approach. This is because the distances between these solutions are not minimum over $\mathbb{T}$. On the other hand, as shown in Figure \ref{fig:lss-trajb}, shadow solution evolves with time transformation, yielding a flexible solution in physical time that remains close to the reference one. In other words, this time transformation applies an orthogonal constraint to the perturbed dynamical system, such that the distance lines, connecting shadow and reference solutions at each time interval, become perpendicular to both trajectories over $\mathbb{T}$. LSS particularly seeks for a new solution with different initial condition that has the lowest loss function, $\frac{1}{T}\int_{0}^{T}\|\textbf{\textit{u}}_{s} - \textbf{\textit{u}} \|_2^{2}dt$, where $[0, T] \in \mathbb{T}$ is a time window with final time $T \in \mathbb{R}_{+}$. Therefore, this approach paves the way to consider sensitivity analysis, uncertainty quantification, and large-scale PDE-constrained optimization of chaotic dynamical systems.

\subsection{Shadow solution and partial derivatives}
\label{sec:partial derivatives}
According to the \textit{shadowing lemma}, a state vector with a significant perturbation can not be considered as the ``shadow'' of the reference state vector. Three criteria should be taken into account for this purpose: (1) when $\delta \mathcal{S} \rightarrow \epsilon$, there is only one unique shadow solution, $\textbf{\textit{u}}_{s}(\tau, \mathcal{S}+\delta\mathcal{S}): \mathbb{T} \rightarrow \mathbb{R}^{n_{u}}$; (2) the behaviour of $\textbf{\textit{u}}_{s}(\tau, \mathcal{S}+\delta\mathcal{S})$ remains in the flux of the augmented dynamical system, $\textbf{\textit{f}}_{s}: \mathbb{R}^{n_{u}} \rightarrow \mathbb{R}^{n_{u}}$, such that it evolves in $\tau \in \mathbb{T}$ with $\tau \neq t$; and (3) the shadow solution satisfies $\| \textbf{\textit{u}}_{s}(\tau, \mathcal{S}+\delta\mathcal{S}) - \textbf{\textit{u}}(t, \mathcal{S}) \|_{2} \leq \varepsilon$, where $\varepsilon \in \mathbb{R}_{+}$ is the maximum distance between the reference and augmented state vectors. These criteria ensure the existence of a shadow for the reference state vector. Therefore, the sensitivity of the reference state with respect to a set of design parameters can be defined as
\begin{equation}
	\textbf{\textit{v}}(t, \mathcal{S}) = \frac{\partial \textbf{\textit{u}}}{\partial \mathcal{S}} \approx \lim_{\delta \mathcal{S} \rightarrow 0} \frac{1}{\delta \mathcal{S}} \Big( \textbf{\textit{u}}_{s}(\tau, \mathcal{S}+\delta\mathcal{S}) - \textbf{\textit{u}}(t, \mathcal{S}) \Big).
	\label{eq:partialstate}
\end{equation}
Rewriting Eq. (\ref{eq:partialstate}) in the form of a Taylor series, neglecting higher-order terms, and assuming $\delta \mathcal{S} \rightarrow 0$, the shadow solution of the reference state vector is
\begin{equation}
	\textbf{\textit{u}}_{s}(\tau, \mathcal{S}+\delta\mathcal{S}) = \textbf{\textit{u}}(t, \mathcal{S}) + \delta\mathcal{S}\, \textbf{\textit{v}}(t, \mathcal{S}).
	\label{eq:shadowstate}
\end{equation}
Finding the derivative of Eq. (\ref{eq:shadowstate}) with respect to $t \in \mathbb{T}$, the velocity of the state vector for the augmented dynamical system is
\begin{equation}
	\lim_{\delta \mathcal{S} \rightarrow 0} \frac{\partial \textbf{\textit{u}}_{s}}{\partial t} = -\nabla.\textbf{\textit{f}}(\textbf{\textit{u}}, t, \mathcal{S}) + \frac{\partial \textbf{\textit{v}}}{\partial t} \delta \mathcal{S},
	\label{eq:velocitystate}
\end{equation}
which shows that the augmented dynamical system is proportional to the velocity of the sensitivity solution, $\partial \textbf{\textit{v}}/\partial t$. Similarly, by definition of the shadow solution, the partial derivative of Eq. (\ref{eq:dynsys}) is
\begin{equation}
	\frac{d \nabla.\textbf{\textit{f}}}{d \mathcal{S}} = \lim_{\delta \mathcal{S} \rightarrow 0} \frac{1}{\delta \mathcal{S}} \Big( \nabla.\textbf{\textit{f}}_{s}(\textbf{\textit{u}}_{s}, \tau, \mathcal{S}+\delta\mathcal{S}) - \nabla.\textbf{\textit{f}}(\textbf{\textit{u}}, t, \mathcal{S}) \Big) = \frac{\partial \nabla.\textbf{\textit{f}}}{\partial \textbf{\textit{u}}} \textbf{\textit{v}} + \frac{\partial \nabla.\textbf{\textit{f}}}{\partial \mathcal{S}},
	\label{eq:dfds}
\end{equation}
and, similar to Eq. (\ref{eq:shadowstate}), the shadow of the dynamical system is
\begin{equation}
	\nabla.\textbf{\textit{f}}_{s}(\textbf{\textit{u}}_{s}, \tau, \mathcal{S}+\delta\mathcal{S}) = \nabla.\textbf{\textit{f}}(\textbf{\textit{u}}, t, \mathcal{S}) + \delta \mathcal{S} \bigg( \frac{\partial \nabla.\textbf{\textit{f}}}{\partial \textbf{\textit{u}}} \textbf{\textit{v}} + \frac{\partial \nabla.\textbf{\textit{f}}}{\partial \mathcal{S}} \bigg), \quad t \in \mathbb{T},
	\label{eq:shadowdynsys}
\end{equation}
where Eq. (\ref{eq:shadowdynsys}) notes that, regardless of $\tau \in \mathbb{T}$, $\nabla.\textbf{\textit{f}}_{s}$ can be computed in the physical time, $t \in \mathbb{T}$.

Based on the \textit{shadowing lemma}, the time transformation of the shadow solution is a function of the physical time, $\tau \equiv \eta(t)$, where $\eta \in \mathbb{R}$ is the time dilation defined by
\begin{equation}
	\eta(t) = \lim_{\delta \mathcal{S} \rightarrow 0} \frac{\delta \big( d\tau/dt - 1 \big)}{\delta \mathcal{S}}.
	\label{eq:eta}
\end{equation}
When $\delta \mathcal{S} \rightarrow 0$, the time dilation becomes negligible (i.e., $\eta \rightarrow 0$), and both the shadow and reference solutions embed in the same time domain (i.e., $\tau \approxeq t \in \mathbb{T}$). Integrating Eq. (\ref{eq:eta}) yields
\begin{equation}
	\frac{d\tau}{dt} = 1 + \eta(t)\, d\mathcal{S}.
	\label{eq:def_eta}
\end{equation}

\subsection{Least-Squares Shadowing formulation}
\label{sec:lss-formulation}
Least-Squares Shadowing (LSS) tries to find $\textbf{\textit{u}}_{s}(\tau, \mathcal{S}+\delta \mathcal{S})$ by minimizing the distance between the trajectories of the shadow solution and the reference state vector over $[0, T] \in \mathbb{T}$. Therefore, the Least-Squares minimization formulated for LSS is given by
\begin{equation}
	\begin{split}
		\underset{\textbf{\textit{u}}_{s} \in \mathbb{R}^{n_{u}}, \tau \in \mathbb{R} }{\text{minimize}} \quad & \lim_{\delta \mathcal{S} \rightarrow 0} \frac{1}{T}\int_{0}^{T} \| \textbf{\textit{u}}_{s}(\tau, \mathcal{S}+\delta \mathcal{S}) - \textbf{\textit{u}}(t, \mathcal{S}) \|^{2} + \alpha_{lss}^{2} \| \frac{d\tau}{dt}-1 \|^{2} dt,\\
		\text{subject to} \quad &\frac{\partial \textbf{\textit{u}}_{s}}{\partial \tau} + \nabla.\textbf{\textit{f}}_{s} =0,  \quad \tau \in \mathbb{T},\\
	\end{split}
	\label{eq:lss1}
\end{equation}
where $\alpha_{lss}$ is the weighting factor of the time dilation. When $\delta \mathcal{S} \rightarrow 0$, Eq. (\ref{eq:partialstate}) and Eq. (\ref{eq:eta}) can be used to reformulate Eq. (\ref{eq:lss1}) as
\begin{equation}
	\begin{split}
		\underset{\textbf{\textit{u}}_{s} \in \mathbb{R}^{n_{u}}, \tau \in \mathbb{R} }{\text{minimize}} \quad & \lim_{\delta \mathcal{S} \rightarrow 0} \frac{1}{T}\int_{0}^{T} \| \textbf{\textit{v}}(t, \mathcal{S}) \|^{2} + \alpha_{lss}^{2} \| \eta(t) \|^{2} dt,\\
		\text{subject to} \quad &\frac{\partial \textbf{\textit{u}}_{s}}{\partial \tau} + \nabla.\textbf{\textit{f}}_{s} =0,  \quad \tau \in \mathbb{T}.\\
	\end{split}
	\label{eq:lss2}
\end{equation}
The main challenge in Eq. (\ref{eq:lss2}) is that the augmented state vector evolves with $\tau \in \mathbb{T}$, and the shadow of the dynamical system is unknown. According to Section \ref{sec:partial derivatives}, the augmented state vector should be solved in the physical time, $t \in \mathbb{T}$. Using chain rule, the velocity of the state vector for the augmented dynamical system can be defined as
\begin{equation}
	\frac{\partial \textbf{\textit{u}}_{s}(\tau, \mathcal{S})}{\partial t}= \frac{\partial \tau}{\partial t} \frac{\partial \textbf{\textit{u}}_{s}(\tau, \mathcal{S})}{\partial \tau}.
	\label{eq:chain-rule}
\end{equation}
According to Eq. (\ref{eq:chain-rule}), Eq. (\ref{eq:shadowdynsys}) and Eq. (\ref{eq:def_eta}), the constraint function in Eq. (\ref{eq:lss2}) are
\begin{equation}
	\begin{split}
		\frac{\partial \textbf{\textit{u}}_{s}}{\partial t} + \nabla.\textbf{\textit{f}} + \eta \delta \mathcal{S} \nabla.\textbf{\textit{f}}
		&+ \delta \mathcal{S} \bigg(\frac{\partial \nabla.\textbf{\textit{f}}}{\partial \textbf{\textit{u}}} \textbf{\textit{v}} + \frac{\partial \nabla.\textbf{\textit{f}}}{\partial \mathcal{S}}\bigg) \\
		&+ \eta \delta \mathcal{S}^{2} \bigg(\frac{\partial \nabla.\textbf{\textit{f}}}{\partial \textbf{\textit{u}}} \textbf{\textit{v}} + \frac{\partial \nabla.\textbf{\textit{f}}}{\partial \mathcal{S}}\bigg)=0. \\
	\end{split}
	\label{eq:constriantlss}
\end{equation}
When $\delta \mathcal{S} \rightarrow 0$, the last term of Eq. (\ref{eq:constriantlss}) will be negligible, and then Eq. (\ref{eq:constriantlss}) can be simplified to
\begin{equation}
	\begin{split}
		\frac{\partial \textbf{\textit{u}}_{s}}{\partial t} &+ \nabla.\textbf{\textit{f}} + \delta \mathcal{S} \bigg(\frac{\partial \nabla.\textbf{\textit{f}}}{\partial \textbf{\textit{u}}} \textbf{\textit{v}} + \frac{\partial \nabla.\textbf{\textit{f}}}{\partial \mathcal{S}} + \eta \nabla.\textbf{\textit{f}}\bigg) + \mathcal{O}(\delta \mathcal{S}^2)=0.
	\end{split}
	\label{eq:constriantlss2}
\end{equation}
Therefore, by substituting Eq. (\ref{eq:constriantlss2}) into Eq. (\ref{eq:velocitystate}), the constraint function becomes
\begin{equation}
	\frac{\partial \textbf{\textit{v}}}{\partial t} + \frac{\partial \nabla.\textbf{\textit{f}}}{\partial \textbf{\textit{u}}} \textbf{\textit{v}} + \frac{\partial \nabla.\textbf{\textit{f}}}{\partial \mathcal{S}} + \eta \nabla.\textbf{\textit{f}}=0, \quad \textbf{\textit{v}}(0)=\textbf{\textit{v}}_{0},  \quad t \in \mathbb{T},\\
	\label{eq:constraintlss3}
\end{equation}
where, $\eta \nabla.\textbf{\textit{f}}$ shows the effect of time dilation on the augmented dynamical system. Note that if we set $\delta \mathcal{S} \rightarrow 0$, or neglect the effect of time dilation on the results (i.e., $\tau \approxeq t$), then $\eta \nabla.\textbf{\textit{f}}$ can be removed from Eq. (\ref{eq:constraintlss3}). In the end, the final formulation of Eq. (\ref{eq:lss2}) is given by
\begin{equation}
	\begin{split}
		\underset{\textbf{\textit{v}} \in \mathbb{R}^{n_{u}}, \eta \in \mathbb{R} }{\text{minimize}} \quad & \lim_{\delta \mathcal{S} \rightarrow 0} \frac{1}{T}\int_{0}^{T} \| \textbf{\textit{v}}(t, \mathcal{S}) \|_{2}^{2} + \alpha_{lss}^{2} \| \eta(t) \|_{2}^{2} dt\\
		\text{subject to} \quad &\frac{\partial \textbf{\textit{v}}}{\partial t} + \frac{\partial \nabla.\textbf{\textit{f}}}{\partial \textbf{\textit{u}}} \textbf{\textit{v}} + \frac{\partial \nabla.\textbf{\textit{f}}}{\partial \mathcal{S}} + \eta \nabla.\textbf{\textit{f}}=0, \quad t \in \mathbb{T}.\\
	\end{split}
	\label{eq:pde-min-cont}
\end{equation}
According to Eq. (\ref{eq:pde-min-cont}), for a given perturbation, $\delta \mathcal{S} \rightarrow \epsilon$, LSS seeks to find the smallest values of $\textbf{\textit{v}}$ and $\eta$, guaranteeing the results of the augmented dynamical system are as close as possible to the shadow solution.
\section{Lagrange Multiplier}
\label{sec:lagrange multiplier}
The Lagrange function for Eq. (\ref{eq:pde-min-cont}) is
\begin{equation}
	\mathcal{L}=
	\frac{1}{T}\int_{0}^{T} \bigg( \textbf{\textit{v}}^{\top} \textbf{K} \textbf{\textit{v}} + \alpha_{lss}^{2}\eta^{2}\bigg)dt
	+\int_{0}^{T}\Lambda^{\top}\bigg( \frac{\partial \textbf{\textit{v}}}{\partial t} + \frac{\partial \nabla.\textbf{\textit{f}}}{\partial \textbf{\textit{u}}} \textbf{\textit{v}} + \frac{\partial \nabla.\textbf{\textit{f}}}{\partial \mathcal{S}} + \eta \nabla.\textbf{\textit{f}} \bigg) dt, \quad t \in \mathbb{T},
	\label{eq:lagrange1}
\end{equation}
where, $\Lambda(t) \in \mathbb{R}^{n_{n}}$ is the Lagrange multiplier, and $\textbf{K} \in \mathbb{R}^{n_{u}\times n_{u}}$ is a weighting matrix. By applying the Karush-Kuhn Tucker (KKT) condition, the solution of the Lagrange function is
\begin{equation}
	\begin{split}
		&\frac{\partial \textbf{\textit{v}}}{\partial t} + \frac{\partial \nabla.\textbf{\textit{f}}}{\partial \textbf{\textit{u}}} \textbf{\textit{v}} + \frac{\partial \nabla.\textbf{\textit{f}}}{\partial \mathcal{S}} + \eta \nabla.\textbf{\textit{f}}=0, \\
		&\frac{\partial \Lambda}{\partial t} - \frac{\partial \nabla.\textbf{\textit{f}}}{\partial \textbf{\textit{u}}}^{\top} \Lambda -\frac{1}{T} \textbf{K}^{\top} \textbf{\textit{v}}=0, \quad \Lambda(0)=\Lambda(t=T)=0, \\
		&\frac{2}{T}\alpha_{lss}^2 \eta + \Lambda^{\top} \nabla.\textbf{\textit{f}}=0. \\
	\end{split}
	\label{eq:kkt-hdm}
\end{equation}
Solving Eq. (\ref{eq:kkt-hdm}) returns the sensitivity solution and the time dilation over $[0,T] \in \mathbb{T}$. However, Eq. (\ref{eq:kkt-hdm}) represents a set of non-linear ODEs that must be solved numerically in the discrete time domain, $\Delta \mathbb{T}$, such that $t^{(n)}=\{t^{(1)}, t^{(2)}, \dots, t^{(m_{u})} \}\in \Delta \mathbb{T}$, where $n$ denotes time index, and $m_{u}$ is the total number of time steps. To solve Eq. (\ref{eq:kkt-hdm}) numerically, we discretize it using the Backward Differentiation Formula (BDF), for which the fully-discrete residual of Eq. (\ref{eq:dynsys}) reads
\begin{equation}
	\textbf{\textit{r}}^{(n)} := \sum_{\textit{j}=0}^{\textit{k}} \zeta_j \textbf{\textit{u}}^{(n-j)} + \Delta t^{(n)} \beta_0 \nabla.\textbf{\textit{f}}^{(n)}\big(\textbf{\textit{u}}^{(n)}, t^{(n)}, \mathcal{S}\big), \quad t^{(n)} \in \Delta\mathbb{T},
	\label{eq:chaos-bdf2}
\end{equation}
where $\zeta_j$ and $\beta_0$ are the coefficients of the $k$-step BDF scheme. Therefore, the discrete form of Eq. (\ref{eq:kkt-hdm}) can be written as
\begin{equation}
	\begin{split}
		&\lim_{\delta \mathcal{S} \rightarrow 0 } \frac{d\textbf{\textit{r}}^{(n)}}{d \mathcal{S}}:=\sum_{\textit{j}=0}^{k}\zeta_j \textbf{\textit{v}}^{(n-j)} +\beta_0 \Delta t \frac{\partial \nabla.\textbf{\textit{f}}^{(n)}}{\partial \textbf{\textit{u}}^{(n)}} \textbf{\textit{v}}^{(n)}+ \Delta t \frac{\partial \nabla.\textbf{\textit{f}}^{(n)}}{\partial \mathcal{S}} + \eta^{(n)} \Delta t \nabla.\textbf{\textit{f}}^{(n)}=0,\\
		&\sum_{\textit{j}=0}^{k} \zeta_{j} \Lambda^{(n+j)} +\beta_0 \Delta t \bigg[\frac{\partial \nabla.\textbf{\textit{f}}^{(n)}}{\partial \textbf{\textit{u}}^{(n)}}\bigg]^{\top}\Lambda^{(n)} + \frac{1}{m_{u}} \textbf{K}^{\top} \textbf{\textit{v}}^{(n)}=0, \quad \Lambda^{(0)}=\Lambda^{(m_{u})}=0,\\
		&2\alpha_{lss}^2 \eta^{(n)} - m_{u} \Delta t \Lambda^{\top(n)} \nabla.\textbf{\textit{f}}^{(n)}=0.\\
	\end{split}
	\label{eq:disc-kkt1}
\end{equation}
Rearranging Eq. (\ref{eq:disc-kkt1}), the following formulation is obtained
\begin{equation}
	\begin{split}
		&\bigg(\zeta_0\mathcal{I} +\beta_0 \Delta t \frac{\partial \nabla.\textbf{\textit{f}}^{(n)}}{\partial \textbf{\textit{u}}^{(n)}}\bigg) \textbf{\textit{v}}^{(n)}
		+ \sum_{\textit{j}=1}^{k} \zeta_j\textbf{\textit{v}}^{(n-j)} + \beta_0 \Delta t\frac{\partial \nabla.\textbf{\textit{f}}^{(n)}}{\partial \mathcal{S}} + \eta^{(n)} \Delta t \nabla.\textbf{\textit{f}}^{(n)}=0,\\
		&\bigg(\zeta_0\mathcal{I}+\beta_0 \Delta t \bigg[\frac{\partial \nabla.\textbf{\textit{f}}^{(n)}}{\partial \textbf{\textit{u}}^{(n)}}\bigg]^{\top} \bigg)\Lambda^{(n)}+\sum_{\textit{j}=1}^{k} \zeta_{j} \Lambda^{(n+j)} + \frac{1}{m_{u}}\textbf{K}^{\top} \textbf{\textit{v}}^{(n)}=0, \quad \Lambda^{(0)}=\Lambda^{(m_{u})}=0,\\
		&2\alpha_{lss}^2 \eta^{(n)} - m_{u} \Delta t \Lambda^{\top(n)} \nabla.\textbf{\textit{f}}^{(n)}=0,\\
	\end{split}
	\label{eq:disc-kkt2}
\end{equation}
where, $\mathcal{I} \in  \mathbb{R}^{n_{u}\times n_{u}}$ is identity matrix. Finally, the residual form of Eq. (\ref{eq:disc-kkt2}) with $\frac{\partial \textbf{\textit{r}}^{(n)}}{\partial \textbf{\textit{u}}^{(n)}}$, $\frac{\partial \textbf{\textit{r}}^{(n)}}{\partial \textbf{\textit{u}}^{(p)}}$ (where, $p\neq n$), and $\frac{\partial \textbf{\textit{r}}^{(n)}}{\partial \mathcal{S}}$ is
\begin{equation}
	\begin{split}
		&\frac{\partial \textbf{\textit{r}}^{(n)}}{\partial \textbf{\textit{u}}^{(n)}} \textbf{\textit{v}}^{(n)} + \sum_{\textit{j}=1}^{k} \frac{\partial \textbf{\textit{r}}^{(n)}}{\partial \textbf{\textit{u}}^{(n-j)}} \textbf{\textit{v}}^{(n-j)} + \frac{\partial \textbf{\textit{r}}^{(n)}}{\partial \mathcal{S}} + \eta^{(n)} \Delta t \nabla.\textbf{\textit{f}}^{(n)}=0,\\
		&\bigg[\frac{\partial \textbf{\textit{r}}^{(n)}}{\partial \textbf{\textit{u}}^{(n)}}\bigg]^{\top}\Lambda^{(n)}+\sum_{\textit{j}=1}^{k} \bigg[\frac{\partial \textbf{\textit{r}}^{(n)}}{\partial \textbf{\textit{u}}^{(n+j)}}\bigg]^{\top} \Lambda^{(n+j)}+\textbf{K}^{\top} \textbf{\textit{v}}^{(n)}=0, \quad \Lambda^{(0)}=\Lambda^{(m_{u})}=0,\\
		&\alpha_{lss}^2 \eta^{(n)} - m_{u} \Delta t \Lambda^{\top(n)} \nabla.\textbf{\textit{f}}^{(n)}=0.\\
	\end{split}
	\label{eq:disc-kkt3}
\end{equation}
Note that because $\alpha_{lss}$ and $\textbf{K}$ are weighting parameters, their factors (i.e., 2 and $\frac{1}{m_{u}}$) are embedded. For simplicity, we solve Eq. (\ref{eq:disc-kkt3}) separately for each $s_{i} \in \mathcal{S}=\{s_{1}, s_{2}, \dots, s_{n_{s}}\}$. Consequently, Eq. (\ref{eq:disc-kkt3}) can be represented as a multi-block matrix
\begin{equation}
	\left[ \begin{array}{ccc}
		\mathcal{B}&\mathcal{C}&\text{\large \textbf{0}}\\
		\text{\large \textbf{0}}&\alpha_{lss}^2\mathcal{I}&m_{u}\mathcal{C}^{\top}\\
		\textbf{K}^{\top}&\text{\large \textbf{0}}&\mathcal{B}^{\top}
	\end{array} \right]
	\left[ \begin{array}{c}
		\textbf{\textit{v}}\\
		\eta\\
		\Lambda\\
	\end{array} \right] =-
	\left[ \begin{array}{c}
		\mathcal{D}\\
		\text{\large \textbf{0}}\\
		\text{\large \textbf{0}}\\
	\end{array} \right],
	\label{eq:kkt-mat2}
\end{equation}
where,
\begin{equation}
	\begin{split}
		&\mathcal{B}=\left[ \begin{array}{c}
			\begin{matrix} \frac{\partial \textbf{\textit{r}}^{(k)}}{\partial \textbf{\textit{u}}^{(1)}}&\frac{\partial \textbf{\textit{r}}^{(k)}}{\partial \textbf{\textit{u}}^{(2)}}&\dots&\frac{\partial \textbf{\textit{r}}^{(k)}}{\partial \textbf{\textit{u}}^{(k)}}\\&\frac{\partial \textbf{\textit{r}}^{(k+1)}}{\partial \textbf{\textit{u}}^{(2)}}&\frac{\partial \textbf{\textit{r}}^{(k+1)}}{\partial \textbf{\textit{u}}^{(3)}}&\dots&\frac{\partial \textbf{\textit{r}}^{(k+1)}}{\partial \textbf{\textit{u}}^{(k+1)}}\\&&\ddots&&\ddots&\ddots\\&&&\frac{\partial \textbf{\textit{r}}^{(m_{u})}}{\partial \textbf{\textit{u}}^{(m_{u}-k)}}&\frac{\partial \textbf{\textit{r}}^{(m_{u})}}{\partial \textbf{\textit{u}}^{(m_{u}-1)}}&\dots&\frac{\partial \textbf{\textit{r}}^{(m_{u})}}{\partial \textbf{\textit{u}}^{(m_{u})}}\end{matrix}
		\end{array} \right],\\
		&\mathcal{C}=\Delta t\left[\begin{array}{c}
			\begin{matrix}\nabla.\textbf{\textit{f}}^{(k)}\\&\nabla.\textbf{\textit{f}}^{(k+1)}\\&\ddots\\&&\nabla.\textbf{\textit{f}}^{(m_{u})} \end{matrix}
		\end{array}\right], \quad
		\mathcal{D}=\left[ \begin{array}{c}
			\begin{matrix}\frac{\partial \textbf{\textit{r}}^{(k)}}{\partial s_{i}}\\\frac{\partial \textbf{\textit{r}}^{(k+1)}}{\partial s_{i}}\\\vdots\\\frac{\partial \textbf{\textit{r}}^{(m_{u})}}{\partial s_{i}} \end{matrix}
		\end{array} \right].
	\end{split}
	\label{eq:kkt-mat1}
\end{equation}
In Eq. (\ref{eq:kkt-mat1}), $\mathcal{B} \in \mathbb{R}^{n_{u}(m_{u}+1-k)\times n_{u}(m_{u}+1)}$, $\mathcal{C} \in \mathbb{R}^{n_{u}(m_{u}+1-k)\times m_{u}}$ are sparse multi-block matrices, and $\mathcal{D} \in \mathbb{R}^{n_{u}m_{u}}$ is a vector. Eq. (\ref{eq:kkt-mat2}) can be represented in lower dimensions using Gaussian elimination
\begin{equation}
	\bigg[ \mathcal{B}\mathcal{B}^{\top} + \frac{1}{\alpha_{lss}^2}\mathcal{C}\mathcal{C}^{\top} \bigg] \Lambda = \mathcal{D},
	\label{eq:kkt-mat3}
\end{equation}
where, $\mathcal{B}\mathcal{B}^{\top} \in \mathbb{R}^{n_{u}(m_{u}+1-k) \times n_{u}(m_{u}+1-k)}$ and $\mathcal{C}\mathcal{C}^{\top} \in \mathbb{R}^{n_{u}(m_{u}+1-k) \times n_{u}(m_{u}+1-k)}$. In general, $\mathcal{B}$ and $\mathcal{C}$ have a large number of non-zeros, which is usually the case for high-fidelity chaotic dynamical systems. Moreover, according to Eq. (\ref{eq:kkt-mat3}), the derivative components, $\frac{\partial \textbf{\textit{r}}}{\partial \textbf{\textit{u}}}$ and $\frac{\partial \textbf{\textit{r}}}{\partial s_{i}}$, should be stored for all time intervals to a hard disk (e.g., Hard Disk Drive (HDD) or Solid-State Drive (SSD)), and Random Access Memory (RAM), which makes Eq. (\ref{eq:kkt-mat3}) prohibitively expensive to solve. For instance, the number of Jacobian matrices that should be stored in the computing device is $m_{u}$, and we require all Jacobian matrices at the same time to build $\mathcal{B}$.

\section{Least Squares Petrov-Galerkin Projection (LSPG)}
\label{sec:lspg}
In the weak form of a FOM, if we apply a discrete time integrator, it will be possible to check and control optimality conditions at all $t^{(n)} \in \Delta \mathbb{T}$. This discrete optimality concept raises up as a key step in the closure modelling, such that $\ell^2$-norm of the residual at each time interval is minimized. According to Eq.(\ref{eq:dynsys}), the residual of the dynamical system at each time interval is
\begin{equation}
	\textbf{\textit{r}}^{(n)} :=\frac{\Delta \textbf{\textit{u}}^{(n)}}{\Delta t^{(n)}} + \nabla.\textbf{\textit{f}}(\textbf{\textit{u}}^{(n)}, t^{(n)}, \mathcal{S}) \approx 0, \quad \textbf{\textit{u}}^{(0)}=\textbf{\textit{u}}_{0},
	\label{eq:lspg1}
\end{equation}
where $\Delta t^{(n)} \in \Delta \mathbb{T}$ is the time-step at $t^{(n)}$. Here, we use a constant time-step for all time intervals, $\Delta t^{(n)}=\Delta t$. To solve for this dynamical system in time, we need to minimize the residual in Eq. (\ref{eq:lspg1}) at each time interval. To this end, we employ the Petrov-Galerkin approach to convert Eq.(\ref{eq:lspg1}) into a minimization problem
\begin{equation}
	\tilde{\Psi}^{\top} \textbf{\textit{r}}^{(n)}= \tilde{\Psi}^{\top} \Bigg(\frac{\Delta \textbf{\textit{u}}^{(n)}}{\Delta t} + \nabla.\textbf{\textit{f}}(\textbf{\textit{u}}^{(n)}, t^{(n)}, \mathcal{S}) \Bigg) = 0, \quad \textbf{\textit{u}}^{(0)}=\textbf{\textit{u}}_{0},
	\label{eq:lspg2}
\end{equation}
where $\tilde{\Psi} \in \mathbb{R}^{n_{u} \times r}$ is the test basis function that enforces $\textbf{\textit{r}}^{(n)}$ to be minimum over $\Delta \mathbb{T}$, and $r \in \mathbb{R}_{+} \ll n_{u}$ is the number of rank in Hilbert space. Solving this minimization problem using the Petrov-Galerkin approach with optimality condition, where $\ell^2$-norm of the residual is considered, is called Least-Squares Petrov-Galerkin (LSPG) approach. Therefore, LSPG formulation is written as follows
\begin{equation}
	\textbf{\textit{u}}^{(n)}=
	\underset{\tiny \makecell{\textbf{\textit{z}} \in \mathbb{R}^{n_{u}}} }{\text{argmin}} \left\Vert \textbf{A}(\textbf{\textit{z}}) \textbf{\textit{r}}^{(n)}(\textbf{\textit{z}}, t^{(n)}, \mathcal{S}) \right\Vert_{2}^2,
	\label{eq:lspg3}
\end{equation}
and $n=1, \dots, m_{u}$, where $m_{u}$ is the number of time intervals, and $\textbf{A} \in \mathbb{R}^{n_{u} \times n_{u}}$ is a weighting matrix that enables definition of weighted norm \cite{carlberg2011low}. The weighting matrix can also be used for hyper-reduction at which we can only select a finite number of the states, and compute the corresponding residual at those states \cite{carlberg2013gnat}. In standard LSPG, the residual for all states are considered, $\textbf{A}=\mathcal{I}$, where $\mathcal{I}$ is identity matrix.

In order to build a ROM using Petrov-Galerkin projection, we define the state vector in the form of a linear combination of only $r$ subspaces in the Hilbert space
\begin{equation}
	\textbf{\textit{u}}^{(n)} \approx \overline{\textbf{\textit{u}}}+ \tilde{\Phi} \tilde{\textbf{\textit{q}}}^{(n)},
	\label{pod}
\end{equation}
where, $\overline{\textbf{\textit{u}}}$ is the reference vector, $\tilde{\Phi} \in \mathbb{R}^{n_{u}\times r}$ is the trial basis function, including orthogonal modes, and $\tilde{\textbf{\textit{q}}} \in \mathbb{R}^{r}$ is the generalized coordinates with $r$-rank. Using this definition, we can extend the application of the LSPG to solve for the generalized coordinates in lower subspaces as follows
\begin{equation}
	\tilde{\textbf{\textit{q}}}^{(n)}=
	\underset{\tiny \makecell{\textbf{\textit{z}} \in \mathbb{R}^{r}} }{\text{argmin}} \left\Vert \textbf{A}(\overline{\textbf{\textit{u}}} + \tilde{\Phi}\textbf{\textit{z}}) \textbf{\textit{r}}^{(n)}(\overline{\textbf{\textit{u}}} + \tilde{\Phi} \textbf{\textit{z}}, t^{(n)}, \mathcal{S}) \right\Vert_{2}^2,
	\label{argmin_q}
\end{equation}
where $n=1, \dots, m_{u}$, and the residual is a full-rank vector, $\textbf{\textit{r}}^{(n)} \in \mathbb{R}^{n_{u}}$, while the vector of the generalized coordinates is $r$-rank, $\tilde{\textbf{\textit{q}}}^{(n)} \in \mathbb{R}^{r}$, which means that this minimization is under-determined problem. Using the Backward Differentiation Formula (BDF) temporal scheme, we discretize Eq. (\ref{eq:lspg2}), and cast it into
\begin{equation}
	\tilde{\Psi}^{\top} \Bigg( \sum_{\textit{j}=0}^{\textit{k}} \zeta_j \textbf{\textit{u}}^{(n-j)} + \Delta t^{(n)} \sum_{\textit{j}=0}^{\textit{k}} \beta_j \nabla.\textbf{\textit{f}}^{(n)} \Big( \textbf{\textit{u}}^{(n-j)}, t^{(n-j)}, \mathcal{S} \Big) \Bigg)=0,
	\label{petrov_res_fom_multistep}
\end{equation}
where, $\zeta_j$ and $\beta_j$ are coefficients in the BDF scheme. We use Eq.(\ref{pod}) to rewrite Eq.(\ref{petrov_res_fom_multistep}) as
\begin{equation}
	\begin{split}
		&\tilde{\Psi}^{\top} \sum_{j=0}^{k} \zeta_j \tilde{\Phi} \tilde{\textbf{\textit{q}}}^{(n-j)}\\
		&+ \tilde{\Psi}^{\top} \Bigg( \Delta t^{(n)} \sum_{\textit{j}=0}^{\textit{k}} \beta_j \nabla.\textbf{\textit{f}}^{(n)} \Big( \overline{\textbf{\textit{u}}}+ \tilde{\Phi} \tilde{\textbf{\textit{q}}}^{(n-j)}, t^{(n-j)}, \mathcal{S} \Big) \Bigg) = 0.
	\end{split}
	\label{petrov_res_rom_multistep}
\end{equation}
Since the trial basis function, $\tilde{\Phi}$, remains constant over $\Delta \mathbb{T}$, then Eq.(\ref{petrov_res_rom_multistep}) is presented as follows
\begin{equation}
	\begin{split}
		&\tilde{\Psi}^{(n)\top}\tilde{\Phi} \sum_{j=0}^{k} \zeta_j \tilde{\textbf{\textit{q}}}^{(n-j)}\\
		&+ \tilde{\Psi}^{(n)\top} \Bigg( \Delta t^{(n)} \sum_{\textit{j}=0}^{\textit{k}} \beta_j \nabla.\textbf{\textit{f}}^{(n)} \Big( \overline{\textbf{\textit{u}}} + \tilde{\Phi} \tilde{\textbf{\textit{q}}}^{(n-j)}, t^{(n-j)}, \mathcal{S} \Big) \Bigg) = 0.
	\end{split}
	\label{petrov_res_rom_multistep3}
\end{equation}
In implicit temporal schemes, such as BDF, the second term on the right-hand side of Eq. (\ref{petrov_res_rom_multistep3}) is not considered, $\forall j>0 \colon \beta_j = 0$. Therefore, Eq. (\ref{petrov_res_rom_multistep3}) can be simplified to
\begin{equation}
	\begin{split}
		&\tilde{\Psi}^{(n)\top}\tilde{\Phi} \sum_{j=0}^{k} \zeta_j \tilde{\textbf{\textit{q}}}^{(n-j)}\\
		&+ \Delta t^{(n)} \beta_0 \tilde{\Psi}^{(n)\top} \nabla.\textbf{\textit{f}}^{(n)} \Big( \overline{\textbf{\textit{u}}} + \tilde{\Phi} \tilde{\textbf{\textit{q}}}^{(n)}, t^{(n)}, \mathcal{S} \Big) = 0,
	\end{split}
	\label{petrov_res_rom_multistep4}
\end{equation}
The relation between the residual of the ROM and that of the FOM in the Hilbert space and high-dimensional physical space is
\begin{equation}
	\mathcal{R}^{(n)}(\tilde{\textbf{\textit{q}}}^{(n)}, t^{(n)}, \mathcal{S})=\tilde{\Psi}^{\top} \textbf{\textit{r}}^{(n)}(\overline{\textbf{\textit{u}}} + \tilde{\Phi} \textbf{\textit{q}}^{(n)}, t^{(n)}, \mathcal{S}),
	\label{gal-lspg2}
\end{equation}
where $\mathcal{R}: \mathbb{R}^{r} \rightarrow \mathbb{R}^{r}$ is the residual of the ROM, and $r \ll n_{u}$. Eq. (\ref{gal-lspg2}) indicates that those terms in the ROM do not necessarily need to be known. In spite of finding unknow terms in this ROM, we can indirectly take them into account using $\mathcal{R}^{(n)}(\tilde{\textbf{\textit{q}}}^{(n)}, t^{(n)}, \mathcal{S})$. We will use Eq. (\ref{gal-lspg2}) later to obtain a closure model, in the form of a ROM, for sensitivity analysis.
\subsection{Descrete Form of Sensitivity Equation}
\label{subsec:disc-sens equation-a}
The residual formulation of the ROM is developed as a set of ODEs. Afterward, these ODEs are discretized, which is referred to as O$\Delta$Es. To this end, we assume that $\mathbb{F}$ is the exact representation of the FOM in the Hilbert space, then we have $\mathbb{F}:\tilde{\textbf{\textit{q}}}\times \mathbb{T} \rightarrow \tilde{\textbf{\textit{q}}}$ with $(\tilde{\textbf{\textit{q}}}, t) \mapsto \mathbb{F}(\tilde{\textbf{\textit{q}}}, t)$ over $t \in \mathbb{T}$. In reality, $\mathbb{F}$ is not available, and we enable only approximate it by $\mathcal{F}: \mathbb{R}^{r}\rightarrow \mathbb{R}^{r}$ as a ROM. Therefore, we assume that $\mathcal{F}\equiv \mathbb{F}$, and can then formulate the discrete form of the ROM using BDF scheme. Hence a set of O$\Delta$Es is obtained by
\begin{equation}
	\begin{split}
		\mathcal{R}^{(n)} &:= \sum_{\textit{j}=0}^{\textit{k}} \zeta_j  \tilde{\textbf{\textit{q}}}^{(n-j)}
		+ \Delta t^{(n)} \beta_0 \mathcal{F}^{(n)}(\tilde{\textbf{\textit{q}}}^{(n)},t^{(n)}, \mathcal{S})\\
		&+ \Omega^{(n)},\quad \tilde{\textbf{\textit{q}}}(t^{(0,\dots,k)})=\tilde{\textbf{\textit{q}}}_{0,\dots,k}, \quad t^{(n)} \in \Delta\mathbb{T},
	\end{split}
	\label{eq:petrov_res_rom_multistep4-a}
\end{equation}
where $\mathcal{R}:\mathbb{R}^{r}\rightarrow \mathbb{R}^{r}$ is the residual of the ROM, $\Omega$ represents a source term, and $\tilde{\textbf{\textit{q}}}_{0,...,k}$ are the initial conditions, such that $\tilde{\Phi}\tilde{\textbf{\textit{q}}}_{0,...,k} \mapsto \textbf{\textit{u}}_{0,...,k}$. Note that $\|\mathcal{R}^{(n)}\|_{2}$ converges to $\|\textbf{\textit{r}}^{(n)}\|_{2}$ when the number of subspaces in the trial and test basis functions is sufficiently large, i.e., $r \rightarrow n_{n}$. Therefore, the residual of the ROM is given by
\begin{equation}
	\lim_{r \rightarrow n_{u}}\mathcal{R}^{(n)}(\tilde{\textbf{\textit{q}}}^{(n)}, t^{(n)}, \mathcal{S}) \approx \textbf{\textit{r}}^{(n)}(\textbf{\textit{u}}^{(n)}, t^{(n)}, \mathcal{S}).
	\label{6E12-a}
\end{equation}

According to the \textit{shadowing lemma} \cite{blonigan2014least}, if a ROM with the reference generalized coordinates, $\tilde{\textbf{\textit{q}}}^{(n)}$, is perturbed by $\mathcal{S} + \delta\mathcal{S}$, then there is a vector of the shadow solution, $\tilde{\textbf{\textit{q}}}_{s}^{(n)}: \mathbb{T} \rightarrow \mathbb{R}^{r}$, that satisfies Eq. (\ref{eq:petrov_res_rom_multistep4-a}) with another time transformation, $\tau^{(n)} \in \Delta \mathbb{T}$. This definition for the shadow of the generalized coordinates is
\begin{equation}
	\lim_{r \rightarrow n_{u}} \lim_{\delta \mathcal{S} \rightarrow 0} \lim_{\delta \tau \rightarrow 0}  \mathcal{R}^{(n)}(\tilde{\textbf{\textit{q}}}_{s}^{(n)}, \tau^{(n)}, \mathcal{S} + \delta \mathcal{S})\approx 0,  \quad \tau^{(n)} \in  \Delta \mathbb{T}.
	\label{6E13-a}
\end{equation}
Therefore, we can find the limit of the generalized coordinates with respect to the perturbation as 
\begin{equation}
	\tilde{\textbf{\textit{h}}}^{(n)} \approx \lim_{\delta \mathcal{S} \rightarrow 0}{\frac{\tilde{\textbf{\textit{q}}}_{s}^{(n)} - \tilde{\textbf{\textit{q}}}^{(n)}}{\delta \mathcal{S}}},
	\label{6E14-a}
\end{equation}
where $\tilde{\textbf{\textit{h}}} \in \mathbb{R}^{r}$ is the sensitivity solution of the generalized coordinates. Similar to Least-Squares Shadowing (LSS) approach provided in Ref.~\cite{blonigan2014least}, we can apply the \textit{shadowing lemma} to Eq. (\ref{eq:petrov_res_rom_multistep4-a}). In this case, the low-dimensional sensitivity function in the form of O$\Delta$Es can be obtained by
\begin{equation}
	\begin{split}
		\lim_{\delta \mathcal{S} \rightarrow 0} \frac{d\mathcal{R}^{(n)}}{d \mathcal{S}} &:= \sum_{\textit{j}=0}^{k} \frac{\partial \mathcal{R}^{(n)}}{\partial \tilde{\textbf{\textit{q}}}^{(n-j)}} \tilde{\textbf{\textit{h}}}^{(n-j)}+ \frac{\partial \mathcal{R}^{(n)}}{\partial \mathcal{S}}\\
		&+ \frac{\partial \mathcal{R}^{(n)}}{\partial \Omega^{(n)}} + \mathcal{E}^{(n)}\Delta t\mathcal{F}^{(n)} = 0, \quad t^{(n)} \in \Delta \mathbb{T},
	\end{split}
	\label{eq:rom-sens-function-a}
\end{equation}
where, $\mathcal{E} \in \mathbb{R}$ is the time dilation of the generalized coordinates. Interestingly, we can apply the Petrov-Galerkin approach to minimize $\|\mathcal{R}^{(n)}\|_{2}$, such that errors, which are produced by dimensionality reduction, remain minimal over $\Delta \mathbb{T}$. Moreover, the unknown parameters in Eq. (\ref{eq:rom-sens-function-a}) are defined as follows
\begin{equation}
	\begin{split}
		\frac{\partial \mathcal{R}^{(n)}}{\partial \tilde{\textbf{\textit{q}}}^{(n)}} &= \Big(\bigg[\frac{\partial \tilde{\Psi}^{(n)}}{\partial \textbf{\textit{u}}^{(n)}}\bigg]^{\top}\textbf{\textit{r}}^{(n)}+\tilde{\Psi}^{(n)\top}\frac{\partial \textbf{\textit{r}}^{(n)}}{\partial \textbf{\textit{u}}^{(n)}}\Big)\frac{\partial \textbf{\textit{u}}^{(n)}}{\partial \tilde{\textbf{\textit{q}}}^{(n)}},\\
		\frac{\partial \mathcal{R}^{(n)}}{\partial \tilde{\textbf{\textit{q}}}^{(p)}} &= \Big(\bigg[\frac{\partial \tilde{\Psi}^{(n)}}{\partial \textbf{\textit{u}}^{(p)}}\bigg]^{\top}\textbf{\textit{r}}^{(n)}+\tilde{\Psi}^{(n)\top}\frac{\partial \textbf{\textit{r}}^{(n)}}{\partial \textbf{\textit{u}}^{(p)}}\Big)\frac{\partial \textbf{\textit{u}}^{(p)}}{\partial \tilde{\textbf{\textit{q}}}^{(p)}}, \quad n \ne p,\\
		\frac{\partial \mathcal{R}^{(n)}}{\partial \mathcal{S}} &= \Big(\bigg[\frac{\partial \tilde{\Psi}^{(n)}}{\partial \mathcal{S}}\bigg]^{\top}\textbf{\textit{r}}^{(n)}+\tilde{\Psi}^{(n)\top}\frac{\partial \textbf{\textit{r}}^{(n)}}{\partial \mathcal{S}}\Big).\\
	\end{split}
	\label{eq:r-to-R-a}
\end{equation}
Additionally, $\mathcal{F}$ and $\frac{\partial \mathcal{R}}{\partial \Omega}$ are given by
\begin{equation}
	\begin{split} 
		\frac{\partial \mathcal{R}^{(n)}}{\partial \Omega^{(n)}} &= \tilde{\Psi}^{(n)\top}\sum_{\textit{j}=0}^{k} \frac{\partial \textbf{\textit{r}}^{(n)}}{\partial \textbf{\textit{u}}^{(n-j)}}\overline{\textbf{\textit{v}}},\\
		\mathcal{E}^{(n)}\Delta t\mathcal{F}^{(n)} &= \eta^{(n)}\Delta t\tilde{\Psi}^{(n)\top}\nabla.\textbf{\textit{f}}^{(n)}.\\
	\end{split}
	\label{eq:rom-sourceterm-a}
\end{equation}

\subsection{Discrete form of the sensitivity equation}
\label{subsec:disc-sens equation}
In this section, the residual formulation of the ROM is developed as a set of ODEs, and then its Lagrange function is found. Afterward, these ODEs are discretized, which is referred to as O$\Delta$Es. To this end, we assume that $\mathbb{F}$ is the exact representation of the FOM in $\mathscr{H}$, then we have $\mathbb{F}:\tilde{\textbf{\textit{q}}}\times \mathbb{T} \rightarrow \tilde{\textbf{\textit{q}}}$ with $(\tilde{\textbf{\textit{q}}}, t) \mapsto \mathbb{F}(\tilde{\textbf{\textit{q}}}, t)$ over $t \in \mathbb{T}$. As a matter of fact, $\mathbb{F}$ is not available, and we can only approximate it using $\mathcal{F}: \mathbb{R}^{r}\rightarrow \mathbb{R}^{r}$ as a ROM. In this case, we suppose that $\mathcal{F}\equiv \mathbb{F}$, and can then formulate the discrete form of the ROM according to Eq. (\ref{eq:chaos-bdf2}). Hence a set of O$\Delta$Es is obtained by
\begin{equation}
	\mathcal{R}^{(n)} := \sum_{\textit{j}=0}^{\textit{k}} \zeta_j  \tilde{\textbf{\textit{q}}}^{(n-j)}
	+ \Delta t^{(n)} \beta_0 \mathcal{F}^{(n)}(\tilde{\textbf{\textit{q}}}^{(n)},t^{(n)}, \mathcal{S}) + \Omega^{(n)},\quad \tilde{\textbf{\textit{q}}}(t^{(0,\dots,k)})=\tilde{\textbf{\textit{q}}}_{0,\dots,k}, \quad t^{(n)} \in \Delta\mathbb{T},
	\label{eq:petrov_res_rom_multistep4}
\end{equation}
where $\mathcal{R}:\mathbb{R}^{r}\rightarrow \mathbb{R}^{r}$ is the residual of the ROM, $\Omega$ represents a source term, and $\tilde{\textbf{\textit{q}}}_{0,...,k}$ are the initial conditions, such that $\tilde{\mathbf{\Phi}}\tilde{\textbf{\textit{q}}}_{0,...,k} \mapsto \textbf{\textit{u}}_{0,...,k}$. Note that $\|\mathcal{R}^{(n)}\|_{2}$ converges to $\|\textbf{\textit{r}}^{(n)}\|_{2}$ when the number of subspaces in the trial and test basis functions is sufficiently large, i.e., $r \rightarrow n_{n}$. Therefore, the residual of the ROM is given by
\begin{equation}
	\lim_{r \rightarrow n_{u}}\mathcal{R}^{(n)}(\tilde{\textbf{\textit{q}}}^{(n)}, t^{(n)}, \mathcal{S}) \approx \textbf{\textit{r}}^{(n)}(\textbf{\textit{u}}^{(n)}, t^{(n)}, \mathcal{S}).
	\label{6E12}
\end{equation}

According to the \textit{shadowing lemma}, if a ROM with the reference generalized coordinates, $\tilde{\textbf{\textit{q}}}^{(n)}$, is perturbed by $\mathcal{S} + \delta\mathcal{S}$, then there is a vector of the shadow solution, $\tilde{\textbf{\textit{q}}}_{s}^{(n)}: \mathbb{T} \rightarrow \mathbb{R}^{r}$, that satisfies Eq. (\ref{eq:petrov_res_rom_multistep4}) with another time transformation, $\tau^{(n)} \in \Delta \mathbb{T}$. This definition for the shadow of the generalized coordinates is
\begin{equation}
	\lim_{r \rightarrow n_{u}} \lim_{\delta \mathcal{S} \rightarrow 0} \lim_{\delta \tau \rightarrow 0}  \mathcal{R}^{(n)}(\tilde{\textbf{\textit{q}}}_{s}^{(n)}, \tau^{(n)}, \mathcal{S} + \delta \mathcal{S})\approx 0,  \quad \tau^{(n)} \in  \Delta \mathbb{T}.
	\label{6E13}
\end{equation}
Recall Eq. (\ref{eq:partialstate}), and we can find the limit of the generalized coordinates with respect to the perturbation as 
\begin{equation}
	\tilde{\textbf{\textit{h}}}^{(n)} \approx \lim_{\delta \mathcal{S} \rightarrow 0}{\frac{\tilde{\textbf{\textit{q}}}_{s}^{(n)} - \tilde{\textbf{\textit{q}}}^{(n)}}{\delta \mathcal{S}}},
	\label{6E14}
\end{equation}
and, similar to Eq. (\ref{eq:constraintlss3}), we can apply the \textit{shadowing lemma} to Eq. (\ref{eq:petrov_res_rom_multistep4}). In this case, the low-dimensional sensitivity function in the form of O$\Delta$Es can be obtained by
\begin{equation}
	\lim_{\delta \mathcal{S} \rightarrow 0} \frac{d\mathcal{R}^{(n)}}{d \mathcal{S}} := \sum_{\textit{j}=0}^{k} \frac{\partial \mathcal{R}^{(n)}}{\partial \tilde{\textbf{\textit{q}}}^{(n-j)}} \tilde{\textbf{\textit{h}}}^{(n-j)} + \frac{\partial \mathcal{R}^{(n)}}{\partial \mathcal{S}} + \frac{\partial \mathcal{R}^{(n)}}{\partial \Omega^{(n)}} + \mathcal{E}^{(n)}\Delta t\mathcal{F}^{(n)} = 0, \quad t^{(n)} \in \Delta \mathbb{T},
	\label{eq:rom-sens-function}
\end{equation}
where, $\mathcal{E} \in \mathbb{R}$ is the time dilation of the generalized coordinates. Finding $\mathcal{F}$ in Eq. (\ref{eq:petrov_res_rom_multistep4}) could be prohibitively expensive. However, the advantage of formulating these O$\Delta$Es into the residual form is that we do not need to find $\mathcal{F}$ directly. Interestingly, we can apply the Petrov-Galerkin approach to minimize $\|\mathcal{R}^{(n)}\|_{2}$, such that errors, which are produced by dimensionality reduction, remain minimal over $\Delta \mathbb{T}$.

The Lagrange function for LSS minimization is given by
\begin{equation}
	\begin{split}
		\mathcal{L} &= \frac{1}{m_{u}} \sum_{\textit{n}=0}^{m_{u}} \Bigg[  \tilde{\textbf{\textit{h}}}^{(n) \top}\textbf{K} \tilde{\textbf{\textit{h}}}^{(n)} + \big(\alpha_{lss}\mathcal{E}^{(n)}\big)^{2}  \Bigg] \\
		&+\sum_{\textit{n}=0}^{m_{u}} \lambda^{(n) \top} \Bigg[  \sum_{\textit{j}=0}^{k} \frac{\partial \mathcal{R}^{(n)}}{\partial \tilde{\textbf{\textit{q}}}^{(n-j)}} \tilde{\textbf{\textit{h}}}^{(n-j)} + \frac{\partial \mathcal{R}^{(n)}}{\partial \mathcal{S}} + \frac{\partial \mathcal{R}^{(n)}}{\partial \Omega^{(n)}} + \mathcal{E}^{(n)}\Delta t\mathcal{F}^{(n)}  \Bigg],
	\end{split}
	\label{eq:lag-mult-rom}
\end{equation}
where $\lambda^{(n)}$ is the discrete Lagrange multiplier. Also, the KKT condition is applied to Eq. (\ref{eq:lag-mult-rom}), which yields
\begin{equation}
	\begin{split}
		&\frac{\partial \mathcal{R}^{(n)}}{\partial \tilde{\textbf{\textit{q}}}^{(n)}} \tilde{\textbf{\textit{h}}}^{(n)} + \sum_{\textit{j}=1}^{k} \frac{\partial \mathcal{R}^{(n)}}{\partial \tilde{\textbf{\textit{q}}}^{(n-j)}} \tilde{\textbf{\textit{h}}}^{(n-j)} + \frac{\partial \mathcal{R}^{(n)}}{\partial \mathcal{S}} + \frac{\partial \mathcal{R}^{(n)}}{\partial \Omega^{(n)}} + \mathcal{E}^{(n)}\Big(\frac{\Delta t}{\beta_0}\mathcal{R}^{(n)} - \sum_{\textit{j}=0}^{k}\frac{\zeta_j}{\beta_0} \tilde{\textbf{\textit{q}}}^{(n-j)}\Big)=0,\\
		&\bigg[\frac{\partial \mathcal{R}^{(n)}}{\partial \tilde{\textbf{\textit{q}}}^{(n)}}\bigg]^{\top}\lambda^{(n)}+\sum_{\textit{j}=1}^{k} \bigg[\frac{\partial \mathcal{R}^{(n)}}{\partial \tilde{\textbf{\textit{q}}}^{(n+j)}}\bigg]^{\top} \lambda^{(n+j)}+\textbf{K}^{\top} \tilde{\textbf{\textit{h}}}^{(n)}=0, \quad \lambda^{(0)}=\lambda^{(m_{u})}=0,\\
		&\alpha_{lss}^2 \mathcal{E}^{(n)} -m_{u}\lambda^{\top(n)} \Big(\frac{\Delta t}{\beta_0}\mathcal{R}^{(n)} - \sum_{\textit{j}=0}^{k}\frac{\zeta_j}{\beta_0} \tilde{\textbf{\textit{q}}}^{(n-j)}\Big)=0.\\
	\end{split}
	\label{eq:lagrange1b}
\end{equation}
Since Eq. (\ref{eq:lagrange1b}) is developed in the form of a low-dimensional closure model, it needs less computational resources, which is a promising factor for applying Eq. (\ref{eq:lagrange1b}) to sensitivity analysis of large-scale chaotic systems. We can solve Eq. (\ref{eq:lagrange1b}) using Eq. (\ref{eq:kkt-mat1}) at which the size of multi-block matrices are reduced to $\mathcal{B} \in \mathbb{R}^{r(m_{u}-k)\times rm_{u}}$, $\mathcal{C} \in \mathbb{R}^{rm_{u}\times m_{u}}$, and $\mathcal{D} \in \mathbb{R}^{rm_{u}}$, where $r \ll n_{u}$.

Although we eliminated $\mathcal{F}$ from Eq. (\ref{eq:lagrange1b}) by rewriting it as $\mathcal{F} = \mathcal{R} - \frac{\partial \textbf{\textit{q}}}{\partial t}$, there is an alternative method to find $\mathcal{F}$. Additionally, there is an unknown term in Eq. (\ref{eq:lagrange1b}), which is the derivative of the residual with respect to the source term, $\frac{\partial \mathcal{R}}{\partial \Omega}$. Therefore, in order to find $\mathcal{F}$ and $\frac{\partial \mathcal{R}}{\partial \Omega}$ directly from the FOM, we project the first equation in Eq. (\ref{eq:disc-kkt1}) into $\mathscr{H}$. then the projected results should be equal to the first O$\Delta$E in Eq. (\ref{eq:lagrange1b}). Therefore, we can write
\begin{equation}
	\begin{split}
		\lim_{\delta \mathcal{S} \rightarrow 0}\tilde{\mathbf{\Psi}}^{(n)\top}\frac{d\textbf{\textit{r}}^{(n)}}{d \mathcal{S}} &= \frac{d\mathcal{R}^{(n)}}{d \mathcal{S}}\\
		&=\sum_{\textit{j}=0}^{k} \frac{\partial \mathcal{R}^{(n)}}{\partial \tilde{\textbf{\textit{q}}}^{(n-j)}} \tilde{\textbf{\textit{h}}}^{(n-j)} + \frac{\partial \mathcal{R}^{(n)}}{\partial \mathcal{S}} + \tilde{\mathbf{\Psi}}^{(n)\top}\sum_{\textit{j}=0}^{k} \frac{\partial \textbf{\textit{r}}^{(n)}}{\partial \textbf{\textit{u}}^{(n-j)}}\overline{\textbf{\textit{v}}} + \eta^{(n)}\Delta t\tilde{\mathbf{\Psi}}^{(n)\top}\nabla.\textbf{\textit{f}}^{(n)},
	\end{split}
	\label{eq:P-to-H}
\end{equation}
and comparing Eq. (\ref{eq:lagrange1b}) and Eq. (\ref{eq:P-to-H}), $\mathcal{F}$ and $\frac{\partial \mathcal{R}}{\partial \Omega}$ can be obtained by
\begin{equation}
	\begin{split} 
		\frac{\partial \mathcal{R}^{(n)}}{\partial \Omega^{(n)}} &= \tilde{\mathbf{\Psi}}^{(n)\top}\sum_{\textit{j}=0}^{k} \frac{\partial \textbf{\textit{r}}^{(n)}}{\partial \textbf{\textit{u}}^{(n-j)}}\overline{\textbf{\textit{v}}},\\
		\mathcal{E}^{(n)}\Delta t\mathcal{F}^{(n)} &= \eta^{(n)}\Delta t\tilde{\mathbf{\Psi}}^{(n)\top}\nabla.\textbf{\textit{f}}^{(n)}.\\
	\end{split}
	\label{eq:rom-sourceterm}
\end{equation}

From Eq. (\ref{eq:rom-sourceterm}), the unknown terms in $\mathscr{H}$ are directly derived from the known terms in $\mathscr{P}$. Therefore, it suggests that the same procedure can be applied to remaining unknown terms in $\mathscr{H}$. In Petrov-Galerkin projection, the relation between the FOM and its corresponding ROM is given by $\mathcal{R}^{(n)} = \tilde{\mathbf{\Psi}}^{(n)\top} \textbf{\textit{r}}^{(n)}$ at all time intervals over $\Delta \mathbb{T}$. Consequently, derivatives of the residuals are
\begin{equation}
	\begin{split}
		\frac{\partial \mathcal{R}^{(n)}}{\partial \tilde{\textbf{\textit{q}}}^{(n)}} &= \Big(\bigg[\frac{\partial \tilde{\mathbf{\Psi}}^{(n)}}{\partial \textbf{\textit{u}}^{(n)}}\bigg]^{\top}\textbf{\textit{r}}^{(n)}+\tilde{\mathbf{\Psi}}^{(n)\top}\frac{\partial \textbf{\textit{r}}^{(n)}}{\partial \textbf{\textit{u}}^{(n)}}\Big)\frac{\partial \textbf{\textit{u}}^{(n)}}{\partial \tilde{\textbf{\textit{q}}}^{(n)}},\\
		\frac{\partial \mathcal{R}^{(n)}}{\partial \tilde{\textbf{\textit{q}}}^{(p)}} &= \Big(\bigg[\frac{\partial \tilde{\mathbf{\Psi}}^{(n)}}{\partial \textbf{\textit{u}}^{(p)}}\bigg]^{\top}\textbf{\textit{r}}^{(n)}+\tilde{\mathbf{\Psi}}^{(n)\top}\frac{\partial \textbf{\textit{r}}^{(n)}}{\partial \textbf{\textit{u}}^{(p)}}\Big)\frac{\partial \textbf{\textit{u}}^{(p)}}{\partial \tilde{\textbf{\textit{q}}}^{(p)}}, \quad n \ne p,\\
		\frac{\partial \mathcal{R}^{(n)}}{\partial \mathcal{S}} &= \Big(\bigg[\frac{\partial \tilde{\mathbf{\Psi}}^{(n)}}{\partial \mathcal{S}}\bigg]^{\top}\textbf{\textit{r}}^{(n)}+\tilde{\mathbf{\Psi}}^{(n)\top}\frac{\partial \textbf{\textit{r}}^{(n)}}{\partial \mathcal{S}}\Big).\\
	\end{split}
	\label{eq:r-to-R}
\end{equation}
Solving the low-dimensional closure model using the LSPG with enough subspaces causes the residual of the FOM to become approximately zero, such that for all $t^{(n)} \in \Delta \mathbb{T}$ we have $\lim_{r\rightarrow n_{u}} \textbf{\textit{r}}^{(n)}(\tilde{\mathbf{\Phi}} \tilde{\textbf{\textit{q}}}^{(n)}) \approx 0$. This definition causes the first term in the derivatives to be negligible. On the other hand, we can unanimously assume that  $\frac{\partial \textbf{\textit{u}}^{(n)}}{\partial \tilde{\textbf{\textit{q}}}^{(n)}} =\tilde{\mathbf{\Phi}}$. Hence, with regard to the definition of the test basis function, $\tilde{\mathbf{\Psi}}^{(n)}=\frac{\partial \textbf{\textit{r}}^{(n)}}{\partial \textbf{\textit{u}}^{(n)}}\tilde{\mathbf{\Phi}}$, we can simplify Eq. (\ref{eq:r-to-R}) as follows
\begin{equation}
	\begin{split}
		\frac{\partial \mathcal{R}^{(n)}}{\partial \tilde{\textbf{\textit{q}}}^{(n)}} &= \tilde{\mathbf{\Psi}}^{(n)\top}\tilde{\mathbf{\Psi}}^{(n)},\\
		\frac{\partial \mathcal{R}^{(n)}}{\partial \tilde{\textbf{\textit{q}}}^{(p)}} &= \zeta_{p}\tilde{\mathbf{\Psi}}^{(n)\top}\tilde{\mathbf{\Phi}}, \quad n \ne p,\\
		\frac{\partial \mathcal{R}^{(n)}}{\partial \mathcal{S}} &= \tilde{\mathbf{\Psi}}^{(n)\top}\frac{\partial \textbf{\textit{r}}^{(n)}}{\partial \mathcal{S}}.\\
	\end{split}
	\label{eq:r-to-R2}
\end{equation}
In the end, substituting Eq. (\ref{eq:r-to-R2}) into Eq. (\ref{eq:lagrange1b}), we can find the the final form of the O$\Delta$Es
\begin{equation}
	\begin{split}
		&\begin{split}
			\tilde{\mathbf{\Psi}}^{(n)\top}\tilde{\mathbf{\Psi}}^{(n)} \tilde{\textbf{\textit{h}}}^{(n)} &+ \sum_{\textit{j}=1}^{k} \zeta_{j}\tilde{\mathbf{\Psi}}^{(n)\top}\tilde{\mathbf{\Phi}} \tilde{\textbf{\textit{h}}}^{(n-j)} + \tilde{\mathbf{\Psi}}^{(n)\top}\frac{\partial \textbf{\textit{r}}^{(n)}}{\partial \mathcal{S}}\\ &+\tilde{\mathbf{\Psi}}^{(n)\top}\Big(\sum_{\textit{j}=1}^{k}\zeta_{j}\mathcal{I} + \frac{\partial \textbf{\textit{r}}^{(n)}}{\partial \textbf{\textit{u}}^{(n)}}\Big)\overline{\textbf{\textit{v}}} - \mathcal{E}^{(n)} \sum_{\textit{j}=0}^{k}\frac{\zeta_j}{\beta_0} \tilde{\textbf{\textit{q}}}^{(n-j)}=0,
		\end{split}\\
		&\tilde{\mathbf{\Psi}}^{(n)\top}\tilde{\mathbf{\Psi}}^{(n)}\lambda^{(n)}+\sum_{\textit{j}=1}^{k} \zeta_{j}\tilde{\mathbf{\Psi}}^{(n)\top}\tilde{\mathbf{\Phi}} \lambda^{(n+j)}+\textbf{K}^{\top} \tilde{\textbf{\textit{h}}}^{(n)}=0, \quad \lambda^{(0)}=\lambda^{(m_{u})}=0,\\
		&\alpha_{lss}^2 \mathcal{E}^{(n)} - m_{u}\lambda^{(n)\top} \sum_{\textit{j}=0}^{k}\frac{\zeta_j}{\beta_0} \tilde{\textbf{\textit{q}}}^{(n-j)}=0,\\
	\end{split}
	\label{eq:rom-kkt-final}
\end{equation}
which results in the final form of the sensitivity function in $\mathscr{H}$. The amount of error in Eq. (\ref{eq:rom-kkt-final}) depends on the error bounds in LSPG \cite{carlberg2017galerkin}, and the error bounds for dimensionality reduction. The total sensitivity of the objective function with respect to a set of design parameters can be represented as
\begin{equation}
	\begin{split}
		\overline{\frac{\Delta \mathcal{J}}{\Delta \mathcal{S}}} &\approx
		\frac{1}{T}
		\int_{0}^{T}\bigg[\frac{\partial \mathcal{J}}{\partial \textbf{\textit{u}}}\bigg]^{\top} (\overline{\textbf{\textit{v}}}+\tilde{\mathbf{\Phi}}\tilde{\textbf{\textit{h}}})dt
		+\frac{1}{T}
		\int_{0}^{T}\mathcal{E}\Big(\mathcal{J}(\overline{\textbf{\textit{u}}}+\tilde{\mathbf{\Phi}}\tilde{\textbf{\textit{q}}},t,\mathcal{S})-\overline{\mathcal{J}}\Big)dt
		+\frac{1}{T}
		\int_{0}^{T}\frac{\partial \mathcal{J}}{\partial \mathcal{S}}dt,\\
	\end{split}
	\label{eq:rom-tang-objsens01}
\end{equation}
and the discrete form of Eq. (\ref{eq:rom-tang-objsens01}) is given by
\begin{equation}
	\begin{split}
		\overline{\frac{\Delta \mathcal{J}}{\Delta \mathcal{S}}} &\approx \frac{1}{m_{u}}\sum_{n=1}^{m_{u}}\bigg[\frac{\partial \mathcal{J}^{(n)}}{\partial \textbf{\textit{u}}^{(n)}}\bigg]^{\top} (\overline{\textbf{\textit{v}}}+\tilde{\mathbf{\Phi}}\tilde{\textbf{\textit{h}}}^{(n)})
		+\frac{1}{m_{u}}\sum_{n=1}^{m_{u}}\mathcal{E}^{(n)}\Big(\mathcal{J}^{(n)}-\overline{\mathcal{J}}\Big) + \frac{1}{m_{u}}\sum_{n=1}^{m_{u}}\frac{\partial \mathcal{J}^{(n)}}{\partial \mathcal{S}}.\\
	\end{split}
	\label{eq:rom-tang-objsens02}
\end{equation}
As shown in Eq.(\ref{eq:rom-tang-objsens02}), the first term approximates the high-dimension sensitivity solution. The second term is the sensitivity of the time transformation. Finally, the last term represents the sensitivity of the objective function with respect to the design parameters, regardless of any perturbation in the state vector.

\section{Physics-Constrained Data-Driven Approach}
\label{sec:machine learning}
\section{Training Strategies}
\label{sec:learning-strategy}
The training procedure is significantly dependent on the data collected from the results of the FOM simulations. A snapshot is a vector of the state, or other related variables, that is obtained at a specific time interval. This snapshot could be multi-dimensional, but it is often converted into a vector (i.e., stack all data in a column). If the dataset is built from entire snapshots at all time intervals, it will often need a large memory allocation. So these types of data are usually built by collecting snapshots, as the samples, from a limited set of time intervals. Therefore, selecting a proper and efficient sampling strategy is an essential matter in building an accurate ROM. The main objective of this study is to employ the ROM for optimization purposes. Hence, this ROM we build should approximate the sensitivity solutions over a finite time for a different set of design parameters.

\begin{figure}[h] 
	\centering
	\includegraphics[width=1\textwidth]{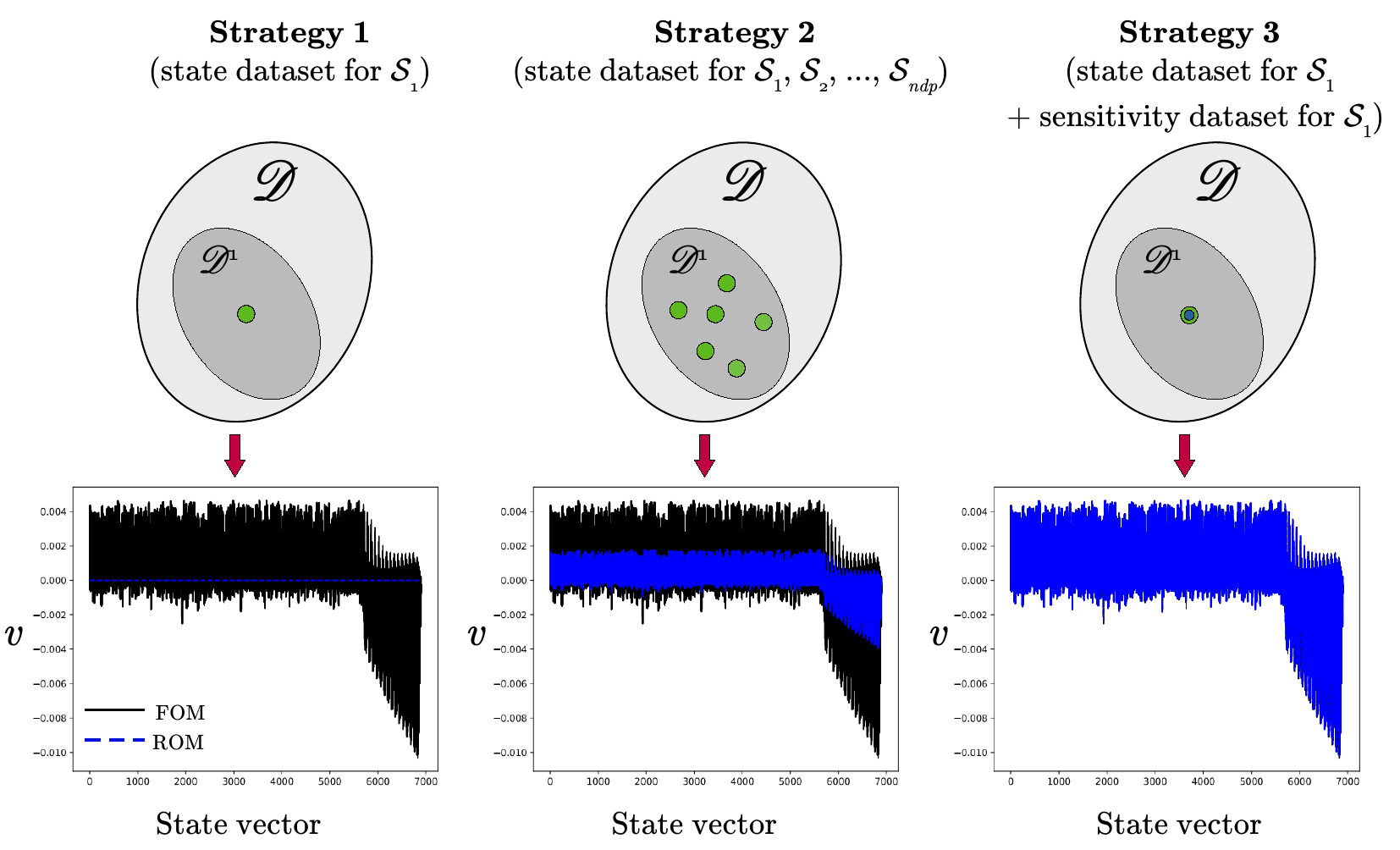}
	\caption{Evaluating different strategies for data sampling in design space.}
	\label{fig:strategy1}
\end{figure}

Figure \ref{fig:strategy1} shows three different data sampling strategies, and their impact on the sensitivity solution approximated via the ROM. In this case, data sampling is provided from a 2D simulation of steady-state flow past a NACA 0012 airfoil at $Re=1000$, $M_{\infty}=0.2$, and $\alpha_{eff}=0^{\circ}$. Let us assume that $\mathcal{S}_{1}$ is in design space, $\mathscr{D}$. We want to build a ROM that approximates the state and sensitivity solutions in a subspace $\mathscr{D}_{1} \subset \mathscr{D}$. We start sampling the data when the FOM is not yet steady-state, and we continue it until this FOM becomes completely steady-state. Therefore, the corresponding dataset contains both quasi-steady and steady solutions, and it can be applied to build the ROB.
In Strategy 1, the ROM is built based on the state dataset collected at $\mathcal{S}_{1}$, which is denoted by green colour in design space. In this case, the sensitivity solution reflects a wrong answer compared to the sensitivity solution computed by FOM, which is computed by $\textbf{\textit{v}}=-(\frac{\partial \nabla.\textbf{\textit{f}}}{\partial \textbf{\textit{u}}})^{-1}\frac{\partial \nabla.\textbf{\textit{f}}}{\partial \mathcal{S}_{1}}$. This situation usually occurs when the dynamical system is susceptible to design parameters. In Strategy 2, several FOMs for different design parameters, $\mathcal{S} \in \{\mathcal{S}_{1}, \dots, \mathcal{S}_{n_{dp}}\} \subset \mathscr{D}_1$, are performed, and afterward, the ROM is built by the state datasets collected at all $\mathcal{S}_{1}, \dots, \mathcal{S}_{n_{dp}}$.  In this case, the sensitivity solution approximated by the ROM is notably improved, but there is still a large discrepancy between the result of the ROM and that of the FOM. The accuracy in Strategy 2 can be increased by adding more data into the dataset, collected from further FOM simulations at different design parameters in $\mathscr{D}_1$. This strategy can be helpful to approximate those solutions in $\mathscr{D}_1$ using thrust region approaches. However, it may not be economical when the number of FOM simulations increases. As shown in Strategy 3, the ROM is built using both the state and the sensitivity datasets.
In this case, the sensitivity function is only solved for the sensitivity solution at $\mathcal{S}_{1}$, denoted by blue colour. The sensitivity solution computed using the ROM and the FOM are in good agreement, and they are almost on top of each other. This is because the ROM is trained via extra information, containing the prominent structure of the sensitivity solution. Therefore, the ROB delivers accurate information to the ROM to approximate the sensitivity solution in lower dimensions. Interestingly, Strategy 3 only needs one simulation of the FOM in subspace $\mathscr{D}_1$. This approach indicates that we can not find dominant structures of the sensitivity solution from those in the flow field. This finding becomes an essential matter when the dynamical system becomes highly non-linear.
\begin{figure}[h] 
	\centering
	\includegraphics[width=1\textwidth]{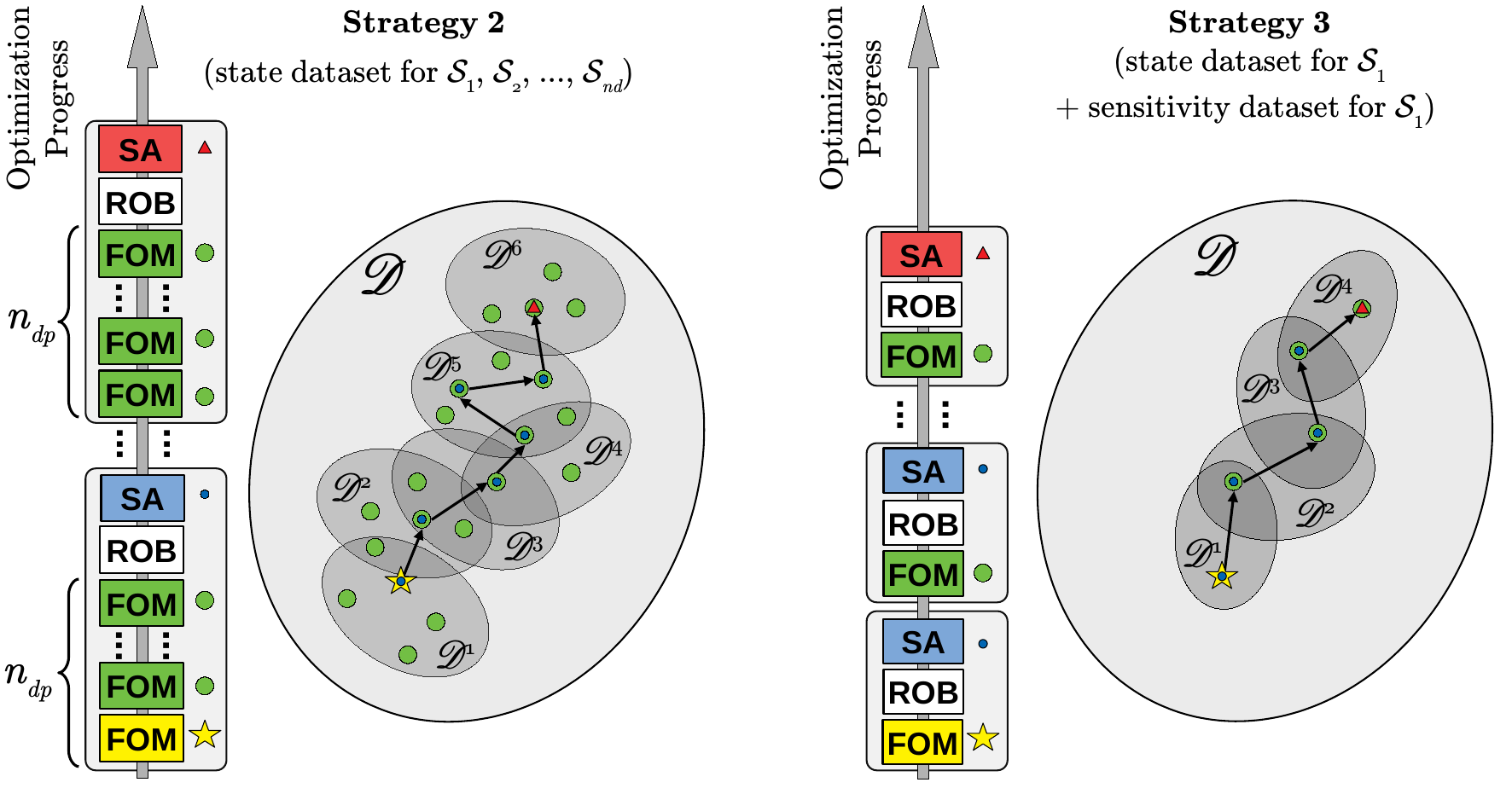}
	\caption{Comparing the performance of Strategies 2 and 3 during the optimization procedure.}
	\label{fig:strategy2}
\end{figure}

Figure \ref{fig:strategy2} is also provided to compare Strategies 2 and 3 during the optimization procedure. Let us assume that $\mathcal{S}_{1}$, as a baseline design, is optimized in design space $\mathscr{D}$. Each design cycle has its subspace, such as $\mathscr{D}_1$ and $\mathscr{D}_2$ for the first and second iterations. In Strategy 2, we select $n_{dp}$ different set of design parameters, $\mathcal{S} \in \{\mathcal{S}_{1}, \dots, \mathcal{S}_{n_{dp}}\} \subset \mathscr{D}_i$, where index $i$ indicates the design iteration. Therefore, $n_{dp}$ different FOMs are simulated, followed by building the ROB, and performing sensitivity analysis (SA). According to Figure \ref{fig:strategy1}, this approach may need a large number of FOMs if the number of design parameters in $\mathcal{S}=[s_1, s_2, \dots, s_{n_s}]$ is very large. Additionally, the updated design may not be very exact in each subspace, since the accuracy of the sensitivity solution increases by adding more information to the dataset. On the other hand, Strategy 3 only needs one FOM simulation for each subspace, followed by building the ROB, and performing SA. In optimization problems with many design parameters, this strategy will significantly reduce the computational cost. Although Strategy 3 may reflect a promising improvement in the accuracy of the ROM, this ROM may lose its accuracy for a different set of design parameters, $\{\mathcal{S}_{2}, \dots, \mathcal{S}_{n_{dp}}\}$, in the same subspace. As a matter of fact, we do not need to use this ROM for different design parameters, since we only require those sensitivities at 
$\mathcal{S}_{1}$. However, in Strategy 2, we learned that adding more information to the dataset will help increase the accuracy of the ROB in each subspace. If we use all of the previous datasets, from $\mathscr{D}_1$ to $\mathscr{D}_{i-1}$, to build the ROB for $\mathscr{D}_i$, we may build a more efficient and effective ROM during the optimization procedure. Consequently, we can combine these two strategies to develop more accurate models. In section \ref{subsec:ROB}, we use these combinations to develop training algorithms. It is worth mentioning that we can change the number of FOMs in each subspace according to the required accuracy in the final design, physics of the problem, and available computational resources. 
\subsection{Architecture}
\label{subsec:architecture}
\begin{figure} 
	\centering
	\includegraphics[width=1\textwidth]{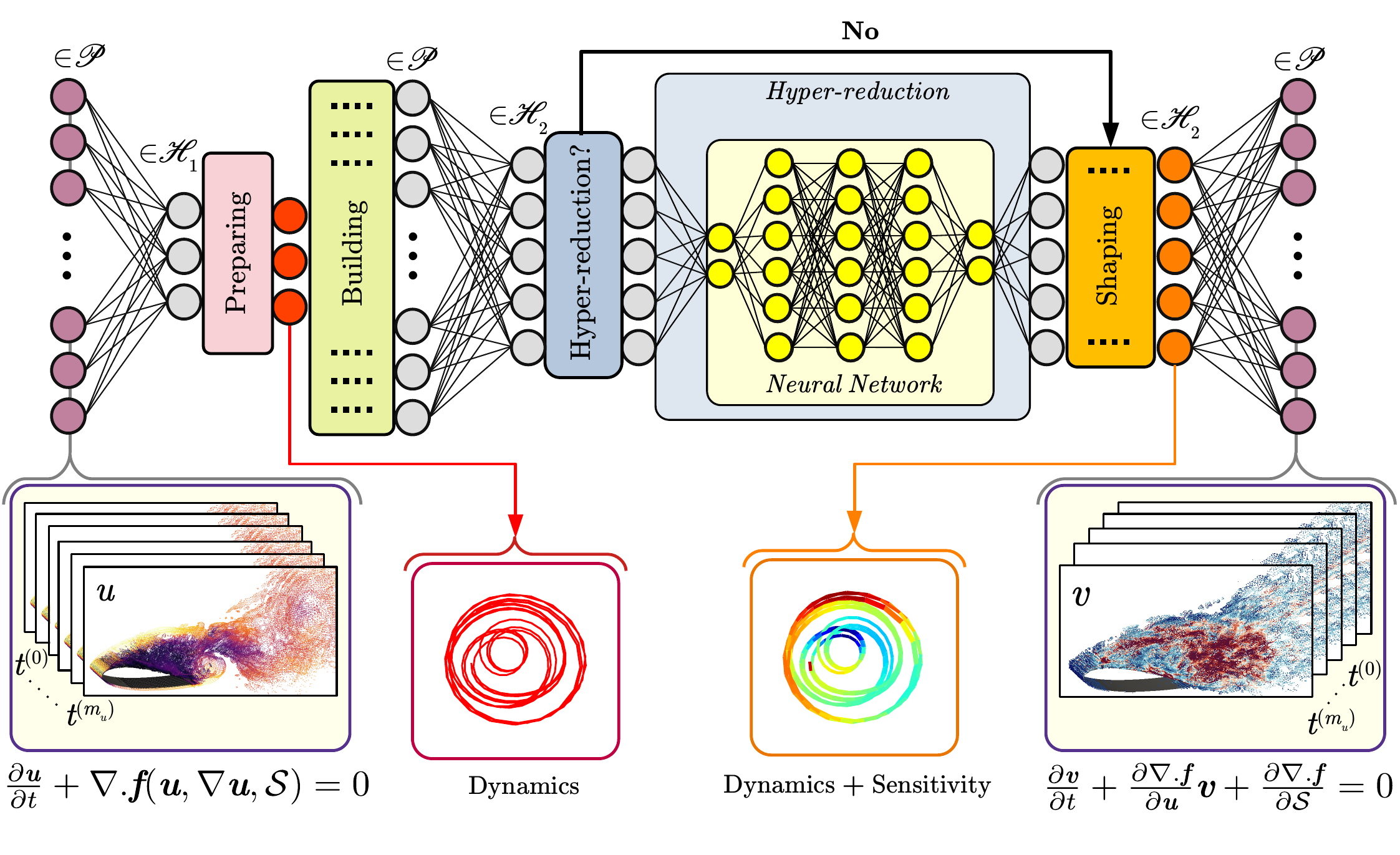}
	\caption{Schematic of the platform designed for sensitivity analysis.}
	\label{fig:nnw}
\end{figure}

Solving the weak form of the FOM, which is converted into a closure model, produces non-physical results unless optimal trial and test basis functions are defined. Therefore, we leverage data science techniques to find these optimal basis functions for the closure model. To this end, we propose a novel algorithm to find the overall structure of the manifolds, and shape them in Hilbert space. This proposed algorithm has four major phases. The first phase is the \textit{preparing} phase, and in this phase, a closure model is developed to obtain the dynamics of the system. The second phase is the \textit{building} phase, during which different data sampling methods are taken into account in order to generate a dataset. In the third phase, \textit{hyper-reduction}, the dynamical behaviour of the reduced-order matrices are found by deep learning. This phase reduces the computational costs attributed to building the closure model for the sensitivity function. The fourth phase is the \textit{shaping} phase, which refers to improvement of the ROB in lower dimensions through an iterative process, and solving for sensitivities. Figure \ref{fig:nnw} illustrates the proposed platform architecture schematically. First, the FOM is solved over $\Delta \mathbb{T}$, and snapshots are collected from it. These snapshots in physical space, $\textbf{\textit{u}} \in \mathscr{P}$, are projected into Hilbert space, $\mathscr{H}_{1}$. Next, in the \textit{preparing} phase, Eq. (\ref{petrov_res_rom_multistep4}) is solved as a closure model via LSPG, such that $\tilde{\textbf{\textit{q}}} \in \mathscr{H}_{1}$ lies in a lower-dimensional space. The red nodes contain the dynamics of the system in $\mathscr{H}_{1}$. In the \textit{building} phase, the ROB are created with sensitivity solutions that are sampled at $\mathbb{N}_{p}(n)=\{\xi_{p}^{(1)}, \xi_{p}^{(2)}, \dots, \xi_{p}^{(m_{p})}\}$, where $\xi_{p}^{(n)}$ is index of each time interval (i.e., $\xi_{p}^{(n)} \Delta t \in \Delta \mathbb{T}$), and $m_{p}$ denotes the number of samples in the \textit{building} phase. Afterward, the overall structure and shape of manifolds in Hilbert space are obtained. Large-scale Jacobian and relative matrices are transformed from $\mathscr{P}$ into $\mathscr{H}_{2}$, where the sensitivity solutions evolve with time in this space. The second phase is optional, while it significantly reduces the computational burden for sensitivity analysis of large-scale chaotic problems. If the closure model does not suffer high computational cost, we can skip this phase. In the \textit{hyper-reduction} phase, costly reduced-order matrices are only computed for a limited set of time intervals, $\mathbb{N}_{s}(n)=\{\xi_{s}^{(1)}, \xi_{s}^{(2)}, \dots, \xi_{s}^{(m_{\text{samp}})}\}$, where $m_{\text{samp}}$ is the number of sample intervals, and the remainder matrices are computed via deep learning. Furthermore, in the \textit{shaping} phase, the trial basis functions are deformed in all directions in $\mathscr{H}_{2}$, and this deformation is optimized with regard to another set of samples at $\mathbb{N}_{q}(n)=\{\xi_{q}^{(1)}, \xi_{q}^{(2)}, \dots, \xi_{q}^{(m_{q})}\}$, where $m_{q}$ is the number of samples in the \textit{shaping} phase. Additionally, LSS minimization is solved for sensitivities in this phase. Afterward, the orange nodes contain the sensitivity solutions of the corresponding dynamics in Hilbert space. At the end of the procedure, the solutions are lifted back from $\mathscr{H}_{2}$ into $\mathscr{P}$ with a proper mapping function. The output of this platform produces the sensitivity solution, $\textbf{\textit{v}}^{(n)} \in \mathscr{P}$, which reflects the solution of Eq. (\ref{eq:get_v1}) over $\Delta \mathbb{T}$. In Sections \ref{subsec:preparing} to \ref{subsec:shaping}, we provide further details to describe the algorithms and related numerical approaches.  
\subsection{Preparing a closure model for the dynamics of the system}
\label{subsec:preparing}
In model-reduction techniques, the trial basis functions are built by a set of ROB matrices. Generally speaking, a ROB is a subset of trial basis functions, containing prominent structures of the dynamical system. Each ROB is generated by different datasets (e.g., state vector, sensitivity solution) that are collected over $\Delta \mathbb{T}$. Additionally, other quantities can help enrich the information we require to shape the manifolds in $\mathscr{H}$. These quantities can be the residuals, fluxes, or Krylov vectors \cite{zahr2015progressive}. High-dimensional datasets can be collected for different sets of design parameters, $\mathcal{S} \in \{\mathcal{S}_{1}, \mathcal{S}_{2}, ..., \mathcal{S}_{n_{d}} \}\subset\mathcal{D}$, where $n_{d}$ denotes the total number of design sets (i.e., sample design parameters). Therefore, snapshots for each design set are assembled as follows
\begin{equation}
	\textbf{U}(\mathcal{S}_{i}) = 
	\begin{bmatrix} 
		| & | & & |\\
		\Delta  \textbf{\textit{u}}^{(1)}(\mathcal{S}_{i}) & \Delta  \textbf{\textit{u}}^{(2)}(\mathcal{S}_{i}) & ... & \Delta  \textbf{\textit{u}}^{(m_{u})}(\mathcal{S}_{i}) \\
		| & | & & |
	\end{bmatrix} \in \mathbb{R}^{n_{u}\times m_{u}}, \quad \Delta  \textbf{\textit{u}}^{(n)}=  \textbf{\textit{u}}^{(n)}-\overline{\textbf{\textit{u}}},
	\label{eq:pod1}
\end{equation}
where, $i=1, 2, ..., n_{d}$. We build all datasets according to Eq. (\ref{eq:pod1}), and then assemble all of them into a united matrix
\begin{equation}
	\textbf{X}_{state}=\bigoplus_{i=1}^{n_{d}} \textbf{U}(\mathcal{S}_{i}),
	\label{eq:pod2}
\end{equation}
where, $\textbf{X}_{state}$ denotes the dataset used to train the ROB. Subsequently, we build the ROB using
\begin{equation}
	\tilde{\mathbf{\Phi}}_{u} = \textbf{POD}(\textbf{X}_{state}), \quad \text{and} \quad 
	\textbf{U}(\mathcal{S}_{i}) = \tilde{\mathbf{\Phi}}_{u} \mathcal{Q}_{u}(\mathcal{S}_{i}), \quad i=1, 2, ..., n_{d},
	\label{eq:pod3}
\end{equation}
where, the \textbf{POD} represents a function that returns the orthogonal basis (i.e., POD modes). This function is described in Algorithm \ref{alg:pod}. 
\begin{algorithm}[h]
	\scriptsize{
		\SetAlgoLined
		\KwIn{$\textbf{X}$, $\beta=\mathcal{O}(10^{-3})$, and $tol=\mathcal{O}(10^{-3})$.}	
		\KwOut{$\tilde{\mathbf{\Phi}}$, $r$, $\mathcal{Q}(\mathcal{S}_{i}), \quad i=1, 2, ..., n_{d}.$}	
		
		Apply SVD such that $\textbf{X} = \texttt{U}\texttt{S}\texttt{V}^{\top}$.\\
		
		\For{$j=1 \quad \text{\textbf{to}} \quad n_{u}$}{
			$Error \leftarrow 0$,\\
			\For{$i=1 \quad \text{\textbf{to}} \quad n_{d}$}{
				$\textbf{U}(\mathcal{S}_{i}) \leftarrow \textbf{X}[:, (i-1)\times m_{u}:i\times m_{u}]$,\\
				$\textbf{U}(\mathcal{S}_{i})\big[\textbf{\textit{u}}^{(0)} - \overline{\textbf{\textit{u}}}, \textbf{\textit{u}}^{(1)} - \overline{\textbf{\textit{u}}}, \dots, \textbf{\textit{u}}^{(m_{u})}- \overline{\textbf{\textit{u}}} \big]$,\\
				$\textbf{\textit{z}} \leftarrow \texttt{V}[:j, :]^{\top}$,\\
				$Error \leftarrow Error + \beta j + \left\Vert \sum_{n=0}^{m_{u}} \textbf{\textit{u}}^{(n)} - \overline{\textbf{\textit{u}}} - \texttt{U}[:, :j]\texttt{S}[:j, :j]\textbf{\textit{z}}^{(n)} \right\Vert_{2}^2$.\\
			}
			\If{$Error\le tol$}{
				$r \leftarrow j$,\\
				$\tilde{\mathbf{\Phi}} \leftarrow \texttt{U}[:, : r]$,\\
				\For{$i=1 \quad \text{\textbf{to}} \quad n_{d}$}{
					$\mathcal{Q}(\mathcal{S}_{i}) \leftarrow \texttt{S}[:r, :r]\texttt{V}[:r, (i-1)\times m_{u}:i\times m_{u}]^{\top}$,\\
				}
				Break (exit from the loop).\\
			}	
		}
		\caption{\textbf{POD} function.}
		\label{alg:pod}
	}
\end{algorithm}
Additionally, $\mathcal{Q}_{u}(\mathcal{S}_{i})$ denotes the generalized coordinates for each set of design parameters
\begin{equation}
	\mathcal{Q}_{u}(\mathcal{S}_{i}) =
	\begin{bmatrix} 
		| & | & & |\\
		\tilde{\textbf{\textit{q}}}_{i}^{(1)} & \tilde{\textbf{\textit{q}}}_{i}^{(2)} & ... & \tilde{\textbf{\textit{q}}}_{i}^{(m_{u})} \\
		| & | & & |
	\end{bmatrix} \in \mathbb{R}^{r_{state}\times m_{u}},
	\label{eq:pod4}
\end{equation}
where $r_{state}$ is the rank of $\tilde{\mathbf{\Phi}}_{u}$. Note that $\tilde{\textbf{\textit{q}}}_{i}^{(n)} \in \mathbb{R}^{r_{state}}$ is actually the generalized coordinates of the weak form of the FOM in Hilbert space. In order to find these coordinates for the closure model, Eq. (\ref{petrov_res_rom_multistep4}) should be solved via LSPG to recompute $\tilde{\textbf{\textit{q}}}_{i}^{(n)}$ over $\Delta \mathbb{T}$, such that $\tilde{\textbf{\textit{q}}} \in \mathscr{H}_{1}$.
\subsection{Building manifolds for the sensitivity function}
\label{subsec:ROB}
In order to improve the robustness of the trial basis function with respect to perturbations in entries, the steady-state forward sensitivity function is solved for a different set of design parameters. At first, snapshots of the steady-state sensitivity solutions for all sets of design parameters are defined as
\begin{equation}
	\textbf{V}_{ss}(\mathcal{S}_{i}) =
	\begin{bmatrix} 
		| & | & & |\\
		\Delta \textbf{\textit{v}}_{ss}^{(\mathbb{N}_{p}(1))}(\mathcal{S}_{i}) & \Delta \textbf{\textit{v}}_{ss}^{(\mathbb{N}_{p}(2))}(\mathcal{S}_{i}) & ... & \Delta \textbf{\textit{v}}_{ss}^{(\mathbb{N}_{p}(m_{p}))}(\mathcal{S}_{i}) \\
		| & | & & |
	\end{bmatrix} \in \mathbb{R}^{n_{u}\times m_{p}}, \quad \Delta \textbf{\textit{v}}_{ss}^{(n)}=\textbf{\textit{v}}_{ss}^{(n)}-\overline{\textbf{\textit{v}}_{ss}},
	\label{eq:pod5}
\end{equation}
where, $\overline{\textbf{\textit{v}}_{ss}} \in \mathbb{R}^{n_{u}}$ represents a reference vector for sensitivity snapshots, and it usually satisfies $\frac{\partial \overline{\textbf{\textit{v}}_{ss}}}{\partial t}=0$. Therefore, the dataset for all sensitivity solutions is represented as
\begin{equation}
	\textbf{X}_{sens}=\bigoplus_{i=1}^{n_{d}} \textbf{V}_{ss}(\mathcal{S}_{i}),
	\label{eq:pod6b}
\end{equation}
and then,
\begin{equation}
	\tilde{\mathbf{\Phi}}_{v} = \textbf{POD}(\textbf{X}_{sens}) \quad \text{and} \quad 
	\textbf{V}_{ss}(\mathcal{S}_{i}) = \tilde{\mathbf{\Phi}}_{v} \mathcal{Q}_{v}(\mathcal{S}_{i}), \quad i=1,2,..., n_{d}.
	\label{eq:pod7}
\end{equation}
It is worth pointing out that the ROB obtained by the sensitivity dataset could have a different number of subspaces, $\tilde{\mathbf{\Phi}}_{v} \in \mathbb{R}^{n_{u} \times r_{sens}}$, where $r_{state}$ is the rank of the ROB that is built by the sensitivity data. Accordingly, the generalized coordinates for the sensitivity solutions can be obtained by
\begin{equation}
	\mathcal{Q}_{v} =
	\begin{bmatrix} 
		| & | & & |\\
		\tilde{\textbf{\textit{h}}}_{i}^{(\mathbb{N}_{p}(1))} &  \tilde{\textbf{\textit{h}}}_{i}^{(\mathbb{N}_{p}(2))} & ... &  \tilde{\textbf{\textit{h}}}_{i}^{(\mathbb{N}_{p}(m_{p}))} \\
		| & | & & |
	\end{bmatrix} \in \mathbb{R}^{r_{sens} \times m_{p}}, \quad \mathbb{N}_{p}(n)=\{\xi_{p}^{(1)}, \xi_{p}^{(2)}, \dots, \xi_{p}^{(m_{p})}\}.
	\label{eq:pod8b}
\end{equation}

Finally, we need to collect data for the unsteady sensitivity solutions to ensure that errors in the ROB for chaotic problems do not grow significantly. However, as mentioned earlier, it is impossible to solve the sensitivity function of chaotic systems using conventional methods. On the other hand, if the dynamical system is chaotic, the steady-state sensitivity datasets are not sufficient to build an accurate closure model. Therefore, we propose a new approach for data collection from the unsteady sensitivity function. The intuition behind this idea is that the solution of the steady-state sensitivity function only gives an acceptable approximation when the system is linear or weakly non-linear. On the other hand, detecting features of the unsteady sensitivity function, and adding them to the trial basis function improves the prediction of the closure model for strongly non-linear systems. Therefore, the unsteady sensitivity solutions are collected according to a procedure provided in Algorithms \ref{alg:dataV_us_ic} and \ref{alg:dataV_us}. 
\begin{algorithm}[h]
	\scriptsize{
		\SetAlgoLined
		\KwIn{$\textbf{U}(\mathcal{S}_{1})$, $\mathcal{S}_{1}$, $k$, and $p$ (given by Algorithm \ref{alg:dataV_us}).}
		\KwOut{$\textbf{\textit{v}}^{(p-k)}$, and $\textbf{\textit{v}}^{(p-k+1)}$.}
		
		\textbf{Approximate I.C.:}\\
		Compute $\frac{\partial \nabla.\textbf{\textit{f}}^{(p-k)}}{\partial  \textbf{\textit{u}}^{(p-k)}}$, and $\frac{\partial \nabla.\textbf{\textit{f}}^{(p-k)}}{\partial \mathcal{S}_{1}}$,\\
		$ \textbf{\textit{v}}^{(p-k)} \leftarrow \underset{\tiny \makecell{\textbf{\textit{z}} \in \mathbb{R}^{n_{u}}} }{\text{argmin}} \left\Vert \frac{\partial \nabla.\textbf{\textit{f}}^{(p-k)}}{\partial  \textbf{\textit{u}}^{(p-k)}}\textbf{\textit{z}}+\frac{\partial \nabla.\textbf{\textit{f}}^{(p-k)}}{\partial \mathcal{S}_{1}} \right\Vert_{2}^2$.\\
		\If{$k>1$}{
			Compute $\frac{\partial \nabla.\textbf{\textit{f}}^{(p-k+1)}}{\partial  \textbf{\textit{u}}^{(p-k+1)}}$, and $\frac{\partial \nabla.\textbf{\textit{f}}^{(p-k+1)}}{\partial \mathcal{S}_{1}}$,\\	
			$\textbf{\textit{v}}^{(p-k+1)} \leftarrow \underset{\tiny \makecell{\textbf{\textit{z}} \in \mathbb{R}^{n_{u}}} }{\text{argmin}} \left\Vert \Big(\mathcal{I}+\Delta t \frac{\partial \nabla.\textbf{\textit{f}}^{(p-k+1)}}{\partial  \textbf{\textit{u}}^{(p-k+1)}}\Big)\textbf{\textit{z}}- \textbf{\textit{v}}^{(p-k)}-\Delta t\frac{\partial \nabla.\textbf{\textit{f}}^{(p-k+1)}}{\partial \mathcal{S}_{1}} \right\Vert_{2}^2$.\\
			Note: a similar approach should be used for higher $k$-steps (i.e., BDF3).
		}
		\caption{Initial condition approximation for unsteady data sampling using BDF scheme.}
		\label{alg:dataV_us_ic}
	}
\end{algorithm}
\begin{algorithm}[h]
	\scriptsize{
		\SetAlgoLined
		\KwIn{$\textbf{U}(\mathcal{S}_{1})$, $m_{k}, n_{u}, \acute{m}_{p}, \Delta t, T_{s}, \mathcal{S}_{1}$, and $\mathbb{N}_{\acute{p}}=\{\xi_{\acute{p}}^{(1)}, \xi_{\acute{p}}^{(2)}, \dots, \xi_{\acute{p}}^{(\acute{m}_{p})}\}$.}
		\KwOut{$\textbf{V}_{us}$.}
		Build an empty matrix $\textbf{V}_{us} \in \mathbb{R}^{n_{u} \times \acute{m}_{p}}$,\\
		Divide $\mathbb{N}_{\acute{p}}$ into $n_{p}$ parts: $\mathbb{N}_{\acute{p}}=\{ \mathbb{N}_{\acute{p}}^{1}, \dots, \mathbb{N}_{\acute{p}}^{n_{p}}   \}$, where $n_{p} \ll \acute{m}_{p}$,\\
		$c \leftarrow 0$,\\
		$m_{s} \leftarrow \Delta t T_{s}$.\\
		\ForEach{$\mathbb{N}_{\acute{p}}^{i} \quad \text{\textbf{in}} \quad \mathbb{N}_{\acute{p}}$}{
			\For{$p \quad \text{\textbf{in}} \quad \mathbb{N}_{\acute{p}}^{i}$}{
				\vspace{5pt}
				Approximate initial condition using Algorithm \ref{alg:dataV_us_ic}\\	
				Set $\acute{m}^{(0)} \leftarrow p$,\\	
				\For{$l= 0 \quad \text{\textbf{to}} \quad m_{k}$}{
					$\acute{m}^{(l+1)} \leftarrow \acute{m}^{(l)}+m_{s}$,\\
					\For{$n = \acute{m}^{(l)} \quad \text{\textbf{to}} \quad \acute{m}^{(l+1)}$}{
						Compute $\frac{\partial \nabla.\textbf{\textit{f}}^{(n)}}{\partial  \textbf{\textit{u}}^{(n)}}$, and $\frac{\partial \nabla.\textbf{\textit{f}}^{(n)}}{\partial \mathcal{S}_{1}}$,\\
						$\frac{\partial \textbf{\textit{r}}^{(n)}}{\partial  \textbf{\textit{u}}^{(n)}} \leftarrow \zeta_0\mathcal{I} +\beta_0 \Delta t \frac{\partial \nabla.\textbf{\textit{f}}^{(n)}}{\partial  \textbf{\textit{u}}^{(n)}}$, \quad
						$\frac{\partial \textbf{\textit{r}}^{(n)}}{\partial  \textbf{\textit{u}}^{(n-j)}} \leftarrow \zeta_j$, \quad
						$\frac{\partial \textbf{\textit{r}}^{(n)}}{\partial \mathcal{S}_{1}} \leftarrow \beta_0 \Delta t\frac{\partial \nabla.\textbf{\textit{f}}^{(n)}}{\partial \mathcal{S}_{1}}$,\\	
						$\textbf{\textit{v}}^{(n)} \leftarrow \underset{\tiny \makecell{\textbf{\textit{z}} \in \mathbb{R}^{n_{u}}} }{\text{argmin}} \left\Vert \frac{\partial \textbf{\textit{r}}^{(n)}}{\partial  \textbf{\textit{u}}^{(n)}} \textbf{\textit{z}} + \sum_{\textit{j}=1}^{k} \frac{\partial \textbf{\textit{r}}^{(n)}}{\partial  \textbf{\textit{u}}^{(n-j)}}  \textbf{\textit{v}}^{(n-j)} + \frac{\partial \textbf{\textit{r}}^{(n)}}{\partial \mathcal{S}_{1}} \right\Vert_{2}^2$.\\	
					}
					$\textbf{V}_{us}[:,c] \leftarrow  \textbf{\textit{v}}^{(\acute{m}^{(l+1)})}$,\\
					$c \leftarrow c +1$,\\
					\For{$j= 0 \quad \text{\textbf{to}} \quad k$}{
						$\textbf{Q}\textbf{R} \leftarrow  \textbf{\textit{v}}^{(\acute{m}^{(l+1)}-k)}$,\\
						$ \textbf{\textit{v}}^{(\acute{m}^{(l+1)}-k)} \leftarrow \textbf{Q}$.\\
					}
				}
			}
		}
		\caption{Data collection from unsteady sensitivity function using BDF scheme.}
		\label{alg:dataV_us}
	}
\end{algorithm}

Let us assume that $\mathcal{S}_{1}$ is the baseline design parameter that reflects the properties of the current problem, and the rest of design parameters $\{\mathcal{S}_{2}, \dots , \mathcal{S}_{n_{d}}\}$ are sample designs for evaluation. Algorithms \ref{alg:dataV_us_ic} and \ref{alg:dataV_us} are only applied to $\mathcal{S}_1$ since this part of data collection is computationally expensive. Keep in mind that Algorithm \ref{alg:dataV_us_ic} sets a proper initial condition to proceed with data sampling using Algorithm \ref{alg:dataV_us}. Therefore, after data sampling using the aforementioned algorithms, the data are collected and represented as
\begin{equation}
	\textbf{V}_{us}(\mathcal{S}_{1}) =
	\begin{bmatrix} 
		| & | & & |\\
		\Delta \textbf{\textit{v}}_{us}^{(\mathbb{N}_{\acute{p}}(1))}(\mathcal{S}_{1}) & \Delta \textbf{\textit{v}}_{us}^{(\mathbb{N}_{\acute{p}}(2))}(\mathcal{S}_{1}) & ... & \Delta \textbf{\textit{v}}_{us}^{(\mathbb{N}_{\acute{p}}(\acute{m}_{p}))}(\mathcal{S}_{1}) \\
		| & | & & |
	\end{bmatrix} \in \mathbb{R}^{n_{u}\times \acute{m}_{p}}, \quad \Delta \textbf{\textit{v}}_{us}^{(n)}=\textbf{\textit{v}}_{us}^{(n)}-\overline{\textbf{\textit{v}}_{us}},
	\label{eq:pod5b}
\end{equation}
where, $\textbf{V}_{us} \in \mathbb{R}^{n_{u} \times \acute{m}_{p}}$ represents the dataset for prototype of the unsteady sensitivity solutions, $\acute{m}_{p} \in \mathbb{R}_{+}$ is the total number of samples. Moreover, $\textbf{\textit{v}}_{us}^{(n)} \in \mathbb{R}^{n_{u}}$ and $\overline{\textbf{\textit{v}}_{us}} \in \mathbb{R}^{n_{u}}$ correspond to the discrete unsteady sensitivity solution and its corresponding reference vector, respectively. Since we only consider the dataset for $\mathcal{S}_{1}$ in the unsteady case, then $\acute{\textbf{X}}_{sens} = \textbf{V}_{us}(\mathcal{S}_{1})$, where $\acute{\textbf{X}}_{sens} \in \mathbb{R}^{n_{u} \times \acute{m}_{p}}$ represents the dataset of the unsteady sensitivity solutions. Therefore, the ROB for the unsteady sensitivity solutions is given by
\begin{equation}
	\acute{\tilde{\mathbf{\Phi}}}_{v} = \textbf{POD}(\acute{\textbf{X}}_{sens}) \quad \text{and} \quad \textbf{V}_{us}(\mathcal{S}_{1}) = \acute{\tilde{\mathbf{\Phi}}}_{v} \acute{\mathcal{Q}}_{v}(\mathcal{S}_{1}),
	\label{eq:pod7c}
\end{equation}
and the generalized coordinates for Eq. (\ref{eq:pod7c}) are
\begin{equation}
	\acute{\mathcal{Q}}_{v} =
	\begin{bmatrix} 
		| & | & & |\\
		\acute{\tilde{\textbf{\textit{h}}}}_{1}^{(\mathbb{N}_{\acute{p}}(1))} & \acute{\tilde{\textbf{\textit{h}}}}_{1}^{(\mathbb{N}_{\acute{p}}(2))} & ... & \acute{\tilde{\textbf{\textit{h}}}}_{1}^{(\mathbb{N}_{\acute{p}}(\acute{m}_{p}))} \\
		| & | & & |
	\end{bmatrix} \in \mathbb{R}^{\acute{r}_{sens} \times \acute{m}_{p}}, \quad \mathbb{N}_{\acute{p}}(n)=\{\xi_{\acute{p}}^{(1)}, \xi_{\acute{p}}^{(2)}, \dots, \xi_{\acute{p}}^{(\acute{m}_{p})}\},
	\label{eq:pod8}
\end{equation}
where $\acute{r}_{sens}$ is the rank of $\acute{\tilde{\mathbf{\Phi}}}_{v}$. The relative scale of the dataset directly influences the effectiveness of the \textbf{POD} function. Therefore, the \textbf{POD} function can not find the most effective trial basis function if we use $\textbf{POD}([\textbf{X}_{state}, \textbf{X}_{sens}, \acute{\textbf{X}}_{sens}])$ \cite{huynh2007reduced, carlberg2008compact}. Consequently, according to Ref.~\cite{zahr2015progressive}, a Gram-Schmidt-like procedure is suggested, such that the ROB are embedded as a subset of trial basis function
\begin{equation}
	\tilde{\mathbf{\Phi}} = \tilde{\mathbf{\Phi}}_{u} \oplus \tilde{\mathbf{\Phi}}_{v} \oplus \acute{\tilde{\mathbf{\Phi}}}_{v}, \quad \text{and} \quad \tilde{\mathbf{\Phi}} \in \mathbb{R}^{n_{u} \times r},
	\label{eq:pod9}
\end{equation}
where, $r = r_{state}+r_{sens}+\acute{r}_{sens}$. The procedure for building the trial basis functions from a different set of ROB is explained in Algorithm \ref{alg:ROB}. 
\begin{algorithm}[h]
	\scriptsize{
		\SetAlgoLined
		\KwIn{$\textbf{U}$, $\textbf{V}_{ss}$, $\textbf{V}_{us}$, $m_{u}$, $m_{p}$, $\acute{m}_{p}$, $r$, and $\{\mathcal{S}_{i}, \mathcal{S}_{2}, \dots, \mathcal{S}_{n_{s}} \}$.}	
		\KwOut{$\tilde{\mathbf{\Phi}}$.}	
		
		\ForEach{$\mathcal{S}_{i} \quad \text{\textbf{in}} \quad \{\mathcal{S}_{1}, \mathcal{S}_{2}, ..., \mathcal{S}_{n_{d}} \}$}{
			Build empty matrices: $\textbf{U}(\mathcal{S}_{i}) \in \mathbb{R}^{n_{u} \times m_{u}}$, \quad $\textbf{V}_{ss}(\mathcal{S}_{i}) \in \mathbb{R}^{n_{u} \times m_{p}}$, \quad and\quad $\textbf{V}_{us}(\mathcal{S}_{i}) \in \mathbb{R}^{n_{u} \times \acute{m}_{p}}$.\\
			
			Compute $\overline{\textbf{\textit{u}}} \leftarrow \frac{1}{m_{u}}\sum_{j=0}^{m_{u}}{  \textbf{\textit{u}}^{(j)} }$, \quad $\overline{\textbf{\textit{v}}_{ss}} \leftarrow \frac{1}{m_{p}}\sum_{j=0}^{m_{p}}{ \textbf{\textit{v}}_{ss}^{(j)} }$, \quad and\quad $\overline{\textbf{\textit{v}}_{us}} \leftarrow \frac{1}{\acute{m}_{p}}\sum_{j=0}^{\acute{m}_{p}}{ \textbf{\textit{v}}_{us}^{(j)} }$.\\
			
			\For{$j=0 \quad \text{\textbf{to}} \quad m_{u}$}{
				$\textbf{U}(\mathcal{S}_{i})[:, j] \leftarrow \textbf{\textit{u}}^{(j)}-\overline{\textbf{\textit{u}}}$,\\
			}
			\For{$j=0 \quad \text{\textbf{to}} \quad m_{p}$}{
				$\textbf{V}_{ss}(\mathcal{S}_{i})[:, j] \leftarrow \textbf{\textit{v}}_{ss}^{(j)}-\overline{\textbf{\textit{v}}_{ss}}$,\\
			}		
		}
		\For{$j=0 \quad \text{\textbf{to}} \quad \acute{m}_{p}$}{
			$\textbf{V}_{us}(\mathcal{S}_{1})[:, j] \leftarrow \textbf{\textit{v}}_{us}^{(j)}-\overline{\textbf{\textit{v}}_{us}}$,\\
		}
		
		$\textbf{X}_{state} \leftarrow \Big[\textbf{U}(\mathcal{S}_{1}), \textbf{U}(\mathcal{S}_{2}), \dots, \textbf{U}(\mathcal{S}_{n_{d}}) \Big]$,\\
		$\textbf{X}_{sens} \leftarrow \Big[\textbf{V}_{ss}(\mathcal{S}_{1}), \textbf{V}_{ss}(\mathcal{S}_{2}), \dots, \textbf{V}_{ss}(\mathcal{S}_{n_{d}}) \Big]$,\\
		$\acute{\textbf{X}}_{sens} \leftarrow \Big[\textbf{V}_{us}(\mathcal{S}_{1})\Big]$,\\
		
		$\tilde{\mathbf{\Phi}}_{u} \leftarrow  \textbf{POD}(\textbf{X}_{state})$ using Algorithm \ref{alg:pod},\\
		$\tilde{\mathbf{\Phi}}_{v} \leftarrow  \textbf{POD}(\textbf{X}_{sens})$ using Algorithm \ref{alg:pod},\\
		$\acute{\tilde{\mathbf{\Phi}}}_{v} \leftarrow \textbf{POD}(\acute{\textbf{X}}_{sens})$ using Algorithm \ref{alg:pod},\\
		$\tilde{\mathbf{\Phi}} \leftarrow \Big[ \tilde{\mathbf{\Phi}}_{u}, \tilde{\mathbf{\Phi}}_{v}, \tilde{\acute{\mathbf{\Phi}}}_{v} \Big]$.
		
		\caption{ROB and trial basis function.}
		\label{alg:ROB}
	}
\end{algorithm}
It is worth mentioning that the LSPG approach, with which the solutions of the ROM are obtained in a minimization framework at each time interval, has the minimum-residual property. In other words, $\tilde{\mathbf{\Phi}}_{u}$ would not influence the results of the sensitivity functions significantly. Furthermore, the POD modes embedded in $\tilde{\mathbf{\Phi}}$ can be rearranged based on their priority. Therefore, the O$\Delta$Es in the closure model are solved based on the priority of prominent features in the dynamical system.
\subsection{Hyper-reduction}
\label{subsec:hyper-reduction}
Developing a closure model for a large-scale chaotic FOM is often computationally expensive. Dimensionality reduction of the Jacobian and relative matrices needs many scalable computational resources. Although the Jacobian of the dynamical system is represented as a sparse matrix, the trial and test basis functions of this dynamical system are dense matrices and, hence, their mathematical operations are computationally expensive. Therefore, building these terms of the closure model, given by Eq. (\ref{eq:rom-kkt-final}), is still limited to available computational resources. These terms are $\tilde{\mathbf{\Psi}}^{(n)\top}\tilde{\mathbf{\Psi}}^{(n)}$, $\tilde{\mathbf{\Psi}}^{(n)\top}\tilde{\mathbf{\Phi}}$, $\tilde{\mathbf{\Psi}}^{(n)\top}\frac{\partial \textbf{\textit{r}}^{(n)}}{\partial \mathcal{S}}$, and $\tilde{\mathbf{\Psi}}^{(n)\top}\Big(\sum_{\textit{j}=1}^{k}\zeta_{j}\mathcal{I} + \frac{\partial \textbf{\textit{r}}^{(n)}}{\partial \textbf{\textit{u}}^{(n)}}\Big)$, which are required to solve Eq. (\ref{eq:rom-kkt-final}) for the sensitivity solutions in Hilbert space. The remainder of this section shows how to reduce this computational burden using another dimensionality reduction, which is referred to as hyper-reduction. 

Deep learning effectively captures non-linear patterns, and characteristics in a dataset. We construct neural network surrogate models to predict patterns in the aforementioned terms. To this end, we use Multi-Layer Perceptron (MLP) architecture to develop a surrogate model using a neural network. First, we compute the aforementioned terms in Eq. (\ref{eq:rom-kkt-final}) over a limited set of time intervals, as the sample dataset. In the next step, using dimensionality reduction techniques, the generalized coordinates of this dataset are obtained. Later, we use these coordinates to train the network. The training objective function, RMSE, is minimized with respect to the sample dataset
\begin{equation}
	\text{RMSE} = \sqrt{\frac{1}{m_{\text{samp}}}\sum_{n=1}^{m_{\text{samp}}}\big(q_{\text{pred}}^{(\mathbb{N}_{s}(n))} - q_{\text{samp}}^{(\mathbb{N}_{s}(n))}\big)^{2}},
	\label{eq:ml-rmse}
\end{equation}
where, $q_{\text{samp}}$, and $q_{\text{pred}}$ are scalar generalized coordinates for sample and prediction, respectively. Since the objective is an accurate prediction of the generalized coordinates over $\Delta \mathbb{T}$, we train the surrogate model in $\mathbb{N}_{s}(n)=\{\xi_{s}^{(1)}, \xi_{s}^{(2)}, \dots, \xi_{s}^{(m_{\text{samp}})}\}$. Consequently, developing surrogate models via deep learning allows us to predict the aforementioned terms in the remainder of time intervals over $\Delta \mathbb{T}$ with minimal computational cost.  
\subsection{Shaping the trial subspaces}
\label{subsec:shaping}
As the FOM becomes non-linear, the mapping functions will be highly sensitive to projection errors. This issue influences the accuracy and robustness of ROMs. Although closure models developed by LSPG are unconditionally stable over $\Delta \mathbb{T}$, the accuracy of the sensitivity solutions are not guaranteed. Most errors in the sensitivity solutions come from the trial basis function, which causes the sensitivity solution diverges from the exact one, i.e., $\forall t^{(n)} \in \Delta \mathbb{T}\colon  \textbf{\textit{v}}^{(n)} \neq \overline{\textbf{\textit{v}}} + \tilde{\mathbf{\Phi}} \tilde{\textbf{\textit{h}}}^{(n)}$. Additionally, a combination of different ROB may deviate the manifolds in $\mathscr{H}$. However, we should note that this is not the usual case.

In the \textit{shaping} step, we perform LSS to compute sensitivities, and if those residuals in LSS are large, we minimize the $\ell^2$-norm of them after embedding the updated sensitivity solution, $\tilde{\textbf{\textit{h}}}_{c} := \tilde{\textbf{\textit{h}}}( \mathbf{\Phi}_{v}\mathbf{\Theta}_{v})$, where $\mathbf{\Theta}_{v} \in \mathbb{R}^{r \times r}$ is a linear operator that reshapes the manifolds for all sets of subspaces. In this study, we set $\mathbf{\Theta}_{v} = \theta \mathcal{I}$ to make it a scaling matrix at which $\theta \in \mathbb{R}$ is a factor to scale the manifolds in each subspace. The updated manifold results in changes to the mapping function, such that $\mathbf{\Phi}\mathbf{\Theta}_{v} \tilde{\textbf{\textit{h}}}_{c}^{(n)} \mapsto \textbf{\textit{v}}^{(n)}$, which indicates that changes in the updated manifolds will influence the sensitivity solution in lower dimensions. Therefore, this numerical procedure can be defined as a minimization problem 
\begin{equation}
	\begin{split}
		\tilde{\textbf{\textit{h}}}_{c}, \mathbf{\Theta}_{v} := \underset{\tiny \makecell{\textbf{\textit{z}} \in \mathbb{R}^{n_{u}},\\ \textbf{Y} \in \mathbb{R}^{r\times r}} }{\text{argmin}} \quad \sum_{n=1}^{m_{q}} \left\Vert \frac{\partial \big(\overline{\textbf{\textit{u}}} + \tilde{\mathbf{\Phi}}_{u} \tilde{\textbf{\textit{q}}}^{(\mathbb{N}_{q}(n))}\big)}{\partial \mathcal{S}} - \tilde{\mathbf{\Phi}}\textbf{Y} \textbf{\textit{z}}^{(\mathbb{N}_{q}(n))}\right\Vert_{2}^2,
	\end{split}
	\label{eq:shape1}
\end{equation}
where $\mathbb{N}_{q}(n)=\{\xi_{q}^{(1)}, \xi_{q}^{(2)}, \dots, \xi_{q}^{(m_{q})} \}$ are a set of random numbers, such that $\xi^{(n)} \Delta t \in \Delta \mathbb{T}$, and $m_{q}$ shows the number of samples chosen randomly for this minimization. We note that $\textbf{\textit{z}}$ is updated using LSS after each minimization iteration in Eq. (\ref{eq:shape1}) to ensure the sensitivity solution remains optimal in $\mathscr{H}$. As mentioned earlier, Eq. (\ref{eq:shape1}) can be considered to ensure the updated sensitivity results are not far from the actual values. The samples we use in Eq. (\ref{eq:shape1}) could be the steady-state sensitivity solutions. We acknowledge that the steady-state sensitivity solutions are not exact for non-linear dynamical systems. However, it gives us useful insights into what the unsteady sensitivity solutions would be, and having these insights may improve the final sensitivity results. Moreover, the \textit{shaping} phase helps fix $\tilde{\mathbf{\Phi}}_{v}$ if it is contaminated with inaccurate results. Nevertheless, we consider Eq. (\ref{eq:shape1}) as an optional step for sensitivity analysis and optimization.

\section{Overview}
\label{sec:overview}
In Sections \ref{sec:least square shadowing}, \ref{sec:lagrange multiplier}, \ref{sec:lspg}, and \ref{sec:learning-strategy}, we discussed the fundamentals required to proceed with convex or non-convex optimization in the presence of the chaotic dynamical system. However, in this section, we focus on how to solve the ROM-constrained optimization for large-scale problems.
\begin{figure}[h] 
	\centering
	\includegraphics[width=0.8\textwidth]{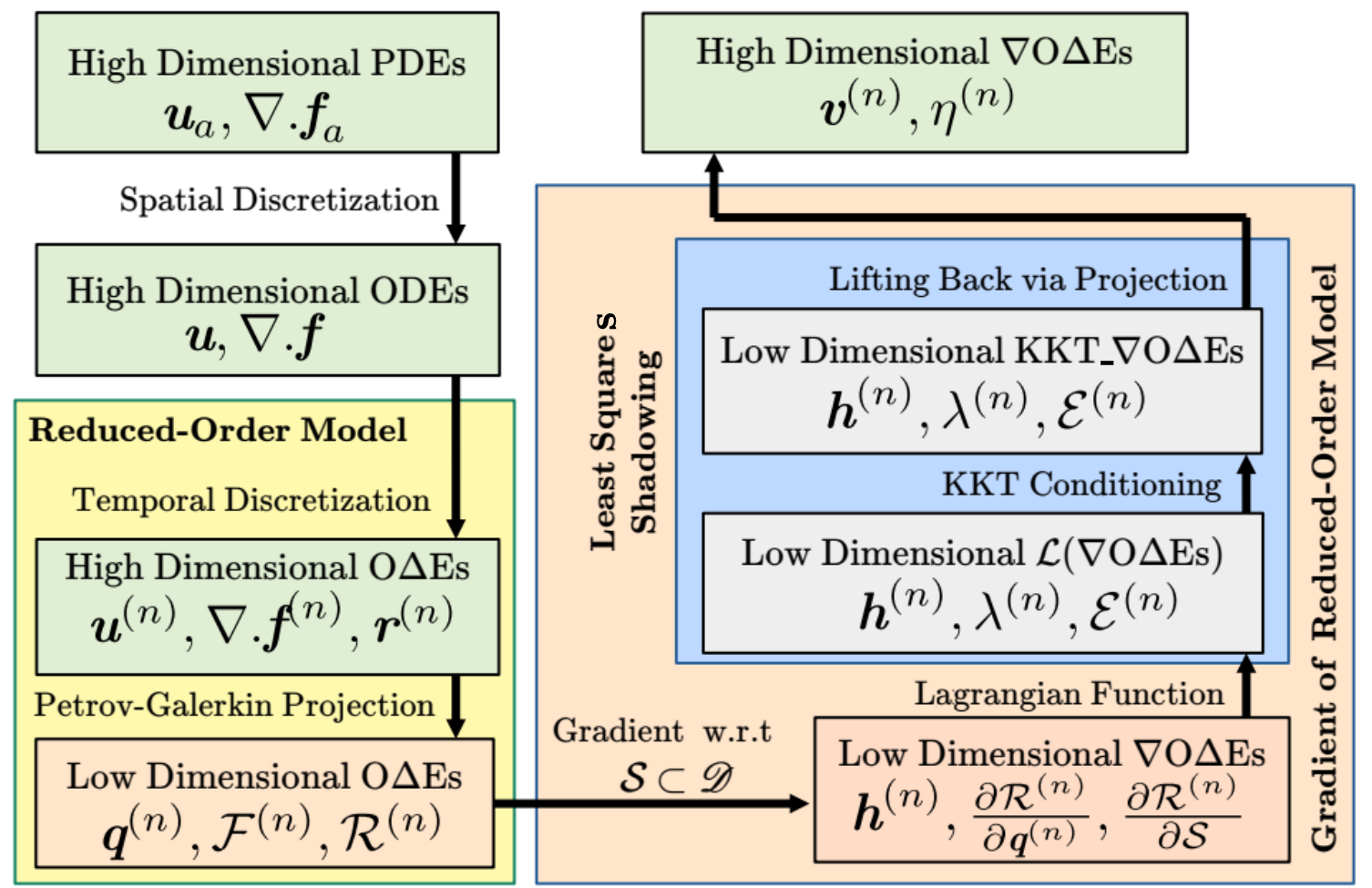}
	\caption{Flow-chart describing the proposed platform for sensitivity analysis.}
	\label{fig:flow-chart}
\end{figure}

Figure \ref{fig:flow-chart} displays this procedure for computing sensitivities of the objective function with respect to all design parameters, $\mathcal{S}$. At first, the high-dimensional Navier-Stokes equations, as a set of PDEs, $\nabla.\textbf{\textit{f}}_{a}$, are spatially discretized using the FR approach, and then converted into a set of ODEs, $\nabla.\textbf{\textit{f}}$. To develop a closure model using \textit{projection-based} approaches, we discretize these high-dimensional ODEs in the time domain, referred to as $\text{O}\Delta\text{Es}$, using an implicit temporal scheme, such as BDF scheme. Next, we project $\text{O}\Delta\text{Es}$ into lower-dimensional spaces, $\mathscr{H}$, using the Petrov-Galerkin projection. This projection yields a lower-dimensional set of $\text{O}\Delta\text{Es}$ that can be used as a reference to derive the sensitivity function. Derivatives of this closure model with respect to $\mathcal{S}$ are referred to as $\nabla\text{O}\Delta\text{Es}$, which is the main challenge in the present study. This is because the Jacobian matrices should be significantly reduced in size, such that they provide sufficient information regarding the dynamics of the corresponding FOM in $\mathscr{H}$. Finally, applying the Lagrange function, $\mathcal{L}(\nabla \text{O} \Delta \text{Es})$, accompanied by Karush-Kuhn-Tucker (KKT) conditioning leads to $\text{KKT-}\nabla \text{O} \Delta \text{Es}$, which is the final set of equations that reflects the sensitivity solution in $\mathscr{H}$. Finally, these sensitivities from $\mathscr{H}$ are lifted back into physical space, $\mathscr{P}$, resulting in the solution of $\nabla\text{O}\Delta\text{Es}$ in high-dimensional space.

\section{Results and Discussion}
\label{sec:results}

\subsection{Shape optimization of a 2D flapping wing}
\label{subsec:2dflappingwing}
\subsubsection{Flow setup and validation}
\label{subsubsec:valid}
An unstructured mesh is used to simulate flow past a 2D symmetric flapping wing. A NACA 0012 airfoil is placed 15$c$ and 25$c$ away from the upstream and downstream boundaries, respectively. Second-order BDF2 temporal scheme with $\Delta t=0.01T^{*}$ is chosen to solve the discretized Navier-Stokes equations. In Table \ref{tab:valid1}, the time-averaged aerodynamic results for different solution polynomial degrees, $p_{s}$, are compared with forces in Ref.~\cite{moriche_flores_garcia-villalba_2017}. In this problem, kinematic parameters are $\text{y}_{0}=c$, $k_{a}=1.41$, $T^{*}=2.22c/u_{\infty}$, $\theta_{m}=10^{\circ}$, $\theta_{0}=30^{\circ}$, and $\theta_{s}=90^{\circ}$. Each simulation initially is run until $10T^{*}$, and then aerodynamic loads are time-averaged over the next $7T^{*}$. The present numerical simulations demonstrate that all polynomial solution degrees give appropriate results, close to those in Ref.~\cite{moriche_flores_garcia-villalba_2017}.
\begin{table}[H]
	\centering
	\begin{tabular}{|c|c|c|c|c|c|}
		\hline
		$p_{s}$&DOF& $\overline{C_{L}}$ & $\overline{C_{T}}$&$C_{L, rms}$&$C_{D, rms}$\\
		\hline
		1&$9.89\times10^{4}$&1.727&0.717&2.827&0.945\\
		2&$2.22\times10^{5}$&1.762&0.712&2.840&0.951\\
		3&$3.95\times10^{5}$&1.7409&0.716&2.833&0.947\\
		\cite{moriche_flores_garcia-villalba_2017}&$18.4\times 10^{6}$&1.5507&0.7245&2.7743&0.9224\\
		\hline
	\end{tabular}
	\caption{Comparing the current results with those in Ref.~\cite{moriche_flores_garcia-villalba_2017}.}
	\label{tab:valid1}
\end{table}

\begin{figure}[H]
	\centering
	\begin{subfigure}[h]{0.22\textwidth}
		\centering
		\includegraphics[width=1\textwidth]{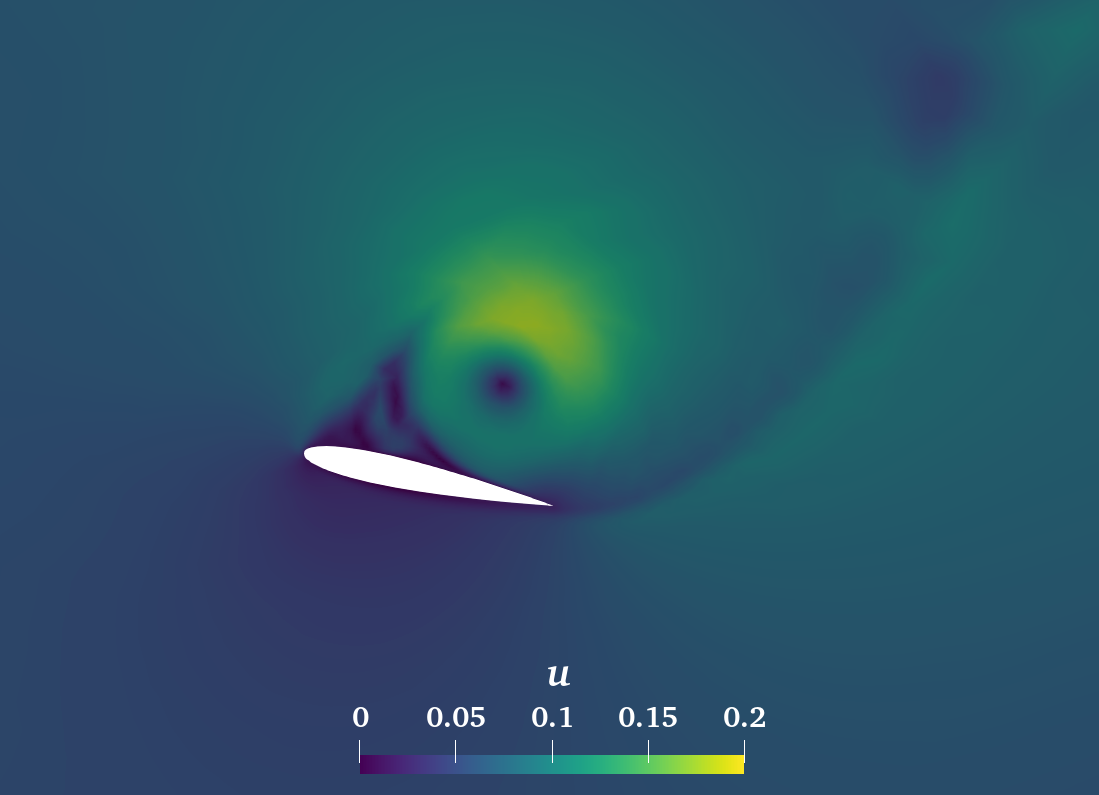}
		\caption{$p_{s}=1$}
		\label{fig-u-contours1a}
	\end{subfigure}
	\begin{subfigure}[h]{0.22\textwidth}
		\centering
		\includegraphics[width=1\textwidth]{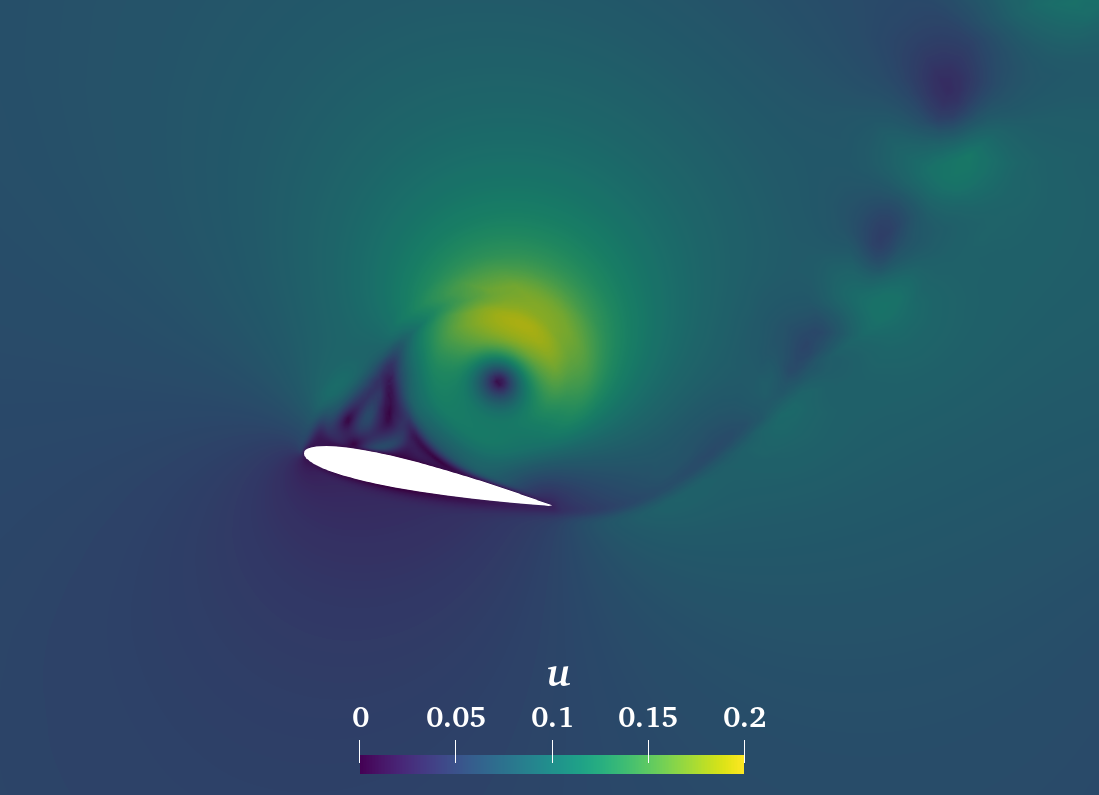}
		\caption{$p_{s}=2$}
		\label{fig-u-contours1b}
	\end{subfigure}
	\begin{subfigure}[h]{0.22\textwidth}
		\centering
		\includegraphics[width=1\textwidth]{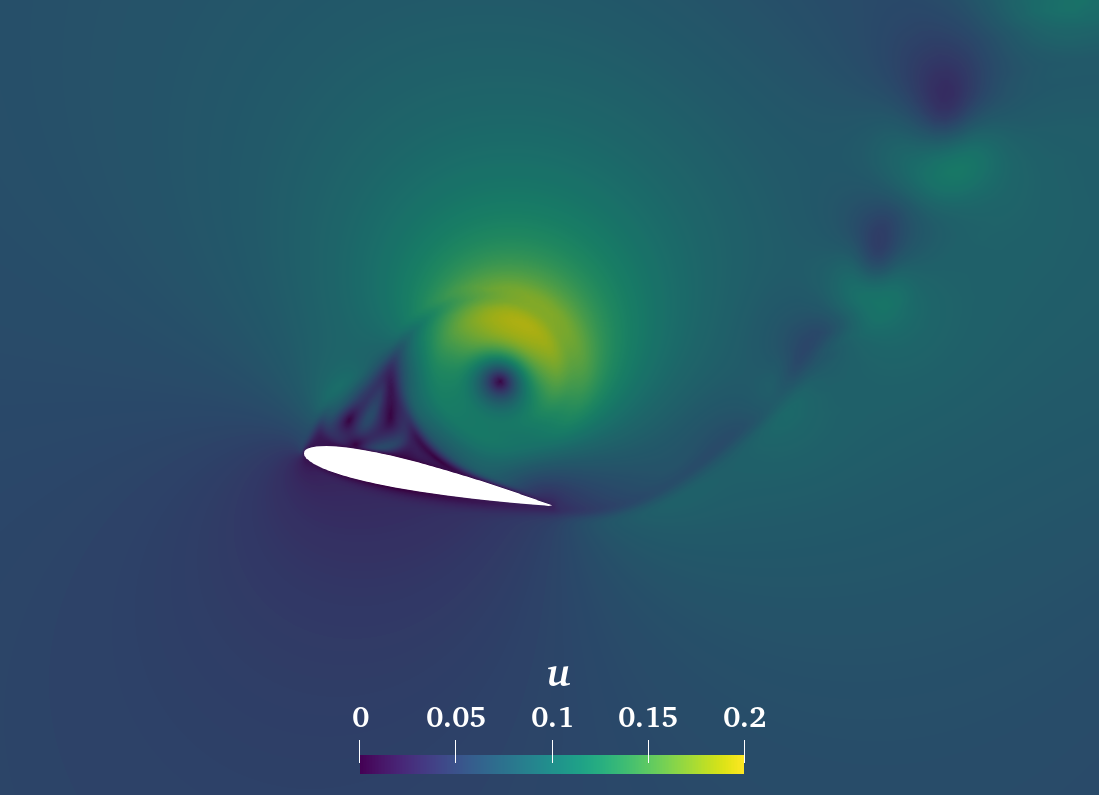}
		\caption{$p_{s}=3$}
		\label{fig-u-contours1c}
	\end{subfigure}
	\caption{Comparing velocity contours for different solution polynomial degrees at the instant $t/T^{*}=0.25$.}
	\label{fig-u-contours1}
\end{figure}

Figure \ref{fig-u-contours1} compares velocity contours for different solution polynomial degrees. In Figure \ref{fig-u-contours1a}, velocity contours for $p_{s}=1$ indicate that large dissipation exists in wake of the wing, compared to those with $p_{s}=2$. On the other hand, Figures \ref{fig-u-contours1b} and \ref{fig-u-contours1c} show identical velocity contours with higher resolutions, indicating local dissipations in wake of the wing remain minimal for these polynomial solution degrees. Since simulation with $p_{s}=2$ is computationally cheaper, compared to $p_{s}=3$, it is decided to choose $p_{s}=2$ for the rest of this study.
\subsubsection{Optimization setup}
\label{subsubsec:opt-setup-2d-flappingwing}
\begin{figure}[t] 
	\centering
	\includegraphics[width=0.6\textwidth]{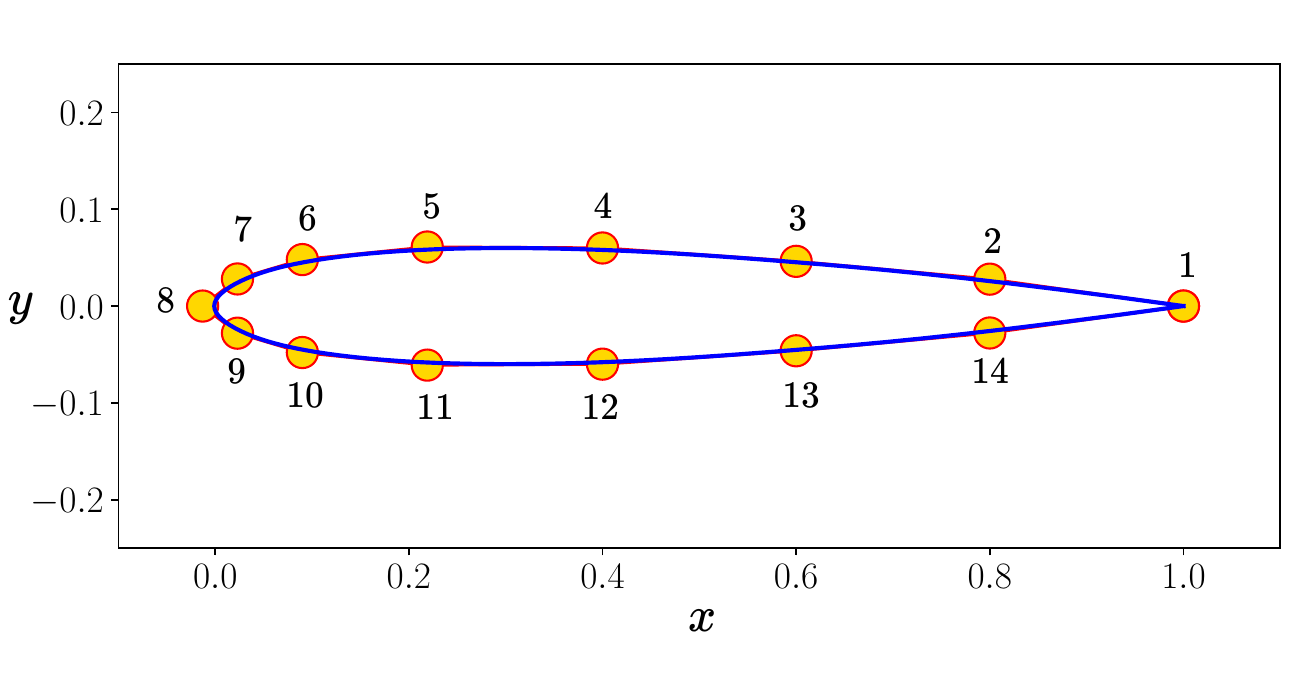}
	\caption{NACA 0012 coordinates defined by control points.}
	\label{fig:cps235}
\end{figure}

This section demonstrates the application of the proposed approach to unsteady aerodynamic optimization. We consider a 2D NACA 0012 flapping wing at $Re=1000$, and $Re=2400$, with a reduced frequency $k_{a}=1.41$. The effect of geometrical constraints on the design is also considered. Four different test cases were selected, as shown in Table \ref{tab:testcases534}. The first two cases, ``C1" and ``C2", were performed at $Re=1000$, and the last two cases, ``C3" and ``C4", at $Re=2400$. Geometrical constraints in ``C1" and ``C2" only limit the minimum thickness. On the other hand, more geometrical constraints are added to ``C3" and ``C4" to control the maximum/minimum thickness of the wing at different sections alongside the chord line.
\begin{table}[H]
	\centering
	\begin{tabular}{|c|c|c|c|c|c|c|c|c|}
		\hline
		Case &$Re$&$\theta_{m}$&$\theta_{0}$&$n_{s}$&Constraints&Bounds\\
		\hline
		C1&1,000&$10^{\circ}$&$30^{\circ}$&11&24&28\\ 
		\hline
		C2&1,000&$10^{\circ}$&$10^{\circ}$&11&24&28\\ 
		\hline
		C3&2,400&$10^{\circ}$&$30^{\circ}$&11&30&28\\ 
		\hline
		C4&2,400&$10^{\circ}$&$10^{\circ}$&11&30&28\\ 
		\hline
	\end{tabular}
	\caption{Test cases for optimization of a flapping wing at $\text{y}_{0}=c$, $k_{a}=1.41$, and $\theta_{s}=90^{\circ}$.}
	\label{tab:testcases534}
\end{table}
Figure \ref{fig:cps235} displays the control points (CPs) assigned to define coordinates, and shape of a NACA 0012 airfoil, as the baseline design. In optimization, CP1 is frozen (i.e., stationary) for all design cycles. Also, CP7, CP8, and CP9 are dependent on the coordinates of CP6 and CP10, which avoids low-quality mesh (i.e., highly skewed mesh, or sharp wall curvatures) during the optimization procedure. In addition to these control points, the pitching amplitude was added to design parameters. Therefore, the total number of design parameters in this optimization reaches $n_{s}=11$.

In general, the objective is to increase the time-averaged thrust coefficient. We also keep the time-averaged lift coefficient unchanged, since it balances with the gravitational force of an external body. Therefore, we define this optimization as
\begin{equation}
	\begin{split}
		&\underset{ \tilde{\textit{\textbf{q}}} \in \mathbb{R}^{r}, \mathcal{S} \in \mathscr{D} }{\text{minimize}} \quad \overline{\mathcal{J}} = \frac{1}{T^{*}}\int_{2T^{*}}^{3T^{*}} -C_{T}^{2} + \big(C_{L} - C_{L, target}\big)^{2}  dt,\\
		&\begin{split}
			\text{subject to}&\\
			&\mathcal{X}(t, \mathcal{S}),\\
			&\mathcal{R}(\overline{\textit{\textbf{u}}} + \tilde{\Phi} \tilde{\textit{\textbf{q}}}, t, \mathcal{X}, \mathcal{S}) = 0,\\
			&\frac{d\mathcal{R}}{d\mathcal{S}}(\overline{\textit{\textbf{u}}} + \tilde{\Phi} \tilde{\textit{\textbf{q}}}, t, \mathcal{X}, \mathcal{S}) = 0,\\
			&\textbf{C}(\mathcal{X}, \mathcal{S}) \leq 0,\\
		\end{split}
	\end{split}
	\label{eq:mini2}
\end{equation}
were $\textbf{C}$ is the constraint function. The target lift coefficient is $C_{L, target}=1.5$. This target lift guarantees additional improvement in the propulsive efficiency as $\overline{C_{T}}$ increases. In this optimization, the sensitivity function is solved using the first-order BDF1 scheme. In the sampling step, the steady-state sensitivity function is solved over 10\% of time intervals in $\Delta \mathbb{T}$ for each design parameter. Moreover, these design parameters were modified using Sequential Least-Squares Quadratic Programming (SLSQP) as an optimizer.
\begin{figure}[t]
	\centering
	\begin{subfigure}[t]{0.45\textwidth}
		\centering
		\includegraphics[width=\textwidth]{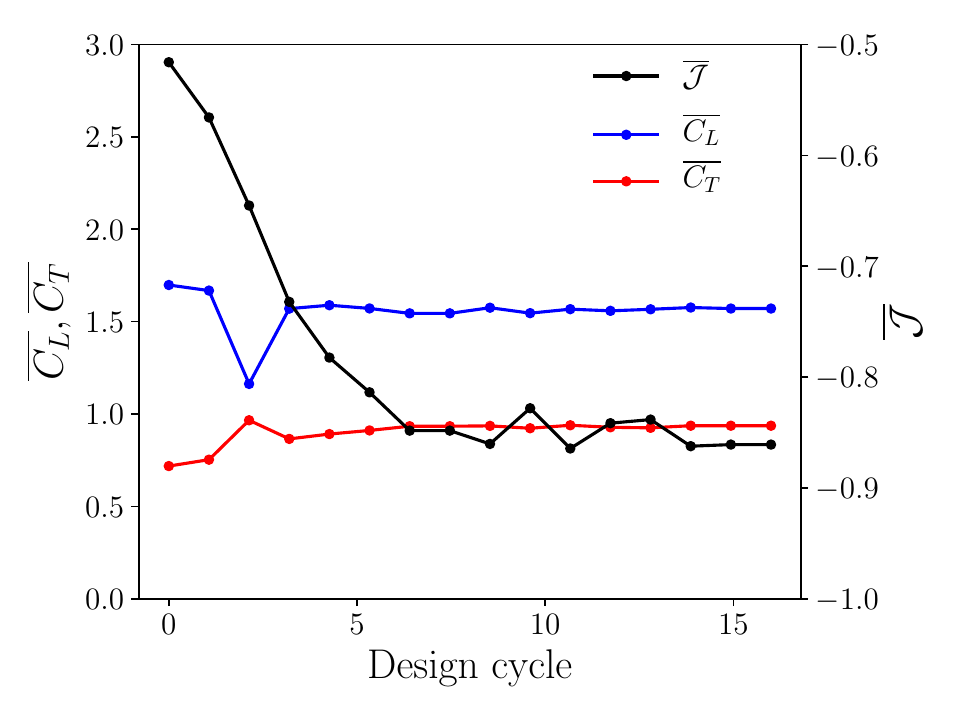}
		\caption{Test case ``C1''}
		\label{fig-inst-perf-case1a}
	\end{subfigure}
	\hfill
	\begin{subfigure}[t]{0.45\textwidth}
		\centering
		\includegraphics[width=\textwidth]{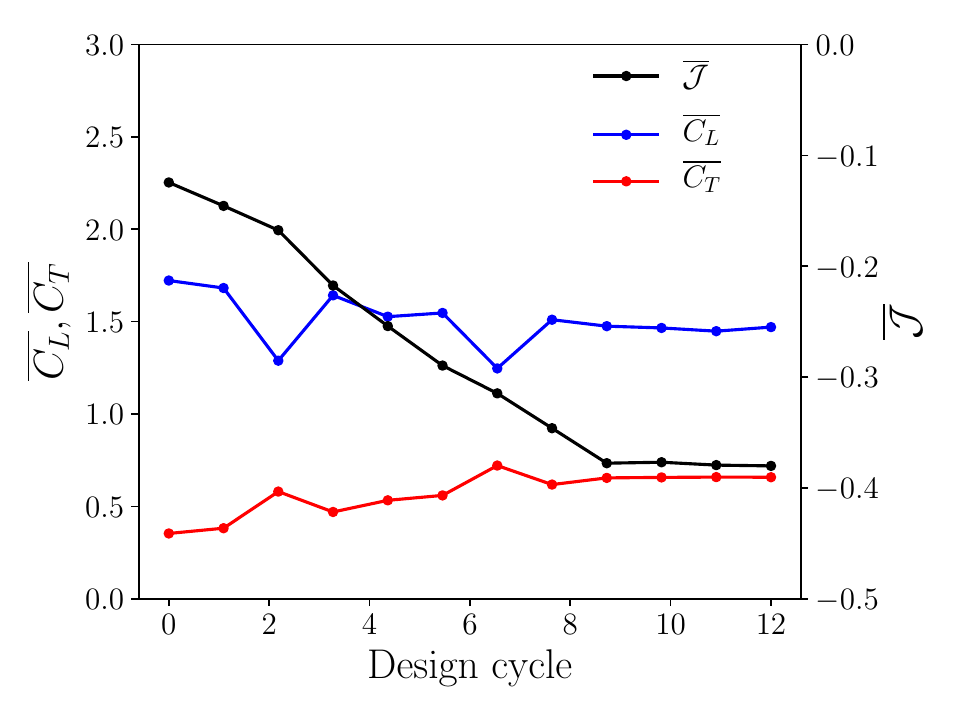}
		\caption{Test case ``C2''}
		\label{fig-inst-perf-case1b}
	\end{subfigure}
	
	\vspace{0.5em}
	\begin{subfigure}[t]{0.45\textwidth}
		\centering
		\includegraphics[width=\textwidth]{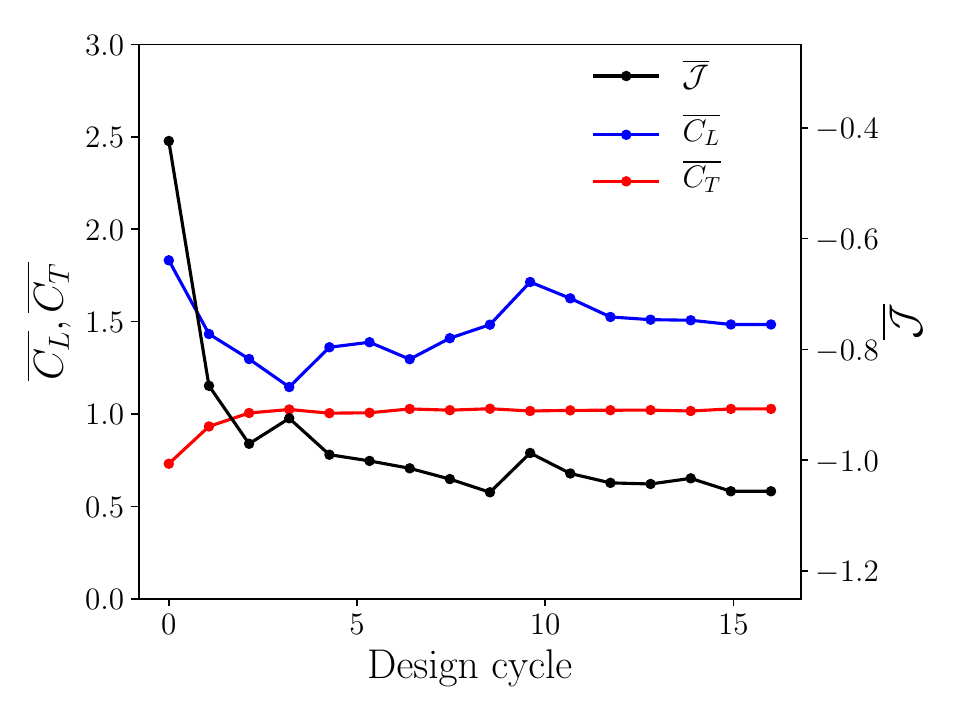}
		\caption{Test case ``C3''}
		\label{fig-inst-perf-case1c}
	\end{subfigure}
	\hfill
	\begin{subfigure}[t]{0.45\textwidth}
		\centering
		\includegraphics[width=\textwidth]{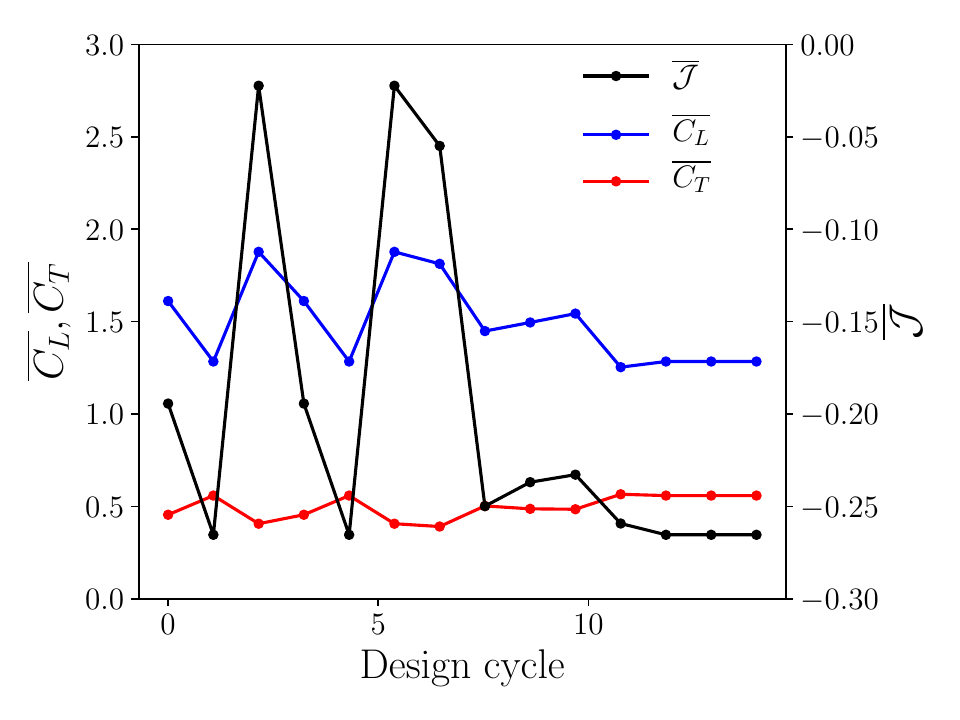}
		\caption{Test case ``C4''}
		\label{fig-inst-perf-case1d}
	\end{subfigure}
	\caption{Progressive results of optimization for different test cases.}
	\label{fig-inst-perf-case1}
\end{figure}

\subsection{Optimization performance}
\label{subsubsec:opt-performance-2d-flappingwing}
Figure \ref{fig-inst-perf-case1} shows the results of optimization for each case. In general, the time-averaged objective function, $\overline{\mathcal{J}}$, decreases after each design cycle. However, we see that the objective function increases at some design cycles. This is because the design space is bounded, or limited, by geometrical constraints. Another reason is also attributed to fluid physics. Ref.~\cite{durante2020bifurcations} showed that flow past a 2D NACA 0012 airfoil at $Re=1000$ has the potential to be chaotic at high angles of attack in post-stall. Additionally, Ref.~\cite{fernandez2017lyapunov} considered chaoticity of fluid physics for the same airfoil, but at $Re=2400$, and they confirmed that flow remains chaotic beyond stall. 
\begin{figure}[t]
	\centering
	\begin{subfigure}[t]{0.45\textwidth}
		\centering
		\includegraphics[width=\textwidth]{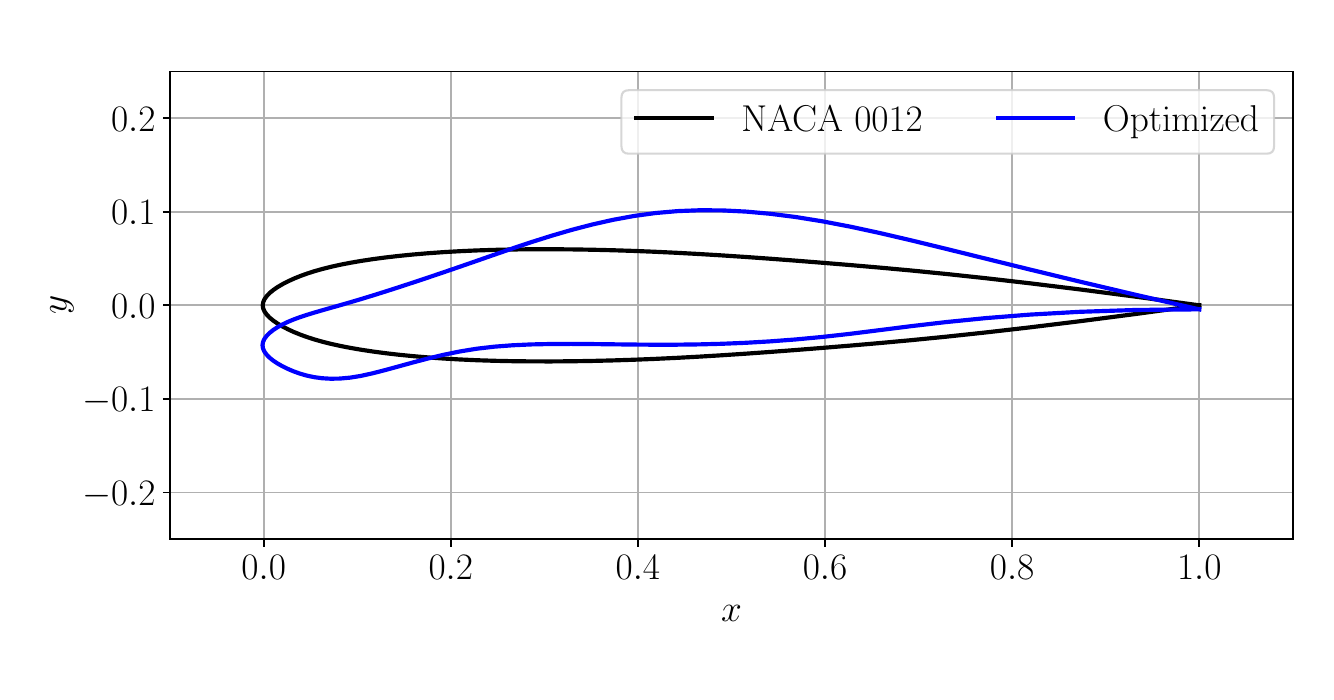}
		\caption{Test case ``C1''}
		\label{fig-shape-coma}
	\end{subfigure}
	\hfill
	\begin{subfigure}[t]{0.45\textwidth}
		\centering
		\includegraphics[width=\textwidth]{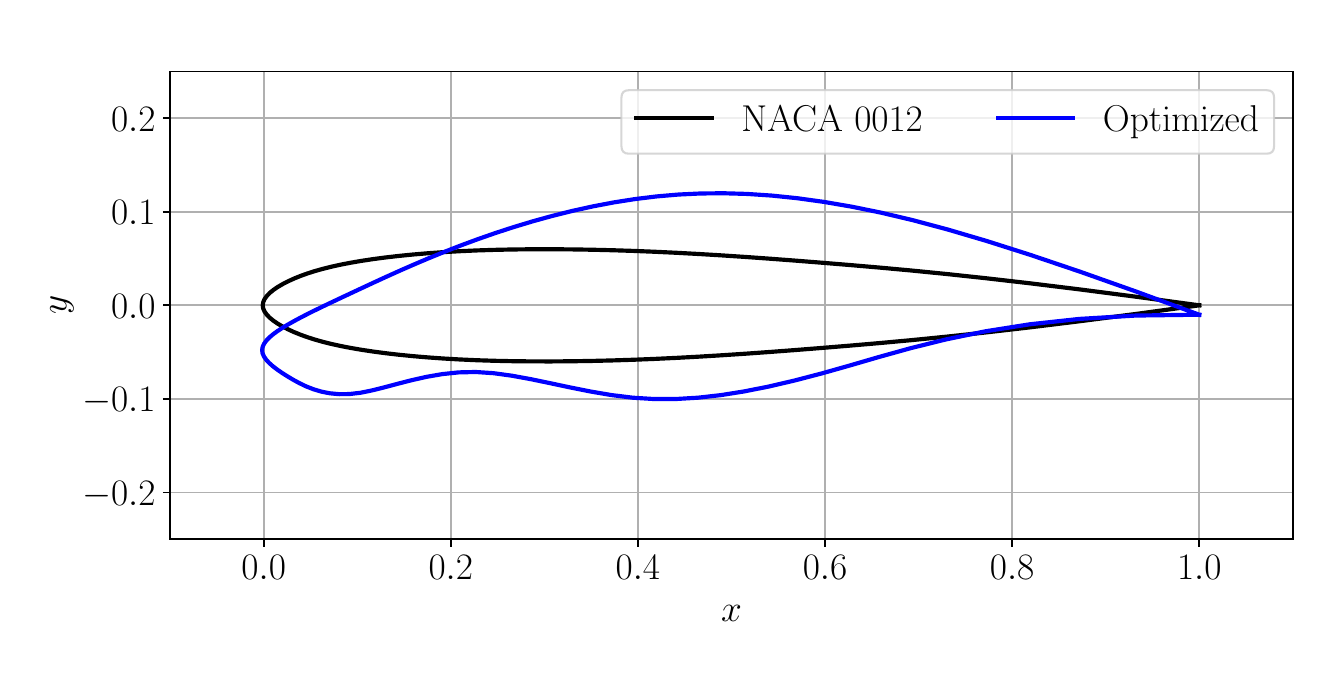}
		\caption{Test case ``C2''}
		\label{fig-shape-comb}
	\end{subfigure}
	
	\vspace{0.5em}
	\begin{subfigure}[t]{0.45\textwidth}
		\centering
		\includegraphics[width=\textwidth]{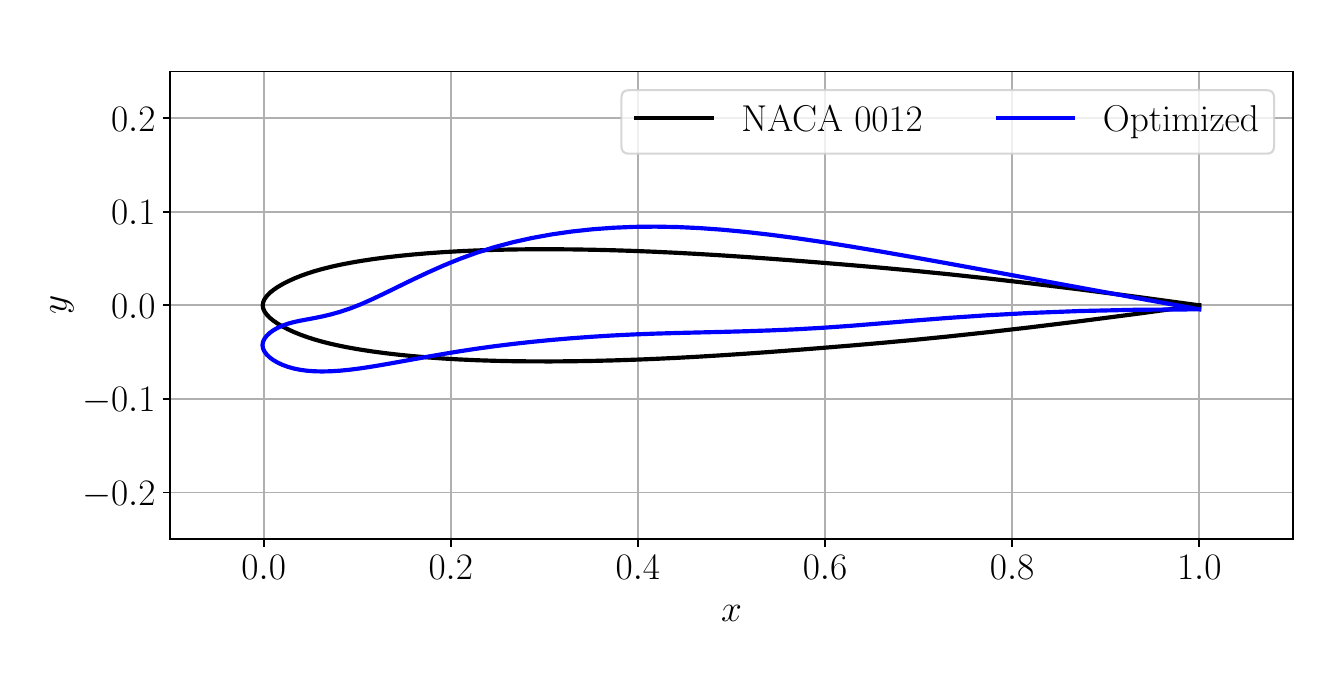}
		\caption{Test case ``C3''}
		\label{fig-shape-comc}
	\end{subfigure}
	\hfill
	\begin{subfigure}[t]{0.45\textwidth}
		\centering
		\includegraphics[width=\textwidth]{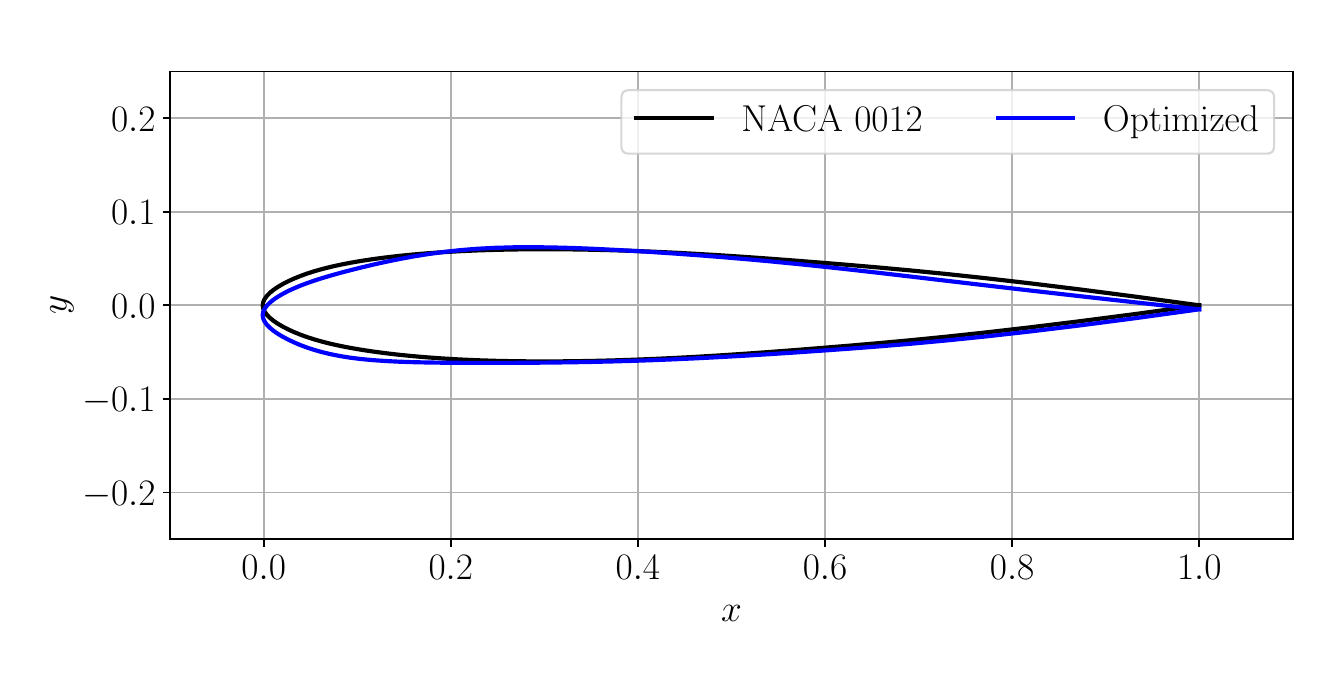}
		\caption{Test case ``C4''}
		\label{fig-shape-comd}
	\end{subfigure}
	\caption{Comparing the shape of the baseline and optimized designs.}
	\label{fig-shape-com}
\end{figure}

\begin{table}[H]
	\centering
	\begin{tabular}{|c|c|c|c|c|c|c|}
		\hline
		Case&Design&$\overline{C_{L}}$&$\overline{C_{T}}$&$\overline{C_{P}}$&$E_p$\\
		\hline
		C1&NACA 0012&1.695&0.7267&2.697&0.2694\\
		&Optimized&1.605&0.9655&2.665&0.3621\\
		\hline
		C2&NACA 0012&1.733&0.3531&4.059&0.0870\\
		&Optimized&1.666&0.6799&4.354&0.1561\\
		\hline
		C3&NACA 0012&1.765&0.7689&2.709&0.2837\\
		&Optimized&1.531&1.0545&2.653&0.3973\\
		\hline
		C4&NACA 0012&1.624&0.4488&4.192&0.1070\\
		&Optimized&1.542&0.5043&4.229&0.1192\\
		\hline
	\end{tabular}
	\caption{Comparing the propulsive performance of the flapping wing for the baseline and optimized designs.}
	\label{tab:opt-perf-imp}
\end{table}

Here, we add more complexity to these problems by considering a flapping wing in the flow field, yielding dominant chaoticity in the fluid dynamics. Therefore, fluctuations in the objective function are expected due to high uncertainties in these chaotic problems. In general, $\overline{C_{T}}$ increases progressively after each design cycle, which is the primary objective of this optimization. According to the constraint, given by the target lift coefficient, $\overline{C_{L}}$ changes significantly at the early stage of the design. Afterward, the optimizer, with the help of sensitivities, try to compensate losses in $\overline{C_{L}}$, and lead it toward $C_{L, target}$. For all cases, except ``C4", $\overline{C_{L}}$ is in an acceptable range. However, in ``C4", the objective function undergoes large fluctuations due to limited routes in design space that is difficult to be found by the optimizer. It is believed that strict constraint functions associated with high uncertainties, are responsible for this deficiency in the design procedure. However, the optimizer provides the best possible design in the presence of these constraints.

Overall results for all test cases are provided in Table \ref{tab:opt-perf-imp}. In ``C1", 32.86\% of improvement is achieved, leading to 34.4\% of increment in the resulting propulsive efficiency. In ``C3", similar performance is observed. In ``C2", $\overline{C_{T}}$ is approximately doubled, yielding 92.55\% and 79.42\% of improvements in the thrust coefficient and propulsive efficiency, respectively. Although ``C4" shows poor optimization performance than other test cases, 12.36\% and 11.4\% of increments in the thrust coefficient and propulsive efficiency are observed, respectively.

\begin{figure}[t]
	\centering
	\begin{subfigure}[t]{0.45\textwidth}
		\centering
		\includegraphics[width=\textwidth]{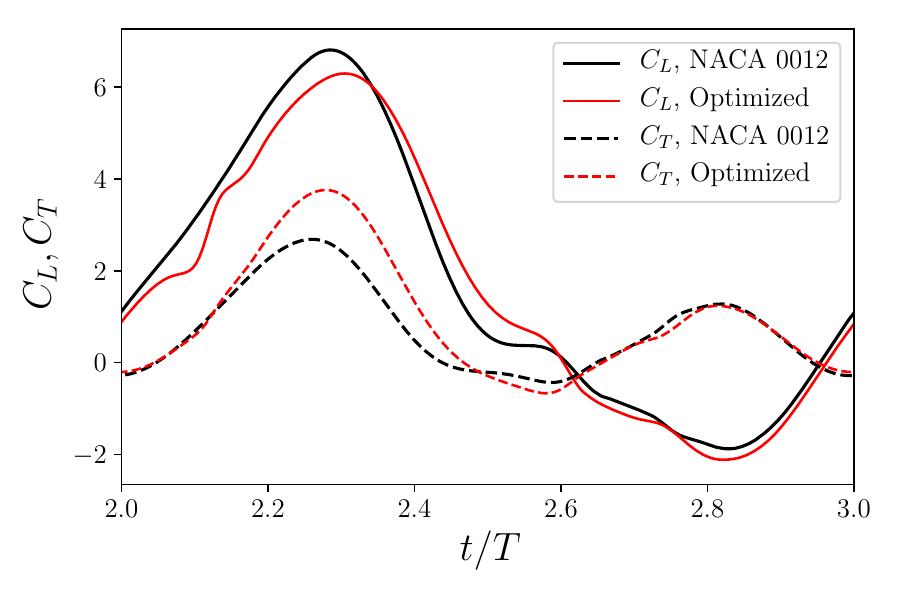}
		\caption{Test case ``C1''}
		\label{fig-inst-perf-flapping2Da}
	\end{subfigure}
	\hfill
	\begin{subfigure}[t]{0.45\textwidth}
		\centering
		\includegraphics[width=\textwidth]{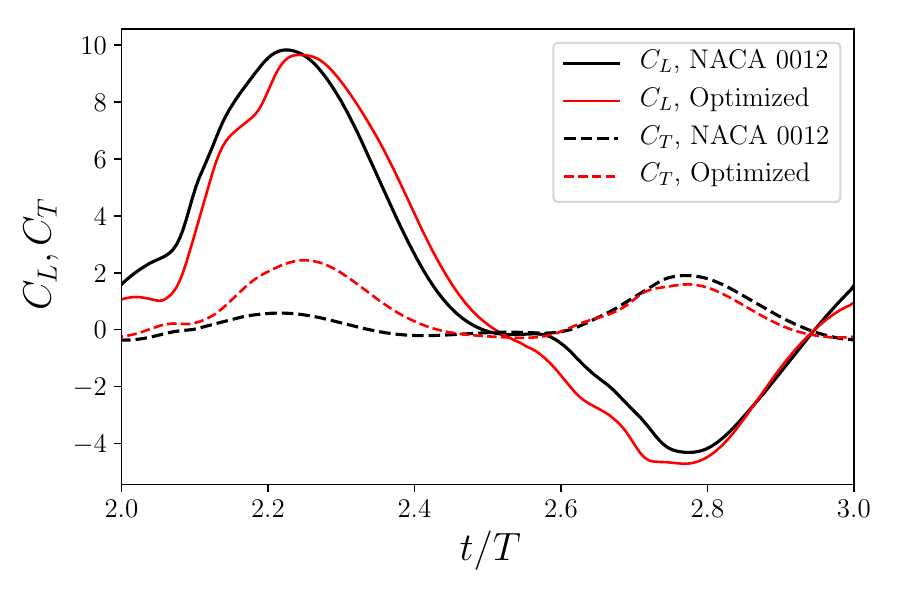}
		\caption{Test case ``C2''}
		\label{fig-inst-perf-flapping2Db}
	\end{subfigure}
	
	\vspace{0.5em}
	\begin{subfigure}[t]{0.45\textwidth}
		\centering
		\includegraphics[width=\textwidth]{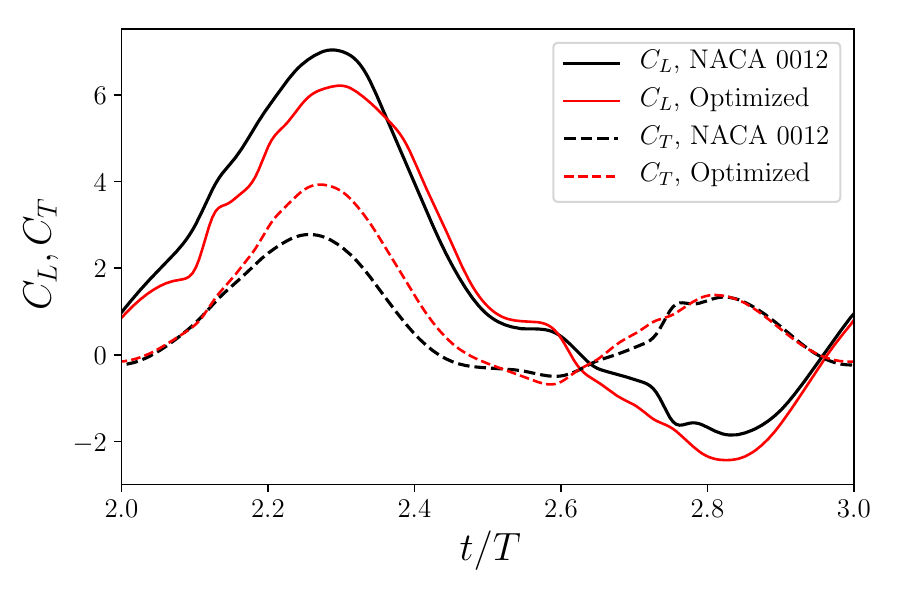}
		\caption{Test case ``C3''}
		\label{fig-inst-perf-flapping2Dc}
	\end{subfigure}
	\hfill
	\begin{subfigure}[t]{0.45\textwidth}
		\centering
		\includegraphics[width=\textwidth]{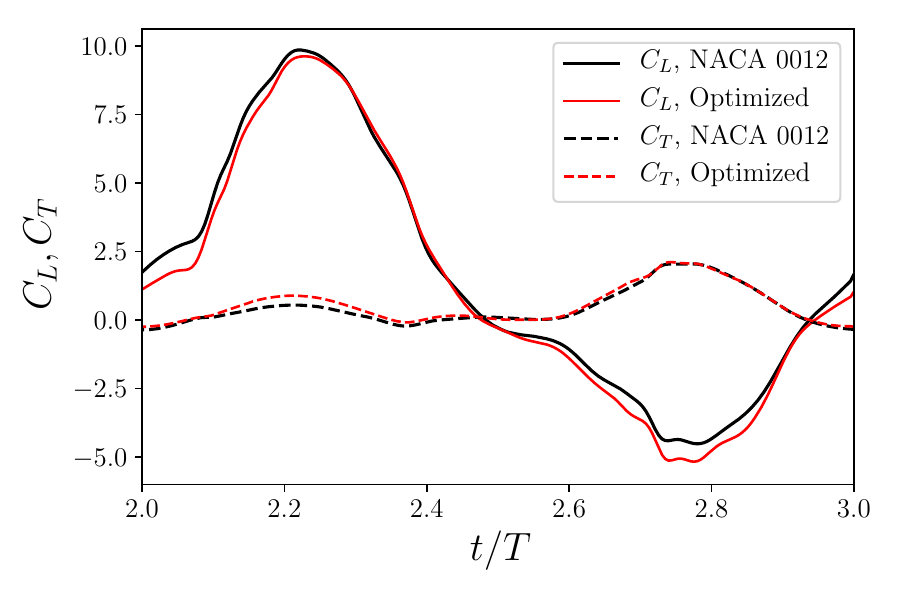}
		\caption{Test case ``C4''}
		\label{fig-inst-perf-flapping2Dd}
	\end{subfigure}
	\caption{Instantaneous lift and thrust coefficients for both baseline and optimized designs.}
	\label{fig-inst-perf-flapping2D}
\end{figure}

Figure \ref{fig-shape-com} shows the shape of both the baseline and optimized designs. A common agreement among all optimized designs is that the nose is deviated downward to produce a larger thrust coefficient. This modification decreases the effective angle of attack, and adds more camber to the wing. It also helps produce higher aerodynamic loads during the down-stroke phase. In ``C1" and ``C2" the optimized wings have more thickness approximately at $\text{x}=0.4c$. This is because there is no user-defined constraint, or bound, to limit the maximum thickness. Cambered wings create a larger Leading Edge Vortex (LEV) during both the up-stroke and down-stroke phases. The LEV produced during the up-stroke phase yields a considerable reduction in the time-averaged lift coefficient, which is against the objective of this optimization. Therefore, the lower side of the wing becomes thicker to weaken this LEV during the up-stroke phase. In ``C3" and ``C4", geometrical constraints limit the maximum thickness of the wing and, hence, the optimizer performs differently according to these new constraints.
\subsection{Flow structures}
\label{subsubsec:opt-flowstr-2d-flappingwing}
\begin{figure}[H] 
	\centering
	\includegraphics[width=0.9\textwidth]{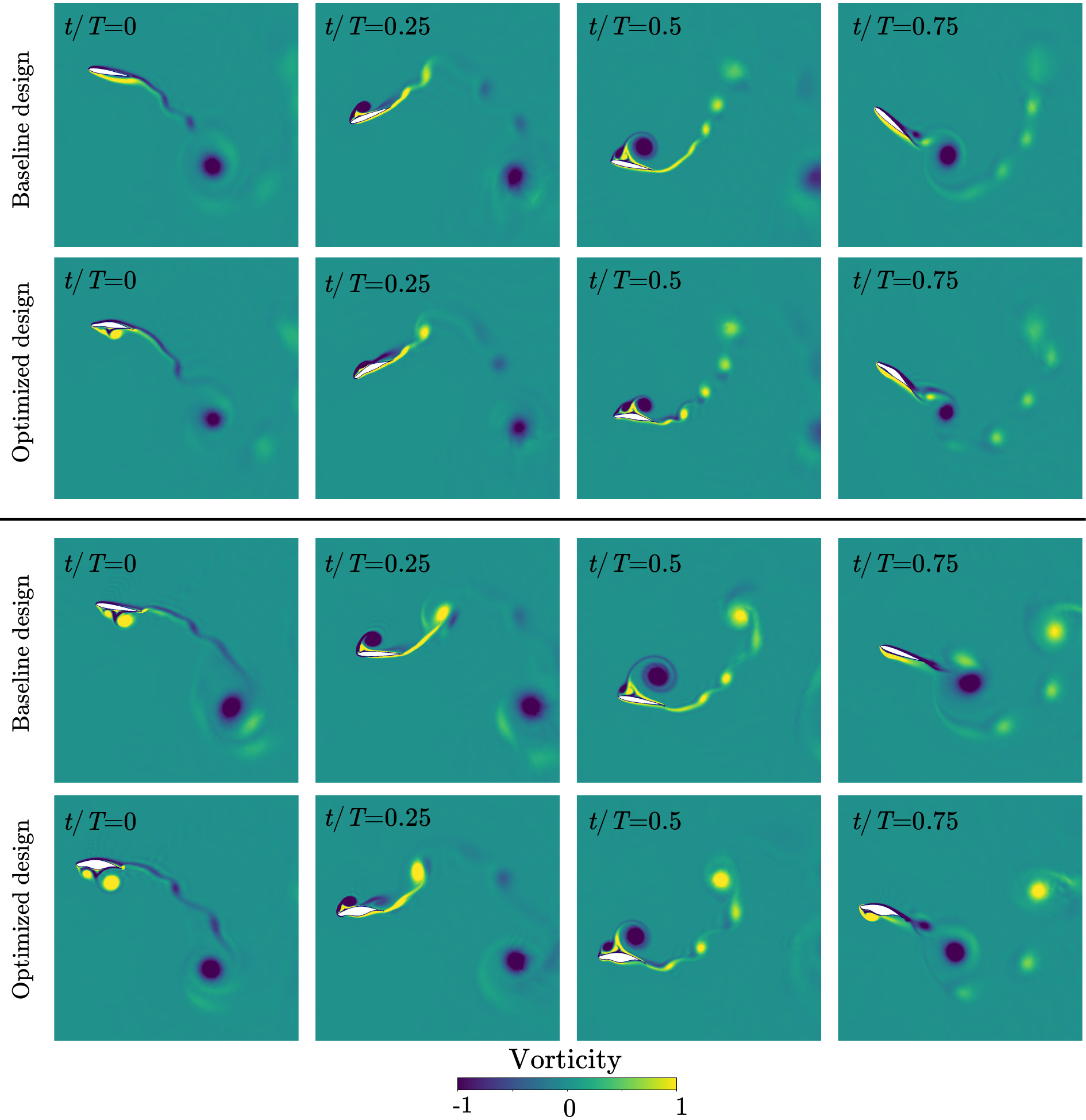}
	\caption{Vorticity contours at different instants: ``C1" (upper), and ``C2" (lower).}
	\label{fig:fsi-testcase12}
\end{figure}

\begin{figure}[H] 
	\centering
	\includegraphics[width=0.9\textwidth]{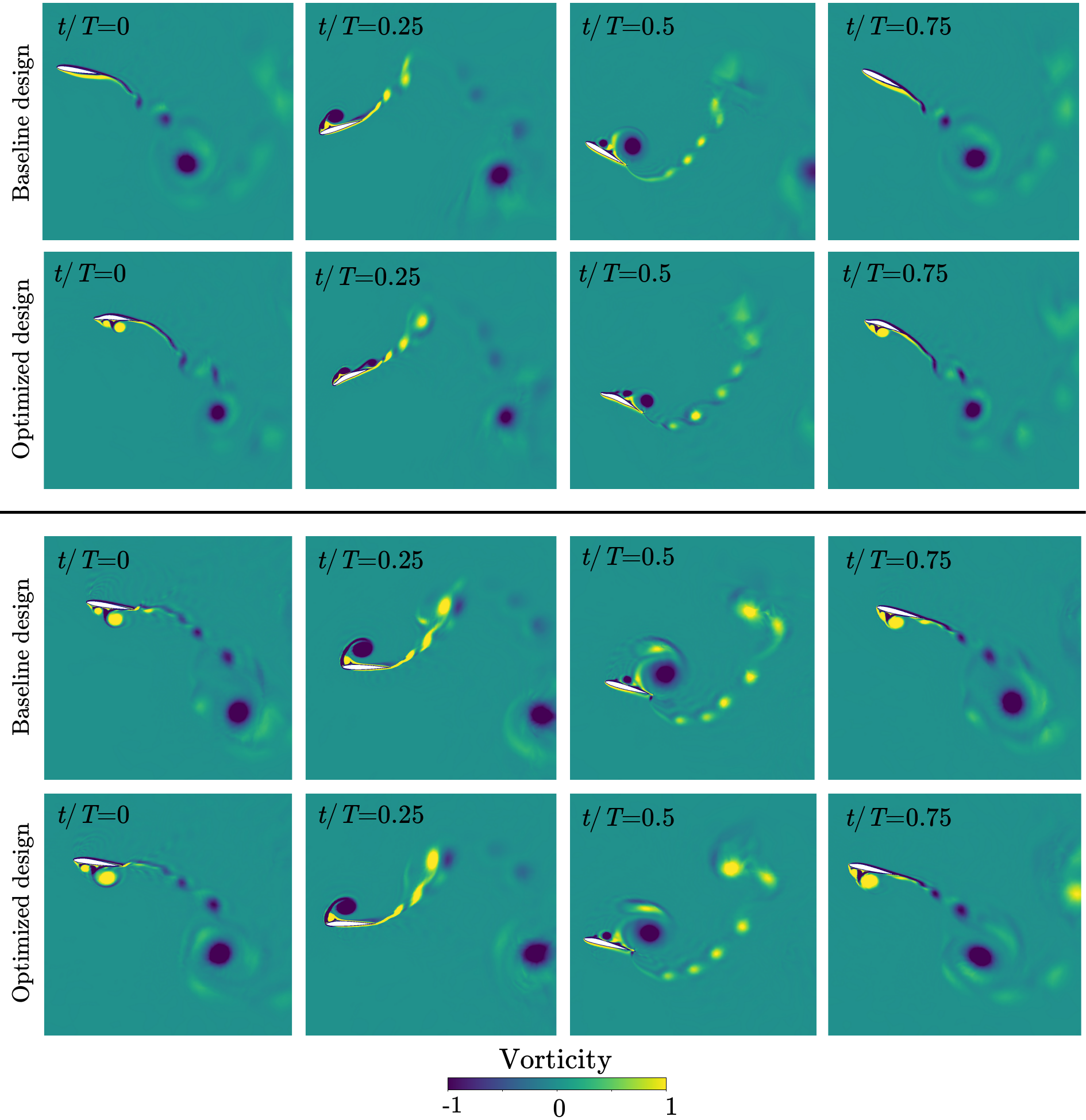}
	\caption{Vorticity contours at different instants: ``C3" (upper), and ``C4" (lower). }
	\label{fig:fsi-testcase67}
\end{figure}

Figures \ref{fig:fsi-testcase12} and \ref{fig:fsi-testcase67} show vorticity contours for different test cases. In Figure \ref{fig:fsi-testcase12} (upper), the optimized wing delays stall during the down-stroke phase. This delay increases the aerodynamic loads by taking advantage of dynamic stall. At instant $t/T^{*}=0.25$, primary LEV for the baseline design is maturely developed, while this LEV for the optimized wing is at the early stage of formation. The same situation is observed in Figure \ref{fig:fsi-testcase12} (lower). At instant $t/T^{*}=0.5$, a small set of vortices, shed from the trailing edge of the baseline wing, appears as a vortex sheet, which is often known as a reason for lift reduction during dynamic stall \cite{gharali2013dynamic, karbasian2016numerical}. However, the optimized wing breaks down this vortex sheet into a set of separated vortices, yielding larger lift.

As shown in Figure \ref{fig:fsi-testcase67} (upper), at instant $t/T^{*}=0.25$, the LEV for the baseline design is fully developed. However, the optimized wing creates two LEVs simultaneously, one at the leading edge, and the other one at approximately $\text{x}=0.5c$. These binary LEVs attach to the upper surface of the wing and, hence, delay stall, yielding higher lift production. At instant $t/T^{*}=0.5$, the baseline wing is beyond stall, and the LEV is about to pinch off. However, at the same instant, the optimized wing produces larger lift due to delayed stall. Furthermore, this optimized design breaks down the vortex sheet into smaller separated vortices. Figure \ref{fig:fsi-testcase67} (lower) presents the results of ``C4". As it is shown, the optimized wing does not change the flow field notably. Only a few minor modifications in the optimized shape are responsible for higher lift production.

Figure \ref{fig-inst-perf-flapping2D} shows the instantaneous lift and trust coefficients during one flapping cycle. An important preference in optimizing the flapping wing is that the aerodynamic loads, both the lift and thrust coefficients, should increase during the down-stroke phase, and the lift coefficient should not decrease notably during the up-stroke phase. In ``C1", the lift coefficient for the baseline and optimized wings have the same trend during one flapping cycle, as shown in Figure \ref{fig-inst-perf-flapping2Da}. However, the optimized wing increases the thrust coefficient significantly during the down-stroke phase. In this case, we can see the impact of dynamic stall on the aerodynamic loads. The same observations exist in ``C2", as shown in Figure \ref{fig-inst-perf-flapping2Db}. It is worth pointing out that the baseline design has a smaller thrust coefficient during the down-stroke phase, while the optimized wing alleviates this deficiency. Figure \ref{fig-inst-perf-flapping2Dc} also dictates the same observations in Figure \ref{fig-inst-perf-flapping2Da}. In ``C4", the optimized wing could only improve the thrust coefficient slightly, as seen from Figure \ref{fig-inst-perf-flapping2Dd}.

According to these results, we showed that shape optimization could significantly improve the aerodynamic performance of a flapping wing. Although strict constraints governed the optimization procedure, the optimizer could search for better parameters in the design space, which is augmented by uncertainties. Additionally, strong non-linearity in fluid physics makes these types of problems more challenging. In the end, we conclude that the proposed ROM-constrained optimization has a particular promise to tackle these kinds of complex and chaotic problems.

\subsection{Kinematics optimization of a 3D flapping wing}
\label{subsec:3d-flappingwing}
\subsubsection{Flow setup and validation}
\label{subsubsec:opt-design-3d-flappingwing}
In this example, an unstructured mesh with hexahedral elements for a 3D symmetric wing is used. A NACA 0020 airfoil, with the aspect ratio of $AR=0.45$, is placed $10c$ and $25c$ away from the upstream and downstream boundaries, respectively. The periodic boundary condition is also set for side-wall boundaries. The discretized Navier-Stokes equations are solved in time using a second-order BDF2 temporal scheme, with $\Delta t=\frac{1}{750}T^{*}$ as the time-step. For validation, we consider a pure plunging motion of the wing at $Re=20,000$, $\text{y}_{0}=0.5$, $k_{a}=2$, and $\theta(t)=0$. Table \ref{tab:valid3dflapp} shows the time-averaged thrust and power coefficients for the present simulations and those in Ref.~\cite{ashraf2011reynolds}. These simulations were run until four flapping cycles, and then the time-averaged values are computed over the last three cycles. The results of the present simulations are provided with 99\% of confidence intervals. As it is shown, the present results are in good agreement with those in Ref.~\cite{ashraf2011reynolds}. Note that Ref.~\cite{ashraf2011reynolds} used a laminar flow solver for a 2D NACA 0020 and, hence, some discrepancies were expected. Moreover, the resulting values of the medium and fine meshes are close to each other, indicating the medium mesh can be used for the rest of this study.
\begin{table}[h]
\centering
\begin{tabular}{|c|c|c|c|}
\hline
Test&DOF&$\overline{C_{T}}$&$\overline{C_{P}}$\\
\hline
Coarse mesh&$1.98\times10^{5}$&$0.38 \pm 0.06$&$0.85 \pm 0.07$\\ 
Medium mesh&$5.9\times10^{5}$&$0.42 \pm 0.07$&$0.84 \pm 0.09$\\ 
Fine mesh&$15.9\times10^{5}$&$0.43 \pm 0.06$&$0.84 \pm 0.06$\\ 
Ref.~\cite{ashraf2011reynolds}&$2.7\times10^{5}$&0.44&0.87\\
\hline
\end{tabular}
\caption{Comparing the time-averaged thrust and power coefficients for the present simulations with those in Ref.~\cite{ashraf2011reynolds}.}
\label{tab:valid3dflapp}
\end{table}
\subsubsection{Optimization setup}
\label{subsubsec:opt-setup-3d-flappingwing}
Kinematics optimization of a 3D NACA 0020 wing at $Re=2\times 10^{4}$, $\text{y}_{0}=0.5$, $k_{a}=2$, $\theta_{m}=0^{\circ}$, and $\theta_{s}=90^{\circ}$ is considered. This optimization helps increase the time-averaged thrust coefficient of a pure plunging motion, $\theta_{0}=0^{\circ}$, by combining it with a pitching motion, $\theta_{0} \neq 0^{\circ}$. Solving this problem is impractocal by conventional PDE-constrained optimization due to the chaoticity of flow at high Reynolds numbers. Moreover, applying conventional LSS requires 1.5TB of memory for solving an extensive system of equations using parallel algorithms. Therefore, here we show the application of the proposed approach to optimization of this large-scale chaotic problem. In this case, we define this optimization as
\begin{equation}
\begin{split}
&\underset{ \tilde{\textit{\textbf{q}}} \in \mathbb{R}^{r}, \mathcal{S} \in \mathscr{D} }{\text{minimize}} \quad \overline{\mathcal{J}} = \frac{1}{T^{*}}\int_{2T^{*}}^{3T^{*}} -\big(C_{T}\big)^{2} dt,\\
&\begin{split}
\text{subject to}&\\
&\mathcal{X}(t, \mathcal{S}),\\
&\mathcal{R}(\overline{\textit{\textbf{u}}} + \tilde{\mathbf{\Phi}} \tilde{\textit{\textbf{q}}}, t, \mathcal{X}, \mathcal{S}) = 0,\\
&\frac{d\mathcal{R}}{d\mathcal{S}}(\overline{\textit{\textbf{u}}} + \tilde{\mathbf{\Phi}} \tilde{\textit{\textbf{q}}}, t, \mathcal{X}, \mathcal{S}) = 0,\\
&\textbf{C}(\mathcal{X}, \mathcal{S}) \leq 0,\\
\end{split}
\end{split}
\label{eq-mini-flapp-3d}
\end{equation}
where the third flapping cycle is considered for evaluating the aerodynamic performance. Therefore, we start optimizing the pitching angle with a set of bounds in design space. In the sampling step, 50\% of time intervals in $\Delta \mathbb{T}$ is chosen to collect state vectors. Moreover, 5\% and 25\% of time intervals in the same time domain are selected to solve steady-state and unsteady sensitivity functions, respectively. Finally, the first eight dominant modes for each sensitivity dataset were kept to build the ROB. The same optimizer in the previous examples is used to perform optimization for ten design cycles.
\subsubsection{Optimization results}
\label{subsubsec:opt-results-3d-flappingwing}
\begin{figure}
\centering
\begin{subfigure}[h]{0.4\textwidth}
\centering
\includegraphics[width=1\textwidth]{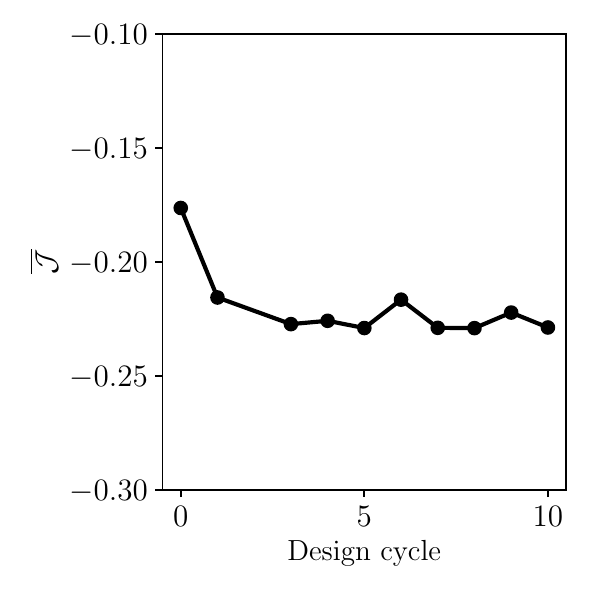}
\caption{Time-averaged objective function.}
\label{fig-perf-flapping3Da}
\end{subfigure}
\begin{subfigure}[h]{0.4\textwidth}
\centering
\includegraphics[width=1\textwidth]{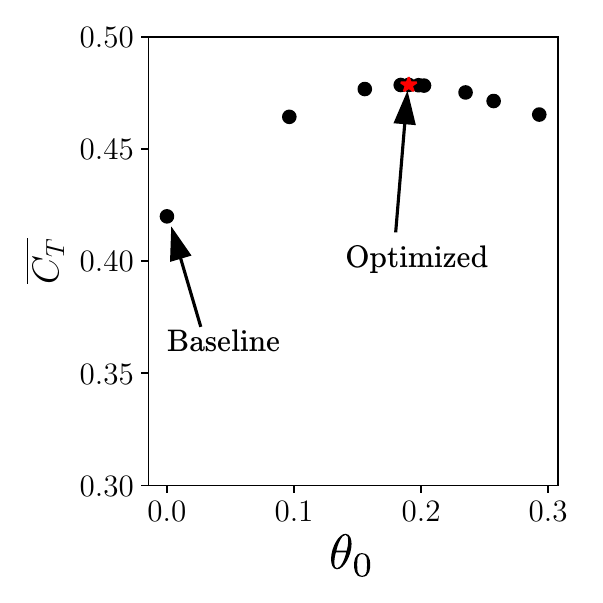}
\caption{Thrust coefficient versus pitching amplitude.}
\label{fig-perf-flapping3Db}
\end{subfigure}
\caption{Optimization results for a flapping wing: The baseline design, which has only pure plunging motion, is optimized over ten design cycle, yielding a combination of plunging and pitching motions with an optimum pitching angle.}
\label{fig-perf-flapping3D}
\end{figure}

\begin{figure}[!t]
\centering
\includegraphics[width=0.8\textwidth]{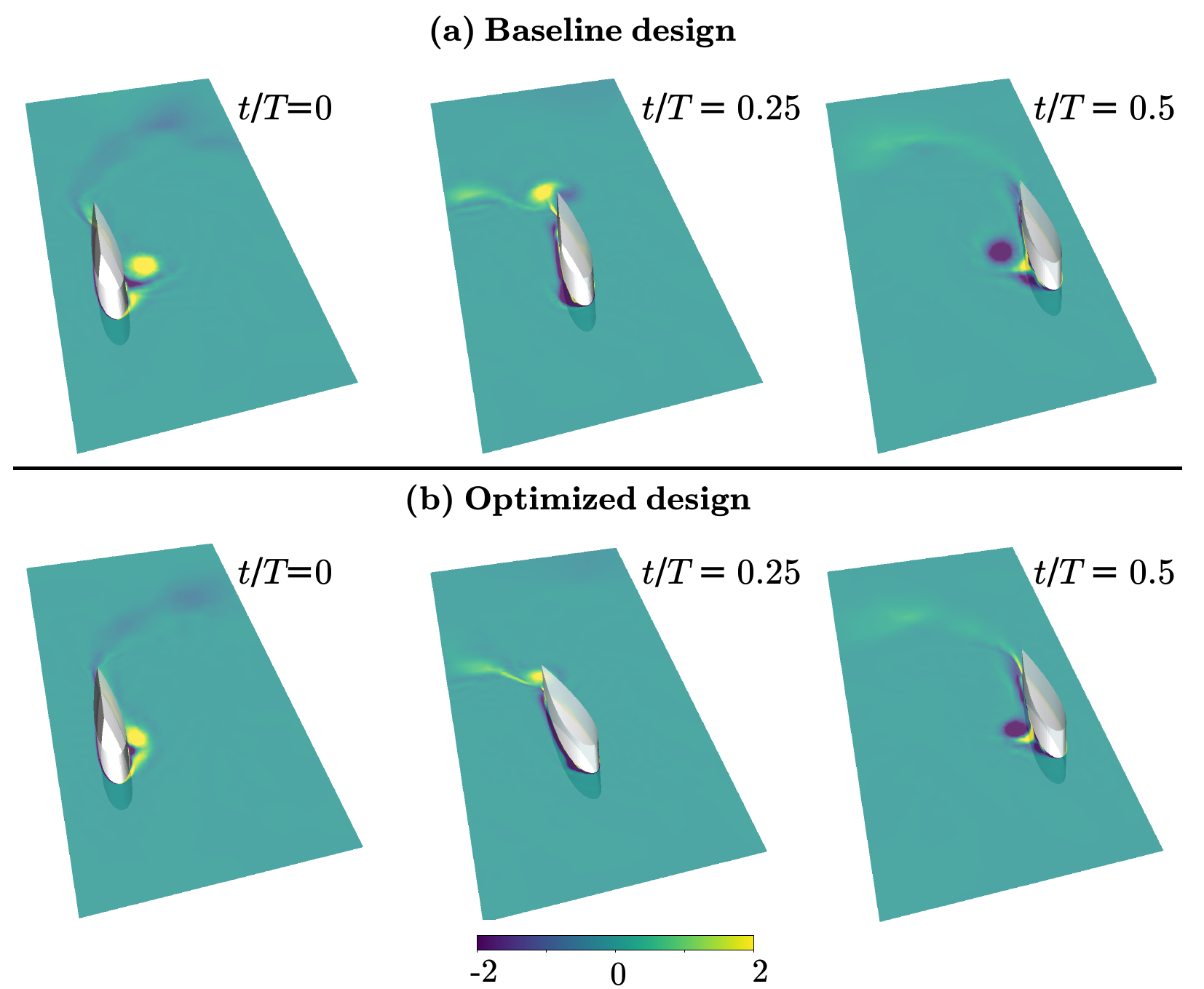}
\caption{Vorticity contours for the baseline and optimized designs.}
\label{fig-flapp-3d-contours1}
\end{figure}

Figure \ref{fig-perf-flapping3D} shows the optimization progress for ten design cycles. In Figure \ref{fig-perf-flapping3Da}, the time-averaged objective function decreases monotonically at the early stage, and then it starts fluctuating once the design parameter reaches around the optimum point. Since the sensitivity of the time-averaged objective function near this point is small, high uncertainty agitates this sensitivity value, yielding fluctuations in the objective function, or some poor approximation in design space. Figure \ref{fig-perf-flapping3Db} shows the time-averaged thrust coefficient versus the pitching angle, $\theta_{0}$. The baseline design starts with pure plunging motion, $\theta_{0}=0^{\circ}$, and finally, the optimized kinematics is obtained by $\theta_{0}=10.54^{\circ}$. This optimized kinematics elevates the time-averaged thrust coefficient by 14\%, which is noted by a red star at the peak.
It is worth mentioning that sensitivity analysis using the present approach only requires 4MB of memory, which is achievable on a single processor. Figure \ref{fig-flapp-3d-contours1} displays vorticity contours on a plane passing through the middle of the wing. It is shown that the optimized design postpones stall, which leads to higher lift compared to the baseline design.

In the end, we conclude that the proposed approach can be applied to unsteady optimization of larger, chaotic, and more complicated fluid dynamics problems. Besides the performance, and robustness of the present approach, it reduces memory requirements significantly, making a particular promise for future engineering design and optimization. Although the examples considered in this study are simple canonical problems, our intention was to used them to demonstrate the feasibility of the present approach to optimization. We will advance the proposed approach to consider it for larger problems with more complicated flows in future work. 
\section{Conclusions}
\label{sec:conclusions}

The present approach, which reduces large-scale chaotic problems into a ROM-constrained optimization framework, makes Least Squares Shadowing sensitivity analysis affordable for unsteady optimization of chaotic flows. Results showed that the proposed approach has a promising potential to be applied to the optimization of complex systems.
 
Here, we performed the shape optimization of the flapping wing, where the wing's motion introduced more complexity into the fluid physics. It was shown that the optimized shape obtained by the proposed ROM-constrained optimization framework increases the time-averaged thrust coefficient and propulsive efficiency significantly, by 12.4--92.6\% and 11.4--79.4\%, respectively, across the four test cases. In the kinematics optimization of a 3D flapping wing at $Re=20{,}000$, adding an optimized pitching motion to the pure plunging motion increased the time-averaged thrust coefficient by 14.0\%, while the memory required for sensitivity analysis was reduced from about 1.5\,TB for conventional LSS to about 4\,MB. The accuracy of the computed sensitivities depends on the data used to build the reduced basis, and the objective functions were averaged over a single flapping cycle, which introduces statistical noise into the optimization. Moreover, the present demonstrations are limited to 2D and spanwise-periodic 3D wings at moderate Reynolds numbers. Applying the approach to aerodynamic optimization using high-fidelity simulation techniques, such as LES, at higher Reynolds numbers is the next step of this work. It is expected that this novel approach can provide more efficient and better aerodynamic designs in the aviation industry.


\bibliography{refs}
\end{document}